\documentclass[a4paper,11pt]{article}

\usepackage{jcappub}
\allowdisplaybreaks
\usepackage{graphicx}
\usepackage{epstopdf}

\newcommand{\be}{\begin{equation}}
\newcommand{\ee}{\end{equation}}
\newcommand{\bea}{\begin{eqnarray}}
\newcommand{\eea}{\end{eqnarray}}

\def\nl{\nonumber \\ &}

\title{Complete conservative dynamics for inspiralling compact binaries with spins 
at the fourth post-Newtonian order}

\author[a,b]{Mich\`ele Levi}
\author[c,d]{and Jan Steinhoff}

\affiliation[a]{Sorbonne Universit\'es, 
Universit\'e Pierre et Marie Curie-Paris VI, CNRS-UMR 7095, \\
Institut d'astrophysique de Paris, 
98 bis Boulevard Arago, 75014 Paris, France} 
\affiliation[b]{Sorbonne Universit\'es, Institut Lagrange de Paris, \\ 
98 bis Boulevard Arago, 75014 Paris, France} 

\affiliation[c]{Max-Planck-Institute for Gravitational Physics 
(Albert-Einstein-Institute),\\ Am M{\"u}hlenberg 1, 14476 Potsdam-Golm, Germany}
\affiliation[d]{Centro Multidisciplinar de Astrofisica, Instituto Superior Tecnico, 
Universidade de Lisboa,\\ Avenida Rovisco Pais 1, 1049-001 Lisboa, Portugal}

\emailAdd{michele.levi@upmc.fr}
\emailAdd{jan.steinhoff@aei.mpg.de}

\abstract{In this work we complete the spin-dependent conservative 
dynamics of inspiralling compact binaries at the fourth post-Newtonian 
order, and in particular the derivation of the 
next-to-next-to-leading order spin-squared interaction potential. We 
derive the physical equations of motion of the position and the spin 
from a direct variation of the action. Further, we derive the 
quadratic-in-spin Hamiltonians, as well as their expressions in the 
center-of-mass frame. We construct the conserved integrals of motion, 
which form the Poincar\'e algebra. This construction provided a 
consistency check for the validity of our result, which is crucial in 
particular in the current absence of another independent derivation of 
the next-to-next-to-leading order spin-squared interaction. Finally, we 
provide here the complete gauge-invariant relations among the binding 
energy, angular momentum, and orbital frequency of an inspiralling 
binary with generic compact spinning components to the fourth 
post-Newtonian order. These high post-Newtonian orders, in particular 
taking into account the spins of the binary constituents, will enable to 
gain more accurate information on the constituents from even more 
sensitive gravitational-wave detections to come.}

\begin{document}

\maketitle

\flushbottom

\section{Introduction}

The first detection of gravitational waves (GWs) from the binary black hole merger event 
GW150914 \cite{Abbott:2016blz} by the twin Advanced LIGO detectors in the US 
\cite{LIGOScientific:2014pky} has opened a new era of observational astronomy.
By now the Advanced LIGO detectors were joined by the Advanced Virgo detector in Europe 
\cite{VIRGO:2014yos} (and more recently also by the KAGRA detector in Japan 
\cite{KAGRA:2020tym}), and have reached a sensitivity allowing for frequent GW detections 
\cite{LIGOScientific:2018mvr,LIGOScientific:2020ibl}. These also include observations of 
neutron stars as components in the binaries, either as neutron-star binaries 
\cite{LIGOScientific:2017vwq}, or in neutron star-black hole binaries 
\cite{LIGOScientific:2021qlt}.
Using the matched filtering 
technique to detect the GW signal, and gain from it as much information as 
possible, requires accurate theoretical waveforms, where the continuous 
signal is modeled via the Effective-One-Body (EOB) approach 
\cite{Buonanno:1998gg}. The initial part of the waveform corresponding 
to the inspiral phase of the binaries' evolution is analytically 
described by the post-Newtonian (PN) approximation of General Relativity 
\cite{Blanchet:2013haa, TheLIGOScientific:2016src}.

In order to enable an improved analysis of the GW events, and 
consequently have an improved parameter estimation 
\cite{TheLIGOScientific:2016wfe, Abbott:2016izl}, i.e.~to gain more 
accurate information about the inner structure of the constituents, high 
order PN corrections are required, taking into account in particular the 
spin of the objects. 
Considering the completion of the 4PN order point-mass correction 
\cite{Damour:2014jta, Bernard:2015njp,Damour:2016abl}, a series of works 
\cite{Levi:2011eq, Levi:2014sba,Levi:2014gsa,Levi:2015msa,Levi:2015uxa,Levi:2015ixa} 
based on an effective field theory (EFT) approach for the binary inspiral 
\cite{Goldberger:2004jt,Goldberger:2007hy} has completed the same PN 
accuracy for binaries with generic compact spinning objects. This line 
of work includes all spin interactions linear- and quadratic-in-spin up 
to the next-to-leading order (NLO) via an EFT for spinning objects, 
which was formulated in \cite{Levi:2015msa}, and further new results 
were obtained up to the next-to-NLO (NNLO), as well as at the leading order (LO) 
cubic- and quartic-in-spin interactions, all via the EFT for spin \cite{Levi:2015msa}.

In this paper we complete this series of works to the 4PN order, and in 
particular the derivation of the NNLO spin-squared interaction 
potential \cite{Levi:2015ixa}. First, we note the physical equations 
of motion (EOMs) of the position and the spin, which are both obtained 
directly via the action approach \cite{Levi:2014sba,Levi:2015msa}. Then 
we derive the Hamiltonian following the procedure outlined in previous 
works \cite{Levi:2014sba,Levi:2015uxa}, and specify it also for the 
center-of-mass frame. 
We provide here also an equivalent potential and Hamiltonian to those of 
\cite{Levi:2011eq,Levi:2014sba} for the NNLO spin1-spin2 sector for a 
total result consistent with the formulation and gauge choices of 
\cite{Levi:2015msa} (which are also different than in  
\cite{Levi:2008nh,Levi:2010zu}). We proceed to construct the conserved 
integrals of motion, which form the Poincar\'e algebra. As there is 
currently no other independent derivation of the NNLO spin-squared 
sector, this construction actually provides a crucial consistency check 
of our result. Finally, we provide complete gauge-invariant relations 
among the binding energy, angular momentum, and orbital frequency of an 
inspiralling binary with generic compact spinning components to the 4PN 
order.

The paper is organized as follows. In section \ref{eomxs} 
we note the equations of motion of the position and the spin. In 
section \ref{ham}, we provide the NNLO quadratic-in-spin 
Hamiltonians consistent with the gauge choices in \cite{Levi:2015msa}, 
and also specify them for the center-of-mass frame. In section 
\ref{poincare} we find the conserved integrals of motion, which 
constitute a strong check of our potential and Hamiltonian resulting from 
\cite{Levi:2015ixa}. In section \ref{GI} we complete the gauge-invariant 
relations among the binding energy, angular momentum, and orbital 
frequency of an inspiralling binary with generic compact spinning 
components to 4PN order, and we conclude in section \ref{theendmyfriend}. 
In addition in appendix \ref{nnlos1s2} we provide 
the NNLO spin1-spin2 potential with the gauge choices of \cite{Levi:2015msa}, and show 
its equivalence to the Hamiltonians provided in \cite{Hartung:2011ea,Levi:2014sba}.

\section{EOMs of position and spin} 
\label{eomxs}

As noted in \cite{Levi:2014sba,Levi:2015msa} the derivation of the 
EOMs of the positions is straightforward using a variation of the action, 
which contains higher-order time derivatives of positions up to $\dot{a}$. 
We recall then that the spin-dependent EOMs are obtained from the 
spin-dependent effective action, given in the form:
\be
S_{\text{eff(spin)}}=\int dt\left[-\frac{1}{2}\sum_{I=1}^2 S_{Iij}
\Omega_I^{ij}-V\left(\vec{x}_I,\dot{\vec{x}}_I, \ddot{\vec{x}}_I, 
\dots, S_{Iij}, \dot{S}_{Iij}, \dots \right)\right],
\ee
where $V$ is the evaluated interaction potential, which complements 
the 3-dimensional physical kinematic terms of the 2 individual objects in 
the binary, given in terms of the physical spin components $S_{ij}$, namely 
the spatial components of the local spin variables in the generalized 
canonical gauge, defined in \cite{Levi:2015msa}.
Hence for the equations of motion of the position, we just make a variation 
with respect to the position for a generalized action with higher-order time 
derivative of the position, as rigorously shown in \cite{Levi:2014sba}:
\begin{align}
\frac{\delta {\cal{S}}(x_i,v_i,a_i,\dot{a}_i,S_{ij},\dot{S}_{ij},
\ddot{S}_{ij})}{\delta x_i}=-m_i a_i-\left(\frac{\partial V}{\partial x_i} 
-\frac{d}{dt}\frac{\partial V}{\partial v_i}
+\frac{d^2}{dt^2}\frac{\partial V}{\partial a_i}
-\frac{d^3}{dt^3}\frac{\partial V}{\partial \dot{a}_i}\right)=0.
\end{align}


The EOMs of the spin are also directly obtained within the EFT of spin 
\cite{Levi:2015msa} in a simple form via a variation of the action. It should 
be highlighted that this is accomplished via an independent variation of the 
action with respect to the spin, and to its conjugate variables, the rotation 
matrices, as rigorously explained uniquely in \cite{Levi:2014sba}. 
Apart from time derivatives of the squared spin-length, which can be
dropped at this stage, the potential contains higher order time 
derivatives of the spin up to $\ddot{S}_i$. Combining the 2 variations of 
the conjugate variables, we then obtain \cite{Levi:2014sba}:  
\begin{align}\label{SEOM}
\dot{S}^{ij} = - 4 S^{k[i} \delta^{j]l} \frac{\delta\int{dt\,V}}{\delta S^{kl}}  
= - 4 S^{k[i} \delta^{j]l} \left[ \frac{\partial V}{\partial S^{kl}} 
- \frac{d}{dt} \frac{\partial V}{\partial \dot{S}^{kl}} 
+ \frac{d^2}{dt^2} \frac{\partial V}{\partial \ddot{S}^{kl}}\right]. 
\end{align}

The public repository \texttt{EFTofPNG} \cite{Levi:2017kzq} contains the final 
total results of the effective action in the various sectors complete up to 4PN 
order (readily reusable in machine-readable files).

\section{Hamiltonian}
\label{ham}

Although it is beneficial to have the potential, given in terms of 
the widely-used harmonic coordinates, an order-reduction at the 
level of the EOMs is required in order to numerically integrate the 
dynamics. In this section we perform an order-reduction at the level
of the action, which enables to transform to a Hamiltonian. The 
Hamiltonian has the advantage that it leads directly to the first-order 
Hamilton's equations. Furthermore, the Hamiltonian is most useful
for implementation in the Effective-One-Body model for GWs from 
compact binary coalescence.

The spin-dependent Hamiltonian and its center-of-mass expression up to 
the 3.5PN order can be found in \cite{Levi:2014sba, Levi:2015uxa}. In 
order to complete the 4PN order Hamiltonian, we need to add the NNLO 
spin1-spin2, NNLO spin-squared, and LO quartic-in-spin Hamiltonians. 
The latter can be found in \cite{Levi:2014gsa}. For the NNLO 
spin-squared part we start with the potential recently derived
in \cite{Levi:2015ixa} using the formulation developed in 
\cite{Levi:2015msa}.
For the NNLO spin1-spin2 part we use the potential given in 
appendix~\ref{nnlos1s2}, which we derived with the formulation and 
gauge choices in \cite{Levi:2015msa} in order to have a consistent 
expression for the spin-dependent Hamiltonian to 4PN order. The NNLO 
spin1-spin2 potentials and Hamiltonians in \cite{Levi:2011eq, 
Hartung:2011ea, Levi:2014sba} are canonically equivalent.

For the NNLO Hamiltonians we follow the computational steps 
outlined in our previous work \cite{Levi:2015msa, Levi:2015uxa}. The 
first step is a reduction of higher-order time derivatives
at the level of the action through suitable variable transformations 
\cite{Damour:1990jh}, with its extension to the spinning case found in 
\cite{Levi:2014sba}.
We are eliminating the higher-order time derivatives successively at 
each PN order.
This reduction is generally not equivalent to a substitution of lower-order EOM, 
which is crucial for the nonlinear-in-spin sectors.
That is, the transformation of the position required at the LO spin-orbit sector
\cite{Tulczyjew:1959, Damour:1982, Damour:1988mr, Levi:2015msa}, given by
\begin{equation} \label{positionshift}
\vec{y}_1 \rightarrow \vec{y}_1 + \frac{1}{2 m_1}\vec{S}_1\times\vec{v}_1 ,
\end{equation}
adds nonlinear contributions.
Thus, all higher-order time derivatives can be removed by performing to quadratic order 
the above redefinition of positions in the action, followed by an iterative 
insertion of lower-order equations of motion into the potential.
The latter is equivalent to variable redefinitions that are taken only to linear order 
\cite{Damour:1990jh,Levi:2014sba}. Furthermore, the redefinitions of variables are also 
available as separate files in the public repository \texttt{EFTofPNG}~\cite{Levi:2017kzq}.

We recall that after the higher-order time derivatives of the position and spin variables 
have been systematically and fully eliminated \cite{Levi:2014sba,Levi:2015msa}, the spin 
variables satisfy the standard Poisson brackets for a 3-dimensional spin tensor, e.g.~for 
spin 1:
\begin{equation}
\{ S_1^{ij}, S_1^{kl} \} = S_1^{ik} \delta_{jl} - S_1^{il} \delta_{jk}
        + S_1^{jl} \delta_{ik} - S_1^{jk} \delta_{il},
\end{equation}
so that the we can rewrite the EOM in eq.~\eqref{SEOM} in the form
\begin{equation}
\dot{S}_1^{ij} = \{ S_1^{ij}, -V_s \},
\end{equation}
where $V_s$ denotes the standard potential, i.e.~without any higher-order time derivatives.

Further, we recall that $\vec{n}\equiv \vec{r}/r$, where 
$\vec{r}\equiv \vec{y}_1 - \vec{y}_2$, with $\vec{y}_1$, $\vec{y}_2$, the positions of 
objects 1 and 2, respectively, and $r\equiv \sqrt{\vec{r}^2}$. We also recall that the 
canonical momentum of object $1$ is defined as usual by
\begin{equation}\label{defpcan}
\vec{p}_1 = \frac{\partial L}{\partial \vec{v}_1} = m_1 \vec{v}_1 - \frac{\partial V_s} 
{\partial \vec{v}_1},
\end{equation} 
where $\vec{v}_1 \equiv \dot{\vec{y}}_1$. 

The next step is a standard Legendre transformation, with the relation 
between velocity and canonical momentum given by
\begin{align}
v_1^i =& \dots
- \frac{3 G}{2 m_2 r^3} \bigg[ 2 S_2^{i} \vec{n}\cdot\vec{\hat{p}}_2 \vec{n}\cdot\vec{S}_2
	 - S_2^{i} \vec{\hat{p}}_2\cdot\vec{S}_2
	 + n^{i} \vec{n}\cdot\vec{S}_2 \vec{\hat{p}}_2\cdot\vec{S}_2
	 + \hat{p}_2^{i} S_2^2
	 - n^{i} \vec{n}\cdot\vec{\hat{p}}_2 S_2^2 \nl
	 - 2 \hat{p}_2^{i} ( \vec{n}\cdot\vec{S}_2 )^{2} \bigg]
+ \frac{C_{2ES^2} G}{4 m_2 r^3} 
\bigg[ 6 S_2^{i} \vec{n}\cdot\vec{\hat{p}}_2 \vec{n}\cdot\vec{S}_2
	 - 2 S_2^{i} \vec{\hat{p}}_2\cdot\vec{S}_2
	 + 6 n^{i} \vec{n}\cdot\vec{S}_2 \vec{\hat{p}}_2\cdot\vec{S}_2
	 - 6 \hat{p}_1^{i} S_2^2 \nl
	 + 9 \hat{p}_2^{i} S_2^2
	 - 3 n^{i} \vec{n}\cdot\vec{\hat{p}}_2 S_2^2
	 + 18 \hat{p}_1^{i} ( \vec{n}\cdot\vec{S}_2 )^{2}
	 - 21 \hat{p}_2^{i} ( \vec{n}\cdot\vec{S}_2 )^{2}
	 - 15 n^{i} \vec{n}\cdot\vec{\hat{p}}_2 ( \vec{n}\cdot\vec{S}_2 )^{2} \bigg] \nl
+ \frac{G}{4 m_1 r^3} \bigg[ -24 S_2^{i} \vec{n}\cdot\vec{\hat{p}}_1 \vec{n}\cdot\vec{S}_1
	 + 21 S_2^{i} \vec{n}\cdot\vec{\hat{p}}_2 \vec{n}\cdot\vec{S}_1
	 + 10 S_2^{i} \vec{\hat{p}}_1\cdot\vec{S}_1
	 - 12 S_2^{i} \vec{\hat{p}}_2\cdot\vec{S}_1 \nl
	 - 6 S_1^{i} \vec{n}\cdot\vec{\hat{p}}_1 \vec{n}\cdot\vec{S}_2
	 + 18 S_1^{i} \vec{n}\cdot\vec{\hat{p}}_2 \vec{n}\cdot\vec{S}_2
	 + 12 \hat{p}_1^{i} \vec{n}\cdot\vec{S}_1 \vec{n}\cdot\vec{S}_2
	 - 21 \hat{p}_2^{i} \vec{n}\cdot\vec{S}_1 \vec{n}\cdot\vec{S}_2 \nl
	 - 30 n^{i} \vec{n}\cdot\vec{\hat{p}}_2 \vec{n}\cdot\vec{S}_1 \vec{n}\cdot\vec{S}_2
	 - 6 n^{i} \vec{\hat{p}}_1\cdot\vec{S}_1 \vec{n}\cdot\vec{S}_2
	 + 21 n^{i} \vec{\hat{p}}_2\cdot\vec{S}_1 \vec{n}\cdot\vec{S}_2
	 + 10 S_1^{i} \vec{\hat{p}}_1\cdot\vec{S}_2 \nl
	 - 24 n^{i} \vec{n}\cdot\vec{S}_1 \vec{\hat{p}}_1\cdot\vec{S}_2
	 - 10 S_1^{i} \vec{\hat{p}}_2\cdot\vec{S}_2
	 + 18 n^{i} \vec{n}\cdot\vec{S}_1 \vec{\hat{p}}_2\cdot\vec{S}_2
	 - 20 \hat{p}_1^{i} \vec{S}_1\cdot\vec{S}_2
	 + 24 \hat{p}_2^{i} \vec{S}_1\cdot\vec{S}_2 \nl
	 + 48 n^{i} \vec{n}\cdot\vec{\hat{p}}_1 \vec{S}_1\cdot\vec{S}_2
	 - 45 n^{i} \vec{n}\cdot\vec{\hat{p}}_2 \vec{S}_1\cdot\vec{S}_2 \bigg]
+ \frac{G m_2}{4 m_1^2 r^3} 
\bigg[ 15 S_1^{i} \vec{n}\cdot\vec{\hat{p}}_1 \vec{n}\cdot\vec{S}_1
	 - 6 S_1^{i} \vec{n}\cdot\vec{\hat{p}}_2 \vec{n}\cdot\vec{S}_1 \nl
	 - 10 S_1^{i} \vec{\hat{p}}_1\cdot\vec{S}_1
	 + 15 n^{i} \vec{n}\cdot\vec{S}_1 \vec{\hat{p}}_1\cdot\vec{S}_1
	 + 6 S_1^{i} \vec{\hat{p}}_2\cdot\vec{S}_1
	 - 12 n^{i} \vec{n}\cdot\vec{S}_1 \vec{\hat{p}}_2\cdot\vec{S}_1
	 + 10 \hat{p}_1^{i} S_1^2
	 - 6 \hat{p}_2^{i} S_1^2 \nl
	 - 9 n^{i} \vec{n}\cdot\vec{\hat{p}}_1 S_1^2
	 + 6 n^{i} \vec{n}\cdot\vec{\hat{p}}_2 S_1^2
	 - 21 \hat{p}_1^{i} ( \vec{n}\cdot\vec{S}_1 )^{2}
	 + 12 \hat{p}_2^{i} ( \vec{n}\cdot\vec{S}_1 )^{2} \bigg] \nl
+ \frac{C_{1ES^2} G m_2}{4 m_1^2 r^3} 
\bigg[ -6 S_1^{i} \vec{n}\cdot\vec{\hat{p}}_1 \vec{n}\cdot\vec{S}_1
	 + 6 S_1^{i} \vec{n}\cdot\vec{\hat{p}}_2 \vec{n}\cdot\vec{S}_1
	 + 4 S_1^{i} \vec{\hat{p}}_1\cdot\vec{S}_1
	 - 6 n^{i} \vec{n}\cdot\vec{S}_1 \vec{\hat{p}}_1\cdot\vec{S}_1 \nl
	 - 2 S_1^{i} \vec{\hat{p}}_2\cdot\vec{S}_1
	 + 6 n^{i} \vec{n}\cdot\vec{S}_1 \vec{\hat{p}}_2\cdot\vec{S}_1
	 - 10 \hat{p}_1^{i} S_1^2
	 + 9 \hat{p}_2^{i} S_1^2
	 + 12 n^{i} \vec{n}\cdot\vec{\hat{p}}_1 S_1^2
	 - 3 n^{i} \vec{n}\cdot\vec{\hat{p}}_2 S_1^2 \nl
	 + 18 \hat{p}_1^{i} ( \vec{n}\cdot\vec{S}_1 )^{2}
	 - 21 \hat{p}_2^{i} ( \vec{n}\cdot\vec{S}_1 )^{2}
	 - 15 n^{i} \vec{n}\cdot\vec{\hat{p}}_2 ( \vec{n}\cdot\vec{S}_1 )^{2} \bigg] ,
\end{align}
where we use the abbreviation $\vec{\hat{p}}_a\equiv\vec{p}_a/m_a$, and the dots denote the 
terms up to 3.5PN order given in eq.~(4.10) of \cite{Levi:2015uxa}.

We remind that the unfixed coefficients $C_{aES^{2n}}$ and $C_{aBS^{2n+1}}$ are the Wilson 
coefficients - from our effective action - of the non-minimal coupling terms, that are linear 
in the curvature, and are even and odd in the power of the spin, and coupled to the even- 
and odd-parity components of the curvature, $E$ and $B$, respectively. These correspond to 
the multipolar deformation constants, familiar in the context of classical gravity, that are 
due to the spin-induced higher multipoles of the objects which give rise to deformations of 
the compact object.

The NNLO spin1-spin2 Hamiltonian now reads
\begin{align}
&H^{\text{NNLO}}_{\text{S$_1$S$_2$}} =
- \frac{G}{16 r^3} \Big[ 22 \hat{p}_{1}^2 \vec{\hat{p}}_{1}\cdot\vec{S}_{1} \vec{\hat{p}}_{1}\cdot\vec{S}_{2}
	 - 8 \vec{\hat{p}}_{1}\cdot\vec{\hat{p}}_{2} \vec{\hat{p}}_{1}\cdot\vec{S}_{1} \vec{\hat{p}}_{1}\cdot\vec{S}_{2}
	 - 18 \hat{p}_{2}^2 \vec{\hat{p}}_{1}\cdot\vec{S}_{1} \vec{\hat{p}}_{1}\cdot\vec{S}_{2} \nl
	 - 10 \hat{p}_{1}^2 \vec{\hat{p}}_{2}\cdot\vec{S}_{1} \vec{\hat{p}}_{1}\cdot\vec{S}_{2}
	 + 14 \vec{\hat{p}}_{1}\cdot\vec{\hat{p}}_{2} \vec{\hat{p}}_{2}\cdot\vec{S}_{1} \vec{\hat{p}}_{1}\cdot\vec{S}_{2}
	 - 10 \hat{p}_{2}^2 \vec{\hat{p}}_{2}\cdot\vec{S}_{1} \vec{\hat{p}}_{1}\cdot\vec{S}_{2}
	 - 20 \hat{p}_{1}^2 \vec{\hat{p}}_{1}\cdot\vec{S}_{1} \vec{\hat{p}}_{2}\cdot\vec{S}_{2} \nl
	 + 46 \vec{\hat{p}}_{1}\cdot\vec{\hat{p}}_{2} \vec{\hat{p}}_{1}\cdot\vec{S}_{1} \vec{\hat{p}}_{2}\cdot\vec{S}_{2}
	 - 20 \hat{p}_{2}^2 \vec{\hat{p}}_{1}\cdot\vec{S}_{1} \vec{\hat{p}}_{2}\cdot\vec{S}_{2}
	 - 18 \hat{p}_{1}^2 \vec{\hat{p}}_{2}\cdot\vec{S}_{1} \vec{\hat{p}}_{2}\cdot\vec{S}_{2}
	 - 8 \vec{\hat{p}}_{1}\cdot\vec{\hat{p}}_{2} \vec{\hat{p}}_{2}\cdot\vec{S}_{1} \vec{\hat{p}}_{2}\cdot\vec{S}_{2} \nl
	 + 22 \hat{p}_{2}^2 \vec{\hat{p}}_{2}\cdot\vec{S}_{1} \vec{\hat{p}}_{2}\cdot\vec{S}_{2}
	 + 42 \hat{p}_{1}^2 \vec{\hat{p}}_{1}\cdot\vec{\hat{p}}_{2} \vec{S}_{1}\cdot\vec{S}_{2}
	 + 8 \hat{p}_{1}^2 \hat{p}_{2}^2 \vec{S}_{1}\cdot\vec{S}_{2}
	 + 42 \vec{\hat{p}}_{1}\cdot\vec{\hat{p}}_{2} \hat{p}_{2}^2 \vec{S}_{1}\cdot\vec{S}_{2} \nl
	 - 34 \vec{S}_{1}\cdot\vec{S}_{2} ( \vec{\hat{p}}_{1}\cdot\vec{\hat{p}}_{2} )^{2}
	 - 22 \vec{S}_{1}\cdot\vec{S}_{2} \hat{p}_{1}^{4}
	 - 22 \vec{S}_{1}\cdot\vec{S}_{2} \hat{p}_{2}^{4} -93 \hat{p}_{1}^2 \vec{\hat{p}}_{1}\cdot\vec{\hat{p}}_{2} \vec{S}_{1}\cdot\vec{n} \vec{S}_{2}\cdot\vec{n}
	 + 42 \hat{p}_{1}^2 \hat{p}_{2}^2 \vec{S}_{1}\cdot\vec{n} \vec{S}_{2}\cdot\vec{n} \nl
	 - 93 \vec{\hat{p}}_{1}\cdot\vec{\hat{p}}_{2} \hat{p}_{2}^2 \vec{S}_{1}\cdot\vec{n} \vec{S}_{2}\cdot\vec{n}
	 - 18 \vec{\hat{p}}_{1}\cdot\vec{n} \hat{p}_{1}^2 \vec{\hat{p}}_{1}\cdot\vec{S}_{1} \vec{S}_{2}\cdot\vec{n}
	 + 36 \hat{p}_{1}^2 \vec{\hat{p}}_{2}\cdot\vec{n} \vec{\hat{p}}_{1}\cdot\vec{S}_{1} \vec{S}_{2}\cdot\vec{n} \nl
	 + 96 \vec{\hat{p}}_{1}\cdot\vec{n} \vec{\hat{p}}_{1}\cdot\vec{\hat{p}}_{2} \vec{\hat{p}}_{1}\cdot\vec{S}_{1} \vec{S}_{2}\cdot\vec{n}
	 - 102 \vec{\hat{p}}_{2}\cdot\vec{n} \vec{\hat{p}}_{1}\cdot\vec{\hat{p}}_{2} \vec{\hat{p}}_{1}\cdot\vec{S}_{1} \vec{S}_{2}\cdot\vec{n}
	 - 60 \vec{\hat{p}}_{1}\cdot\vec{n} \hat{p}_{2}^2 \vec{\hat{p}}_{1}\cdot\vec{S}_{1} \vec{S}_{2}\cdot\vec{n} \nl
	 + 72 \vec{\hat{p}}_{2}\cdot\vec{n} \hat{p}_{2}^2 \vec{\hat{p}}_{1}\cdot\vec{S}_{1} \vec{S}_{2}\cdot\vec{n}
	 - 75 \vec{\hat{p}}_{1}\cdot\vec{n} \hat{p}_{1}^2 \vec{\hat{p}}_{2}\cdot\vec{S}_{1} \vec{S}_{2}\cdot\vec{n}
	 + 66 \hat{p}_{1}^2 \vec{\hat{p}}_{2}\cdot\vec{n} \vec{\hat{p}}_{2}\cdot\vec{S}_{1} \vec{S}_{2}\cdot\vec{n} \nl
	 + 78 \vec{\hat{p}}_{1}\cdot\vec{n} \vec{\hat{p}}_{1}\cdot\vec{\hat{p}}_{2} \vec{\hat{p}}_{2}\cdot\vec{S}_{1} \vec{S}_{2}\cdot\vec{n}
	 - 48 \vec{\hat{p}}_{2}\cdot\vec{n} \vec{\hat{p}}_{1}\cdot\vec{\hat{p}}_{2} \vec{\hat{p}}_{2}\cdot\vec{S}_{1} \vec{S}_{2}\cdot\vec{n}
	 + 33 \vec{\hat{p}}_{1}\cdot\vec{n} \hat{p}_{2}^2 \vec{\hat{p}}_{2}\cdot\vec{S}_{1} \vec{S}_{2}\cdot\vec{n} \nl
	 - 48 \vec{\hat{p}}_{2}\cdot\vec{n} \hat{p}_{2}^2 \vec{\hat{p}}_{2}\cdot\vec{S}_{1} \vec{S}_{2}\cdot\vec{n}
	 - 48 \vec{\hat{p}}_{1}\cdot\vec{n} \hat{p}_{1}^2 \vec{S}_{1}\cdot\vec{n} \vec{\hat{p}}_{1}\cdot\vec{S}_{2}
	 + 33 \hat{p}_{1}^2 \vec{\hat{p}}_{2}\cdot\vec{n} \vec{S}_{1}\cdot\vec{n} \vec{\hat{p}}_{1}\cdot\vec{S}_{2} \nl
	 - 48 \vec{\hat{p}}_{1}\cdot\vec{n} \vec{\hat{p}}_{1}\cdot\vec{\hat{p}}_{2} \vec{S}_{1}\cdot\vec{n} \vec{\hat{p}}_{1}\cdot\vec{S}_{2}
	 + 78 \vec{\hat{p}}_{2}\cdot\vec{n} \vec{\hat{p}}_{1}\cdot\vec{\hat{p}}_{2} \vec{S}_{1}\cdot\vec{n} \vec{\hat{p}}_{1}\cdot\vec{S}_{2}
	 + 66 \vec{\hat{p}}_{1}\cdot\vec{n} \hat{p}_{2}^2 \vec{S}_{1}\cdot\vec{n} \vec{\hat{p}}_{1}\cdot\vec{S}_{2} \nl
	 - 75 \vec{\hat{p}}_{2}\cdot\vec{n} \hat{p}_{2}^2 \vec{S}_{1}\cdot\vec{n} \vec{\hat{p}}_{1}\cdot\vec{S}_{2}
	 + 144 \vec{\hat{p}}_{1}\cdot\vec{n} \vec{\hat{p}}_{2}\cdot\vec{n} \vec{\hat{p}}_{1}\cdot\vec{S}_{1} \vec{\hat{p}}_{1}\cdot\vec{S}_{2}
	 - 360 \vec{\hat{p}}_{1}\cdot\vec{n} \vec{\hat{p}}_{2}\cdot\vec{n} \vec{\hat{p}}_{2}\cdot\vec{S}_{1} \vec{\hat{p}}_{1}\cdot\vec{S}_{2} \nl
	 + 72 \vec{\hat{p}}_{1}\cdot\vec{n} \hat{p}_{1}^2 \vec{S}_{1}\cdot\vec{n} \vec{\hat{p}}_{2}\cdot\vec{S}_{2}
	 - 60 \hat{p}_{1}^2 \vec{\hat{p}}_{2}\cdot\vec{n} \vec{S}_{1}\cdot\vec{n} \vec{\hat{p}}_{2}\cdot\vec{S}_{2}
	 - 102 \vec{\hat{p}}_{1}\cdot\vec{n} \vec{\hat{p}}_{1}\cdot\vec{\hat{p}}_{2} \vec{S}_{1}\cdot\vec{n} \vec{\hat{p}}_{2}\cdot\vec{S}_{2} \nl
	 + 96 \vec{\hat{p}}_{2}\cdot\vec{n} \vec{\hat{p}}_{1}\cdot\vec{\hat{p}}_{2} \vec{S}_{1}\cdot\vec{n} \vec{\hat{p}}_{2}\cdot\vec{S}_{2}
	 + 36 \vec{\hat{p}}_{1}\cdot\vec{n} \hat{p}_{2}^2 \vec{S}_{1}\cdot\vec{n} \vec{\hat{p}}_{2}\cdot\vec{S}_{2}
	 - 18 \vec{\hat{p}}_{2}\cdot\vec{n} \hat{p}_{2}^2 \vec{S}_{1}\cdot\vec{n} \vec{\hat{p}}_{2}\cdot\vec{S}_{2} \nl
	 - 72 \vec{\hat{p}}_{1}\cdot\vec{n} \vec{\hat{p}}_{2}\cdot\vec{n} \vec{\hat{p}}_{1}\cdot\vec{S}_{1} \vec{\hat{p}}_{2}\cdot\vec{S}_{2}
	 + 144 \vec{\hat{p}}_{1}\cdot\vec{n} \vec{\hat{p}}_{2}\cdot\vec{n} \vec{\hat{p}}_{2}\cdot\vec{S}_{1} \vec{\hat{p}}_{2}\cdot\vec{S}_{2}
	 - 153 \vec{\hat{p}}_{1}\cdot\vec{n} \hat{p}_{1}^2 \vec{\hat{p}}_{2}\cdot\vec{n} \vec{S}_{1}\cdot\vec{S}_{2} \nl
	 + 306 \vec{\hat{p}}_{1}\cdot\vec{n} \vec{\hat{p}}_{2}\cdot\vec{n} \vec{\hat{p}}_{1}\cdot\vec{\hat{p}}_{2} \vec{S}_{1}\cdot\vec{S}_{2}
	 - 153 \vec{\hat{p}}_{1}\cdot\vec{n} \vec{\hat{p}}_{2}\cdot\vec{n} \hat{p}_{2}^2 \vec{S}_{1}\cdot\vec{S}_{2}
	 + 120 \vec{\hat{p}}_{2}\cdot\vec{S}_{1} \vec{\hat{p}}_{1}\cdot\vec{S}_{2} ( \vec{\hat{p}}_{1}\cdot\vec{n} )^{2} \nl
	 - 36 \vec{\hat{p}}_{1}\cdot\vec{S}_{1} \vec{\hat{p}}_{2}\cdot\vec{S}_{2} ( \vec{\hat{p}}_{1}\cdot\vec{n} )^{2}
	 - 30 \vec{\hat{p}}_{2}\cdot\vec{S}_{1} \vec{\hat{p}}_{2}\cdot\vec{S}_{2} ( \vec{\hat{p}}_{1}\cdot\vec{n} )^{2}
	 + 48 \hat{p}_{1}^2 \vec{S}_{1}\cdot\vec{S}_{2} ( \vec{\hat{p}}_{1}\cdot\vec{n} )^{2} \nl
	 - 72 \vec{\hat{p}}_{1}\cdot\vec{\hat{p}}_{2} \vec{S}_{1}\cdot\vec{S}_{2} ( \vec{\hat{p}}_{1}\cdot\vec{n} )^{2}
	 + 36 \hat{p}_{2}^2 \vec{S}_{1}\cdot\vec{S}_{2} ( \vec{\hat{p}}_{1}\cdot\vec{n} )^{2}
	 + 78 \vec{S}_{1}\cdot\vec{n} \vec{S}_{2}\cdot\vec{n} ( \vec{\hat{p}}_{1}\cdot\vec{\hat{p}}_{2} )^{2} \nl
	 - 30 \vec{\hat{p}}_{1}\cdot\vec{S}_{1} \vec{\hat{p}}_{1}\cdot\vec{S}_{2} ( \vec{\hat{p}}_{2}\cdot\vec{n} )^{2}
	 + 120 \vec{\hat{p}}_{2}\cdot\vec{S}_{1} \vec{\hat{p}}_{1}\cdot\vec{S}_{2} ( \vec{\hat{p}}_{2}\cdot\vec{n} )^{2}
	 - 36 \vec{\hat{p}}_{1}\cdot\vec{S}_{1} \vec{\hat{p}}_{2}\cdot\vec{S}_{2} ( \vec{\hat{p}}_{2}\cdot\vec{n} )^{2} \nl
	 + 36 \hat{p}_{1}^2 \vec{S}_{1}\cdot\vec{S}_{2} ( \vec{\hat{p}}_{2}\cdot\vec{n} )^{2}
	 - 72 \vec{\hat{p}}_{1}\cdot\vec{\hat{p}}_{2} \vec{S}_{1}\cdot\vec{S}_{2} ( \vec{\hat{p}}_{2}\cdot\vec{n} )^{2}
	 + 48 \hat{p}_{2}^2 \vec{S}_{1}\cdot\vec{S}_{2} ( \vec{\hat{p}}_{2}\cdot\vec{n} )^{2}
	 + 18 \vec{S}_{1}\cdot\vec{n} \vec{S}_{2}\cdot\vec{n} \hat{p}_{1}^{4} \nl
	 + 18 \vec{S}_{1}\cdot\vec{n} \vec{S}_{2}\cdot\vec{n} \hat{p}_{2}^{4}
	 + 120 \vec{\hat{p}}_{1}\cdot\vec{n} \hat{p}_{1}^2 \vec{\hat{p}}_{2}\cdot\vec{n} \vec{S}_{1}\cdot\vec{n} \vec{S}_{2}\cdot\vec{n}
	 - 390 \vec{\hat{p}}_{1}\cdot\vec{n} \vec{\hat{p}}_{2}\cdot\vec{n} \vec{\hat{p}}_{1}\cdot\vec{\hat{p}}_{2} \vec{S}_{1}\cdot\vec{n} \vec{S}_{2}\cdot\vec{n} \nl
	 + 120 \vec{\hat{p}}_{1}\cdot\vec{n} \vec{\hat{p}}_{2}\cdot\vec{n} \hat{p}_{2}^2 \vec{S}_{1}\cdot\vec{n} \vec{S}_{2}\cdot\vec{n}
	 + 120 \vec{\hat{p}}_{1}\cdot\vec{\hat{p}}_{2} \vec{S}_{1}\cdot\vec{n} \vec{S}_{2}\cdot\vec{n} ( \vec{\hat{p}}_{1}\cdot\vec{n} )^{2}
	 - 90 \hat{p}_{2}^2 \vec{S}_{1}\cdot\vec{n} \vec{S}_{2}\cdot\vec{n} ( \vec{\hat{p}}_{1}\cdot\vec{n} )^{2} \nl
	 - 180 \vec{\hat{p}}_{2}\cdot\vec{n} \vec{\hat{p}}_{1}\cdot\vec{S}_{1} \vec{S}_{2}\cdot\vec{n} ( \vec{\hat{p}}_{1}\cdot\vec{n} )^{2}
	 + 330 \vec{\hat{p}}_{2}\cdot\vec{n} \vec{\hat{p}}_{2}\cdot\vec{S}_{1} \vec{S}_{2}\cdot\vec{n} ( \vec{\hat{p}}_{1}\cdot\vec{n} )^{2} \nl
	 - 360 \vec{\hat{p}}_{2}\cdot\vec{n} \vec{S}_{1}\cdot\vec{n} \vec{\hat{p}}_{1}\cdot\vec{S}_{2} ( \vec{\hat{p}}_{1}\cdot\vec{n} )^{2}
	 + 300 \vec{\hat{p}}_{2}\cdot\vec{n} \vec{S}_{1}\cdot\vec{n} \vec{\hat{p}}_{2}\cdot\vec{S}_{2} ( \vec{\hat{p}}_{1}\cdot\vec{n} )^{2}
	 + 240 \vec{\hat{p}}_{2}\cdot\vec{n} \vec{S}_{1}\cdot\vec{S}_{2} ( \vec{\hat{p}}_{1}\cdot\vec{n} )^{3} \nl
	 - 90 \hat{p}_{1}^2 \vec{S}_{1}\cdot\vec{n} \vec{S}_{2}\cdot\vec{n} ( \vec{\hat{p}}_{2}\cdot\vec{n} )^{2}
	 + 120 \vec{\hat{p}}_{1}\cdot\vec{\hat{p}}_{2} \vec{S}_{1}\cdot\vec{n} \vec{S}_{2}\cdot\vec{n} ( \vec{\hat{p}}_{2}\cdot\vec{n} )^{2}
	 + 300 \vec{\hat{p}}_{1}\cdot\vec{n} \vec{\hat{p}}_{1}\cdot\vec{S}_{1} \vec{S}_{2}\cdot\vec{n} ( \vec{\hat{p}}_{2}\cdot\vec{n} )^{2} \nl
	 - 360 \vec{\hat{p}}_{1}\cdot\vec{n} \vec{\hat{p}}_{2}\cdot\vec{S}_{1} \vec{S}_{2}\cdot\vec{n} ( \vec{\hat{p}}_{2}\cdot\vec{n} )^{2}
	 + 330 \vec{\hat{p}}_{1}\cdot\vec{n} \vec{S}_{1}\cdot\vec{n} \vec{\hat{p}}_{1}\cdot\vec{S}_{2} ( \vec{\hat{p}}_{2}\cdot\vec{n} )^{2} \nl
	 - 180 \vec{\hat{p}}_{1}\cdot\vec{n} \vec{S}_{1}\cdot\vec{n} \vec{\hat{p}}_{2}\cdot\vec{S}_{2} ( \vec{\hat{p}}_{2}\cdot\vec{n} )^{2}
	 - 480 \vec{S}_{1}\cdot\vec{S}_{2} ( \vec{\hat{p}}_{1}\cdot\vec{n} )^{2} ( \vec{\hat{p}}_{2}\cdot\vec{n} )^{2}
	 + 240 \vec{\hat{p}}_{1}\cdot\vec{n} \vec{S}_{1}\cdot\vec{S}_{2} ( \vec{\hat{p}}_{2}\cdot\vec{n} )^{3} \nl
         - 210 \vec{S}_{1}\cdot\vec{n} \vec{S}_{2}\cdot\vec{n} ( \vec{\hat{p}}_{1}\cdot\vec{n} )^{2} ( \vec{\hat{p}}_{2}\cdot\vec{n} )^{2} \Big]
- \frac{G^2 m_{1}}{8 r^4} \Big[ 115 \vec{\hat{p}}_{1}\cdot\vec{S}_{1} \vec{\hat{p}}_{1}\cdot\vec{S}_{2}
	 - 110 \vec{\hat{p}}_{2}\cdot\vec{S}_{1} \vec{\hat{p}}_{1}\cdot\vec{S}_{2} \nl
	 - 166 \vec{\hat{p}}_{1}\cdot\vec{S}_{1} \vec{\hat{p}}_{2}\cdot\vec{S}_{2}
	 + 124 \vec{\hat{p}}_{2}\cdot\vec{S}_{1} \vec{\hat{p}}_{2}\cdot\vec{S}_{2}
	 - 163 \hat{p}_{1}^2 \vec{S}_{1}\cdot\vec{S}_{2}
	 + 348 \vec{\hat{p}}_{1}\cdot\vec{\hat{p}}_{2} \vec{S}_{1}\cdot\vec{S}_{2}
	 - 126 \hat{p}_{2}^2 \vec{S}_{1}\cdot\vec{S}_{2} \nl
	 + 123 \hat{p}_{1}^2 \vec{S}_{1}\cdot\vec{n} \vec{S}_{2}\cdot\vec{n}
	 - 304 \vec{\hat{p}}_{1}\cdot\vec{\hat{p}}_{2} \vec{S}_{1}\cdot\vec{n} \vec{S}_{2}\cdot\vec{n}
	 + 86 \hat{p}_{2}^2 \vec{S}_{1}\cdot\vec{n} \vec{S}_{2}\cdot\vec{n}
	 + 17 \vec{\hat{p}}_{1}\cdot\vec{n} \vec{\hat{p}}_{1}\cdot\vec{S}_{1} \vec{S}_{2}\cdot\vec{n} \nl
	 + 256 \vec{\hat{p}}_{2}\cdot\vec{n} \vec{\hat{p}}_{1}\cdot\vec{S}_{1} \vec{S}_{2}\cdot\vec{n}
	 + 126 \vec{\hat{p}}_{1}\cdot\vec{n} \vec{\hat{p}}_{2}\cdot\vec{S}_{1} \vec{S}_{2}\cdot\vec{n}
	 - 276 \vec{\hat{p}}_{2}\cdot\vec{n} \vec{\hat{p}}_{2}\cdot\vec{S}_{1} \vec{S}_{2}\cdot\vec{n} \nl
	 - 314 \vec{\hat{p}}_{1}\cdot\vec{n} \vec{S}_{1}\cdot\vec{n} \vec{\hat{p}}_{1}\cdot\vec{S}_{2}
	 + 267 \vec{\hat{p}}_{2}\cdot\vec{n} \vec{S}_{1}\cdot\vec{n} \vec{\hat{p}}_{1}\cdot\vec{S}_{2}
	 + 226 \vec{\hat{p}}_{1}\cdot\vec{n} \vec{S}_{1}\cdot\vec{n} \vec{\hat{p}}_{2}\cdot\vec{S}_{2}
	 - 84 \vec{\hat{p}}_{2}\cdot\vec{n} \vec{S}_{1}\cdot\vec{n} \vec{\hat{p}}_{2}\cdot\vec{S}_{2} \nl
	 - 565 \vec{\hat{p}}_{1}\cdot\vec{n} \vec{\hat{p}}_{2}\cdot\vec{n} \vec{S}_{1}\cdot\vec{S}_{2}
	 + 342 \vec{S}_{1}\cdot\vec{S}_{2} ( \vec{\hat{p}}_{1}\cdot\vec{n} )^{2}
	 + 276 \vec{S}_{1}\cdot\vec{S}_{2} ( \vec{\hat{p}}_{2}\cdot\vec{n} )^{2} -198 \vec{\hat{p}}_{1}\cdot\vec{n} \vec{\hat{p}}_{2}\cdot\vec{n} \vec{S}_{1}\cdot\vec{n} \vec{S}_{2}\cdot\vec{n} \nl
	 - 108 \vec{S}_{1}\cdot\vec{n} \vec{S}_{2}\cdot\vec{n} ( \vec{\hat{p}}_{1}\cdot\vec{n} )^{2} \Big]
	 - \frac{G^2 m_{2}}{8 r^4} \Big[ 124 \vec{\hat{p}}_{1}\cdot\vec{S}_{1} \vec{\hat{p}}_{1}\cdot\vec{S}_{2}
	 - 110 \vec{\hat{p}}_{2}\cdot\vec{S}_{1} \vec{\hat{p}}_{1}\cdot\vec{S}_{2}
	 - 166 \vec{\hat{p}}_{1}\cdot\vec{S}_{1} \vec{\hat{p}}_{2}\cdot\vec{S}_{2} \nl
	 + 115 \vec{\hat{p}}_{2}\cdot\vec{S}_{1} \vec{\hat{p}}_{2}\cdot\vec{S}_{2}
	 - 126 \hat{p}_{1}^2 \vec{S}_{1}\cdot\vec{S}_{2}
	 + 348 \vec{\hat{p}}_{1}\cdot\vec{\hat{p}}_{2} \vec{S}_{1}\cdot\vec{S}_{2}
	 - 163 \hat{p}_{2}^2 \vec{S}_{1}\cdot\vec{S}_{2}
	 + 86 \hat{p}_{1}^2 \vec{S}_{1}\cdot\vec{n} \vec{S}_{2}\cdot\vec{n} \nl
	 - 304 \vec{\hat{p}}_{1}\cdot\vec{\hat{p}}_{2} \vec{S}_{1}\cdot\vec{n} \vec{S}_{2}\cdot\vec{n}
	 + 123 \hat{p}_{2}^2 \vec{S}_{1}\cdot\vec{n} \vec{S}_{2}\cdot\vec{n}
	 - 84 \vec{\hat{p}}_{1}\cdot\vec{n} \vec{\hat{p}}_{1}\cdot\vec{S}_{1} \vec{S}_{2}\cdot\vec{n}
	 + 226 \vec{\hat{p}}_{2}\cdot\vec{n} \vec{\hat{p}}_{1}\cdot\vec{S}_{1} \vec{S}_{2}\cdot\vec{n} \nl
	 + 267 \vec{\hat{p}}_{1}\cdot\vec{n} \vec{\hat{p}}_{2}\cdot\vec{S}_{1} \vec{S}_{2}\cdot\vec{n}
	 - 314 \vec{\hat{p}}_{2}\cdot\vec{n} \vec{\hat{p}}_{2}\cdot\vec{S}_{1} \vec{S}_{2}\cdot\vec{n}
	 - 276 \vec{\hat{p}}_{1}\cdot\vec{n} \vec{S}_{1}\cdot\vec{n} \vec{\hat{p}}_{1}\cdot\vec{S}_{2} \nl
	 + 126 \vec{\hat{p}}_{2}\cdot\vec{n} \vec{S}_{1}\cdot\vec{n} \vec{\hat{p}}_{1}\cdot\vec{S}_{2}
	 + 256 \vec{\hat{p}}_{1}\cdot\vec{n} \vec{S}_{1}\cdot\vec{n} \vec{\hat{p}}_{2}\cdot\vec{S}_{2}
	 + 17 \vec{\hat{p}}_{2}\cdot\vec{n} \vec{S}_{1}\cdot\vec{n} \vec{\hat{p}}_{2}\cdot\vec{S}_{2}
	 - 565 \vec{\hat{p}}_{1}\cdot\vec{n} \vec{\hat{p}}_{2}\cdot\vec{n} \vec{S}_{1}\cdot\vec{S}_{2} \nl
	 + 276 \vec{S}_{1}\cdot\vec{S}_{2} ( \vec{\hat{p}}_{1}\cdot\vec{n} )^{2}
	 + 342 \vec{S}_{1}\cdot\vec{S}_{2} ( \vec{\hat{p}}_{2}\cdot\vec{n} )^{2} -198 \vec{\hat{p}}_{1}\cdot\vec{n} \vec{\hat{p}}_{2}\cdot\vec{n} \vec{S}_{1}\cdot\vec{n} \vec{S}_{2}\cdot\vec{n} \nl
	 - 108 \vec{S}_{1}\cdot\vec{n} \vec{S}_{2}\cdot\vec{n} ( \vec{\hat{p}}_{2}\cdot\vec{n} )^{2} \Big]
- \frac{G^3 m_{1} m_{2}}{4 r^5} \Big[ 179 \vec{S}_{1}\cdot\vec{S}_{2} -339 \vec{S}_{1}\cdot\vec{n} \vec{S}_{2}\cdot\vec{n} \Big]
	 - \frac{3 G^3 m_{1}^2}{2 r^5} \Big[ 13 \vec{S}_{1}\cdot\vec{S}_{2} \nl
                       -21 \vec{S}_{1}\cdot\vec{n} \vec{S}_{2}\cdot\vec{n} \Big]
	 - \frac{3 G^3 m_{2}^2}{2 r^5} \Big[ 13 \vec{S}_{1}\cdot\vec{S}_{2} -21 \vec{S}_{1}\cdot\vec{n} \vec{S}_{2}\cdot\vec{n} \Big] ,
\label{HNNLOS1S2}
\end{align}
which as we noted is in agreement with previous results 
\cite{Levi:2011eq, Hartung:2011ea, Levi:2014sba} as we detail in appendix~\ref{nnlos1s2}.

The NNLO spin-squared Hamiltonian is given by
\begin{align}
& H^{\text{NNLO}}_{\text{SS}} =
- \frac{G m_{2}}{16 m_{1} r^3} \Big[ 7 \hat{p}_{1}^2 \vec{\hat{p}}_{1}\cdot\vec{S}_{1} \vec{\hat{p}}_{2}\cdot\vec{S}_{1}
	 + 24 \vec{\hat{p}}_{1}\cdot\vec{\hat{p}}_{2} \vec{\hat{p}}_{1}\cdot\vec{S}_{1} \vec{\hat{p}}_{2}\cdot\vec{S}_{1}
	 - 12 \hat{p}_{2}^2 \vec{\hat{p}}_{1}\cdot\vec{S}_{1} \vec{\hat{p}}_{2}\cdot\vec{S}_{1} \nl
	 + 15 \hat{p}_{1}^2 \vec{\hat{p}}_{1}\cdot\vec{\hat{p}}_{2} S_{1}^2
	 - 18 \hat{p}_{1}^2 \hat{p}_{2}^2 S_{1}^2
	 + 12 \vec{\hat{p}}_{1}\cdot\vec{\hat{p}}_{2} \hat{p}_{2}^2 S_{1}^2
	 - 24 S_{1}^2 ( \vec{\hat{p}}_{1}\cdot\vec{\hat{p}}_{2} )^{2}
	 - 11 \hat{p}_{1}^2 ( \vec{\hat{p}}_{1}\cdot\vec{S}_{1} )^{2} \nl
	 - 22 \vec{\hat{p}}_{1}\cdot\vec{\hat{p}}_{2} ( \vec{\hat{p}}_{1}\cdot\vec{S}_{1} )^{2}
	 + 18 \hat{p}_{2}^2 ( \vec{\hat{p}}_{1}\cdot\vec{S}_{1} )^{2}
	 + 11 S_{1}^2 \hat{p}_{1}^{4}
	 + 21 \vec{\hat{p}}_{1}\cdot\vec{n} \hat{p}_{1}^2 \vec{S}_{1}\cdot\vec{n} \vec{\hat{p}}_{1}\cdot\vec{S}_{1} \nl
	 - 9 \hat{p}_{1}^2 \vec{\hat{p}}_{2}\cdot\vec{n} \vec{S}_{1}\cdot\vec{n} \vec{\hat{p}}_{1}\cdot\vec{S}_{1}
	 + 36 \vec{\hat{p}}_{1}\cdot\vec{n} \vec{\hat{p}}_{1}\cdot\vec{\hat{p}}_{2} \vec{S}_{1}\cdot\vec{n} \vec{\hat{p}}_{1}\cdot\vec{S}_{1}
	 - 24 \vec{\hat{p}}_{2}\cdot\vec{n} \vec{\hat{p}}_{1}\cdot\vec{\hat{p}}_{2} \vec{S}_{1}\cdot\vec{n} \vec{\hat{p}}_{1}\cdot\vec{S}_{1} \nl
	 - 30 \vec{\hat{p}}_{1}\cdot\vec{n} \hat{p}_{2}^2 \vec{S}_{1}\cdot\vec{n} \vec{\hat{p}}_{1}\cdot\vec{S}_{1}
	 + 12 \vec{\hat{p}}_{2}\cdot\vec{n} \hat{p}_{2}^2 \vec{S}_{1}\cdot\vec{n} \vec{\hat{p}}_{1}\cdot\vec{S}_{1}
	 + 54 \vec{\hat{p}}_{1}\cdot\vec{n} \hat{p}_{1}^2 \vec{S}_{1}\cdot\vec{n} \vec{\hat{p}}_{2}\cdot\vec{S}_{1} \nl
	 - 96 \vec{\hat{p}}_{1}\cdot\vec{n} \vec{\hat{p}}_{1}\cdot\vec{\hat{p}}_{2} \vec{S}_{1}\cdot\vec{n} \vec{\hat{p}}_{2}\cdot\vec{S}_{1}
	 + 24 \vec{\hat{p}}_{1}\cdot\vec{n} \hat{p}_{2}^2 \vec{S}_{1}\cdot\vec{n} \vec{\hat{p}}_{2}\cdot\vec{S}_{1}
	 - 24 \vec{\hat{p}}_{1}\cdot\vec{n} \vec{\hat{p}}_{2}\cdot\vec{n} \vec{\hat{p}}_{1}\cdot\vec{S}_{1} \vec{\hat{p}}_{2}\cdot\vec{S}_{1} \nl
	 - 33 \vec{\hat{p}}_{1}\cdot\vec{n} \hat{p}_{1}^2 \vec{\hat{p}}_{2}\cdot\vec{n} S_{1}^2
	 + 48 \vec{\hat{p}}_{1}\cdot\vec{n} \vec{\hat{p}}_{2}\cdot\vec{n} \vec{\hat{p}}_{1}\cdot\vec{\hat{p}}_{2} S_{1}^2
	 - 12 \vec{\hat{p}}_{1}\cdot\vec{n} \vec{\hat{p}}_{2}\cdot\vec{n} \hat{p}_{2}^2 S_{1}^2
	 - 66 \vec{\hat{p}}_{1}\cdot\vec{S}_{1} \vec{\hat{p}}_{2}\cdot\vec{S}_{1} ( \vec{\hat{p}}_{1}\cdot\vec{n} )^{2} \nl
	 - 15 \hat{p}_{1}^2 S_{1}^2 ( \vec{\hat{p}}_{1}\cdot\vec{n} )^{2}
	 + 15 \vec{\hat{p}}_{1}\cdot\vec{\hat{p}}_{2} S_{1}^2 ( \vec{\hat{p}}_{1}\cdot\vec{n} )^{2}
	 - 9 \hat{p}_{2}^2 S_{1}^2 ( \vec{\hat{p}}_{1}\cdot\vec{n} )^{2}
	 + 60 \vec{\hat{p}}_{1}\cdot\vec{S}_{1} \vec{\hat{p}}_{2}\cdot\vec{S}_{1} ( \vec{\hat{p}}_{2}\cdot\vec{n} )^{2} \nl
	 + 60 \hat{p}_{1}^2 S_{1}^2 ( \vec{\hat{p}}_{2}\cdot\vec{n} )^{2}
	 - 60 \vec{\hat{p}}_{1}\cdot\vec{\hat{p}}_{2} S_{1}^2 ( \vec{\hat{p}}_{2}\cdot\vec{n} )^{2}
	 - 39 \hat{p}_{1}^2 \vec{\hat{p}}_{1}\cdot\vec{\hat{p}}_{2} ( \vec{S}_{1}\cdot\vec{n} )^{2}
	 + 48 ( \vec{\hat{p}}_{1}\cdot\vec{\hat{p}}_{2} )^{2} ( \vec{S}_{1}\cdot\vec{n} )^{2} \nl
	 + 39 \hat{p}_{1}^2 \hat{p}_{2}^2 ( \vec{S}_{1}\cdot\vec{n} )^{2}
	 - 24 \vec{\hat{p}}_{1}\cdot\vec{\hat{p}}_{2} \hat{p}_{2}^2 ( \vec{S}_{1}\cdot\vec{n} )^{2}
	 + 12 ( \vec{\hat{p}}_{1}\cdot\vec{n} )^{2} ( \vec{\hat{p}}_{1}\cdot\vec{S}_{1} )^{2}
	 + 42 \vec{\hat{p}}_{1}\cdot\vec{n} \vec{\hat{p}}_{2}\cdot\vec{n} ( \vec{\hat{p}}_{1}\cdot\vec{S}_{1} )^{2} \nl
	 - 60 ( \vec{\hat{p}}_{2}\cdot\vec{n} )^{2} ( \vec{\hat{p}}_{1}\cdot\vec{S}_{1} )^{2}
	 + 48 ( \vec{\hat{p}}_{1}\cdot\vec{n} )^{2} ( \vec{\hat{p}}_{2}\cdot\vec{S}_{1} )^{2}
	 - 18 ( \vec{S}_{1}\cdot\vec{n} )^{2} \hat{p}_{1}^{4} -30 \vec{\hat{p}}_{2}\cdot\vec{n} \vec{S}_{1}\cdot\vec{n} \vec{\hat{p}}_{1}\cdot\vec{S}_{1} ( \vec{\hat{p}}_{1}\cdot\vec{n} )^{2} \nl
	 + 15 \vec{\hat{p}}_{2}\cdot\vec{n} S_{1}^2 ( \vec{\hat{p}}_{1}\cdot\vec{n} )^{3}
	 + 180 \vec{\hat{p}}_{1}\cdot\vec{n} \vec{S}_{1}\cdot\vec{n} \vec{\hat{p}}_{1}\cdot\vec{S}_{1} ( \vec{\hat{p}}_{2}\cdot\vec{n} )^{2}
	 - 60 \vec{S}_{1}\cdot\vec{n} \vec{\hat{p}}_{1}\cdot\vec{S}_{1} ( \vec{\hat{p}}_{2}\cdot\vec{n} )^{3} \nl
	 - 120 \vec{\hat{p}}_{1}\cdot\vec{n} \vec{S}_{1}\cdot\vec{n} \vec{\hat{p}}_{2}\cdot\vec{S}_{1} ( \vec{\hat{p}}_{2}\cdot\vec{n} )^{2}
	 - 60 S_{1}^2 ( \vec{\hat{p}}_{1}\cdot\vec{n} )^{2} ( \vec{\hat{p}}_{2}\cdot\vec{n} )^{2}
	 + 60 \vec{\hat{p}}_{1}\cdot\vec{n} S_{1}^2 ( \vec{\hat{p}}_{2}\cdot\vec{n} )^{3} \nl
	 + 15 \vec{\hat{p}}_{1}\cdot\vec{n} \hat{p}_{1}^2 \vec{\hat{p}}_{2}\cdot\vec{n} ( \vec{S}_{1}\cdot\vec{n} )^{2}
	 - 120 \hat{p}_{1}^2 ( \vec{\hat{p}}_{2}\cdot\vec{n} )^{2} ( \vec{S}_{1}\cdot\vec{n} )^{2}
	 + 120 \vec{\hat{p}}_{1}\cdot\vec{\hat{p}}_{2} ( \vec{\hat{p}}_{2}\cdot\vec{n} )^{2} ( \vec{S}_{1}\cdot\vec{n} )^{2} \Big] \nl
	 - \frac{G C_{1(\text{ES}^2)} m_{2}}{16 m_{1} r^3} \Big[ 4 \hat{p}_{1}^2 \vec{\hat{p}}_{1}\cdot\vec{S}_{1} \vec{\hat{p}}_{2}\cdot\vec{S}_{1}
	 - 24 \vec{\hat{p}}_{1}\cdot\vec{\hat{p}}_{2} \vec{\hat{p}}_{1}\cdot\vec{S}_{1} \vec{\hat{p}}_{2}\cdot\vec{S}_{1}
	 + 8 \hat{p}_{2}^2 \vec{\hat{p}}_{1}\cdot\vec{S}_{1} \vec{\hat{p}}_{2}\cdot\vec{S}_{1}
	 - 24 \hat{p}_{1}^2 \vec{\hat{p}}_{1}\cdot\vec{\hat{p}}_{2} S_{1}^2 \nl
	 + 11 \hat{p}_{1}^2 \hat{p}_{2}^2 S_{1}^2
	 - 4 \vec{\hat{p}}_{1}\cdot\vec{\hat{p}}_{2} \hat{p}_{2}^2 S_{1}^2
	 + 26 S_{1}^2 ( \vec{\hat{p}}_{1}\cdot\vec{\hat{p}}_{2} )^{2}
	 - 4 \hat{p}_{1}^2 ( \vec{\hat{p}}_{1}\cdot\vec{S}_{1} )^{2}
	 + 24 \vec{\hat{p}}_{1}\cdot\vec{\hat{p}}_{2} ( \vec{\hat{p}}_{1}\cdot\vec{S}_{1} )^{2} \nl
	 - 10 \hat{p}_{2}^2 ( \vec{\hat{p}}_{1}\cdot\vec{S}_{1} )^{2}
	 + 2 \hat{p}_{1}^2 ( \vec{\hat{p}}_{2}\cdot\vec{S}_{1} )^{2}
	 - S_{1}^2 \hat{p}_{1}^{4}
	 - 5 S_{1}^2 \hat{p}_{2}^{4}
	 + 18 \vec{\hat{p}}_{1}\cdot\vec{n} \hat{p}_{1}^2 \vec{S}_{1}\cdot\vec{n} \vec{\hat{p}}_{1}\cdot\vec{S}_{1} \nl
	 - 12 \hat{p}_{1}^2 \vec{\hat{p}}_{2}\cdot\vec{n} \vec{S}_{1}\cdot\vec{n} \vec{\hat{p}}_{1}\cdot\vec{S}_{1}
	 - 72 \vec{\hat{p}}_{1}\cdot\vec{n} \vec{\hat{p}}_{1}\cdot\vec{\hat{p}}_{2} \vec{S}_{1}\cdot\vec{n} \vec{\hat{p}}_{1}\cdot\vec{S}_{1}
	 + 72 \vec{\hat{p}}_{2}\cdot\vec{n} \vec{\hat{p}}_{1}\cdot\vec{\hat{p}}_{2} \vec{S}_{1}\cdot\vec{n} \vec{\hat{p}}_{1}\cdot\vec{S}_{1} \nl
	 + 24 \vec{\hat{p}}_{1}\cdot\vec{n} \hat{p}_{2}^2 \vec{S}_{1}\cdot\vec{n} \vec{\hat{p}}_{1}\cdot\vec{S}_{1}
	 - 24 \vec{\hat{p}}_{2}\cdot\vec{n} \hat{p}_{2}^2 \vec{S}_{1}\cdot\vec{n} \vec{\hat{p}}_{1}\cdot\vec{S}_{1}
	 - 24 \vec{\hat{p}}_{1}\cdot\vec{n} \hat{p}_{1}^2 \vec{S}_{1}\cdot\vec{n} \vec{\hat{p}}_{2}\cdot\vec{S}_{1} \nl
	 - 12 \hat{p}_{1}^2 \vec{\hat{p}}_{2}\cdot\vec{n} \vec{S}_{1}\cdot\vec{n} \vec{\hat{p}}_{2}\cdot\vec{S}_{1}
	 + 72 \vec{\hat{p}}_{1}\cdot\vec{n} \vec{\hat{p}}_{1}\cdot\vec{\hat{p}}_{2} \vec{S}_{1}\cdot\vec{n} \vec{\hat{p}}_{2}\cdot\vec{S}_{1}
	 - 24 \vec{\hat{p}}_{1}\cdot\vec{n} \hat{p}_{2}^2 \vec{S}_{1}\cdot\vec{n} \vec{\hat{p}}_{2}\cdot\vec{S}_{1} \nl
	 - 24 \vec{\hat{p}}_{1}\cdot\vec{n} \vec{\hat{p}}_{2}\cdot\vec{n} \vec{\hat{p}}_{1}\cdot\vec{S}_{1} \vec{\hat{p}}_{2}\cdot\vec{S}_{1}
	 - 24 \vec{\hat{p}}_{1}\cdot\vec{n} \hat{p}_{1}^2 \vec{\hat{p}}_{2}\cdot\vec{n} S_{1}^2
	 - 12 \vec{\hat{p}}_{1}\cdot\vec{n} \vec{\hat{p}}_{2}\cdot\vec{n} \vec{\hat{p}}_{1}\cdot\vec{\hat{p}}_{2} S_{1}^2
	 + 12 \vec{\hat{p}}_{1}\cdot\vec{n} \vec{\hat{p}}_{2}\cdot\vec{n} \hat{p}_{2}^2 S_{1}^2 \nl
	 + 12 \vec{\hat{p}}_{1}\cdot\vec{S}_{1} \vec{\hat{p}}_{2}\cdot\vec{S}_{1} ( \vec{\hat{p}}_{1}\cdot\vec{n} )^{2}
	 - 12 \hat{p}_{1}^2 S_{1}^2 ( \vec{\hat{p}}_{1}\cdot\vec{n} )^{2}
	 + 60 \vec{\hat{p}}_{1}\cdot\vec{\hat{p}}_{2} S_{1}^2 ( \vec{\hat{p}}_{1}\cdot\vec{n} )^{2}
	 - 27 \hat{p}_{2}^2 S_{1}^2 ( \vec{\hat{p}}_{1}\cdot\vec{n} )^{2} \nl
	 + 9 \hat{p}_{1}^2 S_{1}^2 ( \vec{\hat{p}}_{2}\cdot\vec{n} )^{2}
	 - 12 \hat{p}_{1}^2 \vec{\hat{p}}_{1}\cdot\vec{\hat{p}}_{2} ( \vec{S}_{1}\cdot\vec{n} )^{2}
	 - 6 ( \vec{\hat{p}}_{1}\cdot\vec{\hat{p}}_{2} )^{2} ( \vec{S}_{1}\cdot\vec{n} )^{2}
	 - 9 \hat{p}_{1}^2 \hat{p}_{2}^2 ( \vec{S}_{1}\cdot\vec{n} )^{2} \nl
	 - 12 \vec{\hat{p}}_{1}\cdot\vec{\hat{p}}_{2} \hat{p}_{2}^2 ( \vec{S}_{1}\cdot\vec{n} )^{2}
	 - 6 ( \vec{\hat{p}}_{1}\cdot\vec{n} )^{2} ( \vec{\hat{p}}_{1}\cdot\vec{S}_{1} )^{2}
	 + 24 \vec{\hat{p}}_{1}\cdot\vec{n} \vec{\hat{p}}_{2}\cdot\vec{n} ( \vec{\hat{p}}_{1}\cdot\vec{S}_{1} )^{2}
	 - 6 ( \vec{\hat{p}}_{2}\cdot\vec{n} )^{2} ( \vec{\hat{p}}_{1}\cdot\vec{S}_{1} )^{2} \nl
	 - 6 ( \vec{\hat{p}}_{1}\cdot\vec{n} )^{2} ( \vec{\hat{p}}_{2}\cdot\vec{S}_{1} )^{2}
	 + 15 ( \vec{S}_{1}\cdot\vec{n} )^{2} \hat{p}_{1}^{4}
	 + 15 ( \vec{S}_{1}\cdot\vec{n} )^{2} \hat{p}_{2}^{4}
         - 60 \vec{\hat{p}}_{2}\cdot\vec{n} \vec{S}_{1}\cdot\vec{n} \vec{\hat{p}}_{1}\cdot\vec{S}_{1} ( \vec{\hat{p}}_{1}\cdot\vec{n} )^{2} \nl
	 + 60 \vec{\hat{p}}_{2}\cdot\vec{n} \vec{S}_{1}\cdot\vec{n} \vec{\hat{p}}_{2}\cdot\vec{S}_{1} ( \vec{\hat{p}}_{1}\cdot\vec{n} )^{2}
	 + 60 \vec{\hat{p}}_{2}\cdot\vec{n} S_{1}^2 ( \vec{\hat{p}}_{1}\cdot\vec{n} )^{3}
	 + 60 \vec{\hat{p}}_{1}\cdot\vec{n} \vec{S}_{1}\cdot\vec{n} \vec{\hat{p}}_{1}\cdot\vec{S}_{1} ( \vec{\hat{p}}_{2}\cdot\vec{n} )^{2} \nl
	 - 45 S_{1}^2 ( \vec{\hat{p}}_{1}\cdot\vec{n} )^{2} ( \vec{\hat{p}}_{2}\cdot\vec{n} )^{2}
	 + 60 \vec{\hat{p}}_{1}\cdot\vec{n} \hat{p}_{1}^2 \vec{\hat{p}}_{2}\cdot\vec{n} ( \vec{S}_{1}\cdot\vec{n} )^{2}
	 + 15 \hat{p}_{1}^2 ( \vec{\hat{p}}_{2}\cdot\vec{n} )^{2} ( \vec{S}_{1}\cdot\vec{n} )^{2} \nl
	 - 180 \vec{\hat{p}}_{1}\cdot\vec{n} \vec{\hat{p}}_{2}\cdot\vec{n} \vec{\hat{p}}_{1}\cdot\vec{\hat{p}}_{2} ( \vec{S}_{1}\cdot\vec{n} )^{2}
	 + 15 \hat{p}_{2}^2 ( \vec{\hat{p}}_{1}\cdot\vec{n} )^{2} ( \vec{S}_{1}\cdot\vec{n} )^{2}
	 + 60 \vec{\hat{p}}_{1}\cdot\vec{n} \vec{\hat{p}}_{2}\cdot\vec{n} \hat{p}_{2}^2 ( \vec{S}_{1}\cdot\vec{n} )^{2} \nl
         - 105 ( \vec{\hat{p}}_{1}\cdot\vec{n} )^{2} ( \vec{\hat{p}}_{2}\cdot\vec{n} )^{2} ( \vec{S}_{1}\cdot\vec{n} )^{2} \Big]
+ \frac{G^2 C_{1(\text{ES}^2)} m_{2}^2}{8 m_{1} r^4} \Big[ 72 \vec{\hat{p}}_{1}\cdot\vec{S}_{1} \vec{\hat{p}}_{2}\cdot\vec{S}_{1}
	 + 64 \hat{p}_{1}^2 S_{1}^2
	 - 216 \vec{\hat{p}}_{1}\cdot\vec{\hat{p}}_{2} S_{1}^2 \nl
	 + 143 \hat{p}_{2}^2 S_{1}^2
	 + 4 ( \vec{\hat{p}}_{1}\cdot\vec{S}_{1} )^{2}
	 - 73 ( \vec{\hat{p}}_{2}\cdot\vec{S}_{1} )^{2}
         - 40 \vec{\hat{p}}_{1}\cdot\vec{n} \vec{S}_{1}\cdot\vec{n} \vec{\hat{p}}_{1}\cdot\vec{S}_{1}
	 + 48 \vec{\hat{p}}_{2}\cdot\vec{n} \vec{S}_{1}\cdot\vec{n} \vec{\hat{p}}_{1}\cdot\vec{S}_{1} \nl
	 - 402 \vec{\hat{p}}_{1}\cdot\vec{n} \vec{S}_{1}\cdot\vec{n} \vec{\hat{p}}_{2}\cdot\vec{S}_{1}
	 + 356 \vec{\hat{p}}_{2}\cdot\vec{n} \vec{S}_{1}\cdot\vec{n} \vec{\hat{p}}_{2}\cdot\vec{S}_{1}
	 + 368 \vec{\hat{p}}_{1}\cdot\vec{n} \vec{\hat{p}}_{2}\cdot\vec{n} S_{1}^2
	 - 64 S_{1}^2 ( \vec{\hat{p}}_{1}\cdot\vec{n} )^{2} \nl
	 - 332 S_{1}^2 ( \vec{\hat{p}}_{2}\cdot\vec{n} )^{2}
	 - 120 \hat{p}_{1}^2 ( \vec{S}_{1}\cdot\vec{n} )^{2}
	 + 280 \vec{\hat{p}}_{1}\cdot\vec{\hat{p}}_{2} ( \vec{S}_{1}\cdot\vec{n} )^{2}
	 - 148 \hat{p}_{2}^2 ( \vec{S}_{1}\cdot\vec{n} )^{2}
	 + 48 ( \vec{\hat{p}}_{1}\cdot\vec{n} )^{2} ( \vec{S}_{1}\cdot\vec{n} )^{2} \nl
	 + 162 \vec{\hat{p}}_{1}\cdot\vec{n} \vec{\hat{p}}_{2}\cdot\vec{n} ( \vec{S}_{1}\cdot\vec{n} )^{2}
	 - 96 ( \vec{\hat{p}}_{2}\cdot\vec{n} )^{2} ( \vec{S}_{1}\cdot\vec{n} )^{2} \Big]
- \frac{G^2 m_{2}^2}{16 m_{1} r^4} \Big[ 138 \vec{\hat{p}}_{1}\cdot\vec{S}_{1} \vec{\hat{p}}_{2}\cdot\vec{S}_{1}
	 + 139 \hat{p}_{1}^2 S_{1}^2 \nl
	 - 146 \vec{\hat{p}}_{1}\cdot\vec{\hat{p}}_{2} S_{1}^2
	 - 52 \hat{p}_{2}^2 S_{1}^2
	 - 134 ( \vec{\hat{p}}_{1}\cdot\vec{S}_{1} )^{2}
	 + 60 ( \vec{\hat{p}}_{2}\cdot\vec{S}_{1} )^{2}
	 + 406 \vec{\hat{p}}_{1}\cdot\vec{n} \vec{S}_{1}\cdot\vec{n} \vec{\hat{p}}_{1}\cdot\vec{S}_{1} \nl
	 - 48 \vec{\hat{p}}_{2}\cdot\vec{n} \vec{S}_{1}\cdot\vec{n} \vec{\hat{p}}_{1}\cdot\vec{S}_{1}
	 - 404 \vec{\hat{p}}_{1}\cdot\vec{n} \vec{S}_{1}\cdot\vec{n} \vec{\hat{p}}_{2}\cdot\vec{S}_{1}
	 - 128 \vec{\hat{p}}_{2}\cdot\vec{n} \vec{S}_{1}\cdot\vec{n} \vec{\hat{p}}_{2}\cdot\vec{S}_{1}
	 + 368 \vec{\hat{p}}_{1}\cdot\vec{n} \vec{\hat{p}}_{2}\cdot\vec{n} S_{1}^2 \nl
	 - 142 S_{1}^2 ( \vec{\hat{p}}_{1}\cdot\vec{n} )^{2}
	 - 100 S_{1}^2 ( \vec{\hat{p}}_{2}\cdot\vec{n} )^{2}
	 - 269 \hat{p}_{1}^2 ( \vec{S}_{1}\cdot\vec{n} )^{2}
	 + 224 \vec{\hat{p}}_{1}\cdot\vec{\hat{p}}_{2} ( \vec{S}_{1}\cdot\vec{n} )^{2}
	 + 136 \hat{p}_{2}^2 ( \vec{S}_{1}\cdot\vec{n} )^{2} \nl
         - 132 \vec{\hat{p}}_{1}\cdot\vec{n} \vec{\hat{p}}_{2}\cdot\vec{n} ( \vec{S}_{1}\cdot\vec{n} )^{2}
	 + 84 ( \vec{\hat{p}}_{2}\cdot\vec{n} )^{2} ( \vec{S}_{1}\cdot\vec{n} )^{2} \Big]
- \frac{G^2 m_{2}}{8 r^4} \Big[ 92 \vec{\hat{p}}_{1}\cdot\vec{S}_{1} \vec{\hat{p}}_{2}\cdot\vec{S}_{1}
	 + 89 \hat{p}_{1}^2 S_{1}^2 \nl
	 - 72 \vec{\hat{p}}_{1}\cdot\vec{\hat{p}}_{2} S_{1}^2
	 - 30 \hat{p}_{2}^2 S_{1}^2
	 - 95 ( \vec{\hat{p}}_{1}\cdot\vec{S}_{1} )^{2}
	 + 24 ( \vec{\hat{p}}_{2}\cdot\vec{S}_{1} )^{2}
	 + 300 \vec{\hat{p}}_{1}\cdot\vec{n} \vec{S}_{1}\cdot\vec{n} \vec{\hat{p}}_{1}\cdot\vec{S}_{1} \nl
	 - 84 \vec{\hat{p}}_{2}\cdot\vec{n} \vec{S}_{1}\cdot\vec{n} \vec{\hat{p}}_{1}\cdot\vec{S}_{1}
	 - 184 \vec{\hat{p}}_{1}\cdot\vec{n} \vec{S}_{1}\cdot\vec{n} \vec{\hat{p}}_{2}\cdot\vec{S}_{1}
	 - 72 \vec{\hat{p}}_{2}\cdot\vec{n} \vec{S}_{1}\cdot\vec{n} \vec{\hat{p}}_{2}\cdot\vec{S}_{1}
	 + 104 \vec{\hat{p}}_{1}\cdot\vec{n} \vec{\hat{p}}_{2}\cdot\vec{n} S_{1}^2 \nl
	 - 77 S_{1}^2 ( \vec{\hat{p}}_{1}\cdot\vec{n} )^{2}
	 + 42 S_{1}^2 ( \vec{\hat{p}}_{2}\cdot\vec{n} )^{2}
	 - 163 \hat{p}_{1}^2 ( \vec{S}_{1}\cdot\vec{n} )^{2}
	 + 148 \vec{\hat{p}}_{1}\cdot\vec{\hat{p}}_{2} ( \vec{S}_{1}\cdot\vec{n} )^{2}
	 + 30 \hat{p}_{2}^2 ( \vec{S}_{1}\cdot\vec{n} )^{2} \nl
         - 102 ( \vec{\hat{p}}_{1}\cdot\vec{n} )^{2} ( \vec{S}_{1}\cdot\vec{n} )^{2}
	 + 12 \vec{\hat{p}}_{1}\cdot\vec{n} \vec{\hat{p}}_{2}\cdot\vec{n} ( \vec{S}_{1}\cdot\vec{n} )^{2}
	 + 18 ( \vec{\hat{p}}_{2}\cdot\vec{n} )^{2} ( \vec{S}_{1}\cdot\vec{n} )^{2} \Big] \nl
+ \frac{G^2 C_{1(\text{ES}^2)} m_{2}}{8 r^4} \Big[ 8 \vec{\hat{p}}_{1}\cdot\vec{S}_{1} \vec{\hat{p}}_{2}\cdot\vec{S}_{1}
	 + 49 \hat{p}_{1}^2 S_{1}^2
	 - 68 \vec{\hat{p}}_{1}\cdot\vec{\hat{p}}_{2} S_{1}^2
	 + 20 \hat{p}_{2}^2 S_{1}^2
	 - 13 ( \vec{\hat{p}}_{1}\cdot\vec{S}_{1} )^{2}
	 + 4 ( \vec{\hat{p}}_{2}\cdot\vec{S}_{1} )^{2} \nl
	 + 32 \vec{\hat{p}}_{1}\cdot\vec{n} \vec{S}_{1}\cdot\vec{n} \vec{\hat{p}}_{1}\cdot\vec{S}_{1}
	 - 16 \vec{\hat{p}}_{2}\cdot\vec{n} \vec{S}_{1}\cdot\vec{n} \vec{\hat{p}}_{1}\cdot\vec{S}_{1}
	 - 24 \vec{\hat{p}}_{2}\cdot\vec{n} \vec{S}_{1}\cdot\vec{n} \vec{\hat{p}}_{2}\cdot\vec{S}_{1}
	 + 28 \vec{\hat{p}}_{1}\cdot\vec{n} \vec{\hat{p}}_{2}\cdot\vec{n} S_{1}^2 \nl
	 - 38 S_{1}^2 ( \vec{\hat{p}}_{1}\cdot\vec{n} )^{2}
	 - 2 S_{1}^2 ( \vec{\hat{p}}_{2}\cdot\vec{n} )^{2}
	 - 102 \hat{p}_{1}^2 ( \vec{S}_{1}\cdot\vec{n} )^{2}
	 + 164 \vec{\hat{p}}_{1}\cdot\vec{\hat{p}}_{2} ( \vec{S}_{1}\cdot\vec{n} )^{2}
	 - 64 \hat{p}_{2}^2 ( \vec{S}_{1}\cdot\vec{n} )^{2} \nl
         - 6 ( \vec{\hat{p}}_{1}\cdot\vec{n} )^{2} ( \vec{S}_{1}\cdot\vec{n} )^{2}
	 + 12 \vec{\hat{p}}_{1}\cdot\vec{n} \vec{\hat{p}}_{2}\cdot\vec{n} ( \vec{S}_{1}\cdot\vec{n} )^{2}
	 + 30 ( \vec{\hat{p}}_{2}\cdot\vec{n} )^{2} ( \vec{S}_{1}\cdot\vec{n} )^{2} \Big] \nl
- \frac{7 G^3 m_{2}^3}{2 m_{1} r^5} \Big[ S_{1}^2
	 - ( \vec{S}_{1}\cdot\vec{n} )^{2} \Big]
	 - \frac{8 G^3 m_{2}^2}{r^5} \Big[ S_{1}^2 -2 ( \vec{S}_{1}\cdot\vec{n} )^{2} \Big]
	 - \frac{23 G^3 C_{1(\text{ES}^2)} m_{1} m_{2}}{28 r^5} \Big[ S_{1}^2 -3 ( \vec{S}_{1}\cdot\vec{n} )^{2} \Big] \nl
	 - \frac{5 G^3 C_{1(\text{ES}^2)} m_{2}^2}{4 r^5} \Big[ 13 S_{1}^2
                          -19 ( \vec{S}_{1}\cdot\vec{n} )^{2} \Big]
	 - \frac{G^3 C_{1(\text{ES}^2)} m_{2}^3}{4 m_{1} r^5} \Big[ 19 S_{1}^2 -37 ( \vec{S}_{1}\cdot\vec{n} )^{2} \Big] \nl
	 - \frac{G^3 m_{1} m_{2}}{14 r^5} \Big[ 97 S_{1}^2 -123 ( \vec{S}_{1}\cdot\vec{n} )^{2} \Big]
	 + [1 \leftrightarrow 2] .
\end{align}
This completes the spin-dependent Hamiltonian to 4PN order, together with the LO 
quartic-in-spin Hamiltonian
at 4PN order from \cite{Levi:2014gsa}, and the Hamiltonian up to 3.5PN order in 
\cite{Levi:2014sba, Levi:2015uxa}.

It is useful for applications to further specify to the center-of-mass 
frame, which considerably simplifies the Hamiltonians. We proceed with 
the conventions and notations of our previous work, where the total momentum, $\vec{P}\equiv\vec{p}_1+\vec{p}_2$, is set to $0$, and 
$\vec{p}\equiv\vec{p}_1=-\vec{p}_2$ \cite{Levi:2014sba,Levi:2015uxa}. 
That is, we employ dimensionless variables denoted by a tilde,
with units of length given by the total mass $G m = G(m_1 + m_2)$, and 
masses given in terms of the reduced mass $\mu = m_1 m_2 / m$, 
i.e.~\cite{Levi:2014sba}
\be 
\tilde{L}\equiv \frac{L}{Gm\mu}, \qquad 
\tilde{S}_I\equiv \frac{S_I}{Gm\mu}, \qquad 
\tilde{r}\equiv \frac{r}{Gm}, \qquad 
\tilde{p}\equiv \frac{p}{\mu},
\ee
where $\vec{L} = r \vec{n} \times \vec{p}$.
Furthermore, we utilize a triad $\vec{n}$, $\vec{\lambda}$, $\vec{l}$, where
$\vec{l} = \vec{L} / L$, $\vec{\lambda} = \vec{l} \times \vec{n}$, so that we have 
\cite{Levi:2015uxa}
\be
\vec{p}=p_r\vec{n}+\frac{L}{r}\vec{\lambda}.
\ee
The result is expressed using the mass ratio $q \equiv m_1 / m_2$,
and the symmetric mass ratio $\nu \equiv \mu / m$. These conventions 
lead us to the following spin-dependent Hamiltonians at 4PN order:
\begin{align}
&H_\text{NNLO}^\text{S$_1$S$_2$} =
\bigg[
	 \frac{\tilde{L}^4}{8 \tilde{r}^7} (11 - 23 \nu - 7 \nu^2)
	 + \frac{7 \tilde{L}^2}{8 \tilde{r}^6} (18 + 19 \nu)
	 - \frac{1}{4 \tilde{r}^5} (78 + 23 \nu)
         - \frac{\tilde{L}^2 \tilde{p}_r^2}{16 \tilde{r}^5} (4 + 125 \nu + 52 \nu^2) \nl
	 - \frac{3 \tilde{p}_r^2}{4 \tilde{r}^4} (25 - 9 \nu)
	 - \frac{\tilde{p}_r^4}{16 \tilde{r}^3} (26 - 161 \nu + 38 \nu^2)
\bigg] \nu \vec{\tilde{S}}_{1}\cdot\vec{\tilde{S}}_{2}
+ \bigg[
	 - \frac{3 \tilde{L}^4}{16 \tilde{r}^7} (6 + 7 \nu - 10 \nu^2) \nl
	 - \frac{\tilde{L}^2}{8 \tilde{r}^6} (86 + 169 \nu)
	 + \frac{3}{4 \tilde{r}^5} (42 + 29 \nu)
	 + \frac{3 \tilde{p}_r^2}{4 \tilde{r}^4} (25 - 31 \nu)
         + \frac{\tilde{L}^2 \tilde{p}_r^2}{8 \tilde{r}^5} (4 + 49 \nu + 49 \nu^2) \nl
	 + \frac{\tilde{p}_r^4}{16 \tilde{r}^3} (26 - 193 \nu + 134 \nu^2)
\bigg] \nu \vec{n}\cdot\vec{\tilde{S}}_{1} \vec{n}\cdot\vec{\tilde{S}}_{2}
+ \nu^2 \bigg[
         \frac{15 \tilde{L}^3 \tilde{p}_r}{8 \tilde{r}^6} (1 - 2 \nu)
	 + \frac{\tilde{L} \tilde{p}_r}{2 \tilde{r}^5} (48 + 7 \nu) \nl
	 + \frac{15 \tilde{L} \tilde{p}_r^3}{8 \tilde{r}^4} (1 - 8 \nu)
\bigg] \bigg[ q \vec{n}\cdot\vec{\tilde{S}}_{2} \vec{\tilde{S}}_{1}\cdot\vec{\lambda}
              + \frac{\vec{n}\cdot\vec{\tilde{S}}_{1} \vec{\tilde{S}}_{2}\cdot\vec{\lambda}}{q} \bigg]
- \nu \bigg[
         \frac{\tilde{L}^3 \tilde{p}_r}{16 \tilde{r}^6} (4 - 35 \nu + 82 \nu^2) \nl
	 + \frac{\tilde{L} \tilde{p}_r}{8 \tilde{r}^5} (40 - 205 \nu - 28 \nu^2)
	 + \frac{\tilde{L} \tilde{p}_r^3}{16 \tilde{r}^4} (4 - 83 \nu + 316 \nu^2)
\bigg] \Big[ \vec{n}\cdot\vec{\tilde{S}}_{2} \vec{\tilde{S}}_{1}\cdot\vec{\lambda}
            + \vec{n}\cdot\vec{\tilde{S}}_{1} \vec{\tilde{S}}_{2}\cdot\vec{\lambda} \Big] \nl
- \bigg[
         \frac{\tilde{L}^4}{8 \tilde{r}^7} (11 - 25 \nu - 4 \nu^2)
	 + \frac{\tilde{L}^2}{8 \tilde{r}^6} (124 + 19 \nu)
	 + \frac{\tilde{L}^2 \tilde{p}_r^2}{8 \tilde{r}^5} (11 - 139 \nu - 22 \nu^2)
\bigg] \nu \vec{\tilde{S}}_{1}\cdot\vec{\lambda} \vec{\tilde{S}}_{2}\cdot\vec{\lambda} ,
\end{align}
\begin{align}
& H_\text{NNLO}^\text{S$^2$} =
\nu^2 \bigg\{
    \bigg[
	 \frac{\tilde{L}^4}{16 \tilde{r}^7} (11 - 25 \nu)
	 + \frac{\tilde{L}^2}{16 \tilde{r}^6} (199 + 33 \nu)
	 - \frac{24}{7 \tilde{r}^5}
         + \frac{\tilde{L}^2 \tilde{p}_r^2}{16 \tilde{r}^5} (7 - 74 \nu)
	 - \frac{9 \tilde{p}_r^2}{16 \tilde{r}^4} (3 - 5 \nu) \nl
	 - \frac{\tilde{p}_r^4}{4 \tilde{r}^3} (1 + \nu)
    \bigg] \tilde{S}_{1}^2
    - \bigg[
	 \frac{3 \tilde{L}^4}{16 \tilde{r}^7} (6 - 17 \nu)
	 + \frac{\tilde{L}^2}{16 \tilde{r}^6} (329 + 61 \nu)
	 - \frac{37}{7 \tilde{r}^5}
         + \frac{\tilde{L}^2 \tilde{p}_r^2}{16 \tilde{r}^5} (26 - 145 \nu) \nl
	 - \frac{\tilde{p}_r^2}{16 \tilde{r}^4} (3 - 85 \nu)
         - \frac{\tilde{p}_r^4}{4 \tilde{r}^3} (1 + \nu)
    \bigg] ( \vec{n}\cdot\vec{\tilde{S}}_{1} )^{2}
    - \bigg[
         \frac{\tilde{L}^4}{16 \tilde{r}^7} (11 - 25 \nu)
	 + \frac{\tilde{L}^2}{8 \tilde{r}^6} (91 + 11 \nu) \nl
	 - \frac{\tilde{L}^2 \tilde{p}_r^2}{16 \tilde{r}^5} (1 + 85 \nu)
    \bigg] ( \vec{\tilde{S}}_{1}\cdot\vec{\lambda} )^{2}
    - \bigg[
         \frac{\tilde{L}^3 \tilde{p}_r}{16 \tilde{r}^6} (1 + 37 \nu)
	 - \frac{\tilde{L} \tilde{p}_r}{8 \tilde{r}^5} (93 + 25 \nu) \nl
	 - \frac{\tilde{L} \tilde{p}_r^3}{16 \tilde{r}^4} (23 - 67 \nu)
    \bigg] \vec{n}\cdot\vec{\tilde{S}}_{1} \vec{\tilde{S}}_{1}\cdot\vec{\lambda}
\bigg\}
+ \nu^2 C_{1(\text{ES}^2)} \bigg\{
    \bigg[
	 \frac{\tilde{L}^4}{4 \tilde{r}^7} (1 + 3 \nu)
	 - \frac{\tilde{L}^2}{4 \tilde{r}^6} (22 + 5 \nu)
	 + \frac{55}{14 \tilde{r}^5} \nl
         - \frac{\tilde{L}^2 \tilde{p}_r^2}{4 \tilde{r}^5} (1 - 6 \nu)
	 + \frac{\tilde{p}_r^2}{4 \tilde{r}^4} (9 - 13 \nu)
	 - \frac{\tilde{p}_r^4}{2 \tilde{r}^3} (1 + 6 \nu)
    \bigg] \tilde{S}_{1}^2
    + \bigg[
         \frac{7 \tilde{L}^2}{4 \tilde{r}^6} (4 + \nu)
	 - \frac{95}{14 \tilde{r}^5}
         + \frac{\tilde{L}^2 \tilde{p}_r^2}{8 \tilde{r}^5} (7 + 6 \nu) \nl
	 + \frac{3 \tilde{p}_r^2}{4 \tilde{r}^4} (9 + 11 \nu)
	 + \frac{\tilde{p}_r^4}{2 \tilde{r}^3} (1 + 6 \nu)
    \bigg] ( \vec{n}\cdot\vec{\tilde{S}}_{1} )^{2}
    - \bigg[
         \frac{\tilde{L}^4}{4 \tilde{r}^7} (1 + 3 \nu)
	 - \frac{\nu \tilde{L}^2}{2 \tilde{r}^6}
	 + \frac{\tilde{L}^2 \tilde{p}_r^2}{8 \tilde{r}^5} (5 + 18 \nu)
    \bigg] ( \vec{\tilde{S}}_{1}\cdot\vec{\lambda} )^{2} \nl
    + \bigg[
         \frac{5 \tilde{L}^3 \tilde{p}_r}{8 \tilde{r}^6}
	 + \frac{\tilde{L} \tilde{p}_r}{4 \tilde{r}^5} (8 - 11 \nu)
	 - \frac{\tilde{L} \tilde{p}_r^3}{8 \tilde{r}^4} (1 - 6 \nu)
    \bigg] \vec{n}\cdot\vec{\tilde{S}}_{1} \vec{\tilde{S}}_{1}\cdot\vec{\lambda}
\bigg\}
+ \frac{\nu}{q} \bigg\{
    \bigg[
	 - \frac{\tilde{L}^4}{16 \tilde{r}^7} (11 - 59 \nu + 21 \nu^2) \nl
	 - \frac{\tilde{L}^2}{16 \tilde{r}^6} (139 - 93 \nu - 33 \nu^2)
	 - \frac{1}{2 \tilde{r}^5} (7 + 2 \nu)
         - \frac{\tilde{L}^2 \tilde{p}_r^2}{16 \tilde{r}^5} (7 - 40 \nu + 60 \nu^2)
	 + \frac{3 \tilde{p}_r^2}{16 \tilde{r}^4} (1 + 63 \nu + 15 \nu^2) \nl
	 + \frac{\tilde{p}_r^4}{16 \tilde{r}^3} (4 - 4 \nu - 9 \nu^2)
    \bigg] \tilde{S}_{1}^2
    + \bigg[
	 \frac{3 \tilde{L}^4}{16 \tilde{r}^7} (6 - 37 \nu + 15 \nu^2)
	 + \frac{\tilde{L}^2}{16 \tilde{r}^6} (269 - 257 \nu - 61 \nu^2) \nl
         + \frac{\tilde{L}^2 \tilde{p}_r^2}{16 \tilde{r}^5} (26 - 101 \nu + 126 \nu^2)
	 + \frac{1}{2 \tilde{r}^5} (7 + 18 \nu)
	 - \frac{\tilde{p}_r^2}{16 \tilde{r}^4} (3 + 93 \nu + 85 \nu^2) \nl
	 - \frac{\tilde{p}_r^4}{16 \tilde{r}^3} (4 - 4 \nu - 9 \nu^2)
    \bigg] ( \vec{n}\cdot\vec{\tilde{S}}_{1} )^{2}
    + \bigg[
         \frac{\tilde{L}^4}{16 \tilde{r}^7} (11 - 59 \nu + 21 \nu^2)
	 + \frac{\tilde{L}^2}{8 \tilde{r}^6} (67 - 37 \nu - 11 \nu^2) \nl
	 - \frac{\tilde{L}^2 \tilde{p}_r^2}{16 \tilde{r}^5} (1 + 35 \nu - 69 \nu^2)
    \bigg] ( \vec{\tilde{S}}_{1}\cdot\vec{\lambda} )^{2}
    + \bigg[
         \frac{\tilde{L}^3 \tilde{p}_r}{16 \tilde{r}^6} (1 + 47 \nu - 33 \nu^2)
	 - \frac{\tilde{L} \tilde{p}_r}{8 \tilde{r}^5} (69 - 9 \nu - 25 \nu^2) \nl
	 - \frac{\tilde{L} \tilde{p}_r^3}{16 \tilde{r}^4} (23 - 65 \nu + 57 \nu^2)
    \bigg] \vec{n}\cdot\vec{\tilde{S}}_{1} \vec{\tilde{S}}_{1}\cdot\vec{\lambda}
\bigg\}
+ \frac{\nu}{q} C_{1(\text{ES}^2)} \bigg\{
    \bigg[
	 \frac{\tilde{L}^4}{16 \tilde{r}^7} (1 - 28 \nu + 9 \nu^2) \nl
	 + \frac{\tilde{L}^2}{8 \tilde{r}^6} (64 + 73 \nu - 10 \nu^2)
	 - \frac{1}{4 \tilde{r}^5} (19 + 27 \nu)
         + \frac{\tilde{L}^2 \tilde{p}_r^2}{8 \tilde{r}^5} (7 - 34 \nu + 6 \nu^2)
	 - \frac{\nu \tilde{p}_r^2}{8 \tilde{r}^4} (141 + 26 \nu) \nl
	 + \frac{\tilde{p}_r^4}{16 \tilde{r}^3} (13 + 20 \nu - 72 \nu^2)
    \bigg] \tilde{S}_{1}^2
    - \bigg[
         \frac{\tilde{L}^3 \tilde{p}_r}{8 \tilde{r}^6} (5 + 6 \nu + 9 \nu^2)
	 + \frac{\tilde{L} \tilde{p}_r}{4 \tilde{r}^5} (16 - 156 \nu + 11 \nu^2) \nl
	 + \frac{\tilde{L} \tilde{p}_r^3}{8 \tilde{r}^4} (-1 + 24 \nu + 27 \nu^2)
    \bigg] \vec{n}\cdot\vec{\tilde{S}}_{1} \vec{\tilde{S}}_{1}\cdot\vec{\lambda}
    - \bigg[
	 \frac{3 \tilde{L}^4}{16 \tilde{r}^7} (5 - 16 \nu - 3 \nu^2)
	 + \frac{\tilde{L}^2}{4 \tilde{r}^6} (60 + 11 \nu - 7 \nu^2) \nl
	 - \frac{1}{4 \tilde{r}^5} (37 + 21 \nu)
         + \frac{\tilde{L}^2 \tilde{p}_r^2}{8 \tilde{r}^5} (22 - 66 \nu - 21 \nu^2)
	 + \frac{3 \tilde{p}_r^2}{8 \tilde{r}^4} (36 - 25 \nu - 22 \nu^2) \nl
	 + \frac{\tilde{p}_r^4}{16 \tilde{r}^3} (23 - 36 \nu - 120 \nu^2)
    \bigg] ( \vec{n}\cdot\vec{\tilde{S}}_{1} )^{2}
    + \bigg[
         \frac{\tilde{L}^4}{4 \tilde{r}^7} (1 + 3 \nu - 3 \nu^2)
	 + \frac{\tilde{L}^2}{8 \tilde{r}^6} (4 - 97 \nu + 4 \nu^2) \nl
	 + \frac{\tilde{L}^2 \tilde{p}_r^2}{8 \tilde{r}^5} (5 + 12 \nu - 15 \nu^2)
   \bigg] ( \vec{\tilde{S}}_{1}\cdot\vec{\lambda} )^{2}
\bigg\}  + [ 1 \leftrightarrow 2 ] ,
\end{align}
\begin{align}
&H_\text{LO}^\text{S$^4$} = \frac{\nu^2}{8 \tilde{r}^5} \bigg\{
         \frac{4 C_{1(\text{BS}^3)}}{q} \big[ 15 \vec{n}\cdot\vec{\tilde{S}}_{1} \tilde{S}_{1}^2 \vec{n}\cdot\vec{\tilde{S}}_{2}
	 - 3 \tilde{S}_{1}^2 \vec{\tilde{S}}_{1}\cdot\vec{\tilde{S}}_{2}
	 - 35 \vec{n}\cdot\vec{\tilde{S}}_{2} ( \vec{n}\cdot\vec{\tilde{S}}_{1} )^{3}
	 + 15 \vec{\tilde{S}}_{1}\cdot\vec{\tilde{S}}_{2} ( \vec{n}\cdot\vec{\tilde{S}}_{1} )^{2} \big] \nl
	 - \frac{C_{1(\text{ES}^4)}}{q^2} \big[ 35 ( \vec{n}\cdot\vec{\tilde{S}}_{1} )^{4}
	 - 30 \tilde{S}_{1}^2 ( \vec{n}\cdot\vec{\tilde{S}}_{1} )^{2}
	 + 3 \tilde{S}_{1}^{4} \big]
	 + C_{1(\text{ES}^2)} C_{2(\text{ES}^2)} \big[ 60 \vec{n}\cdot\vec{\tilde{S}}_{1} \vec{n}\cdot\vec{\tilde{S}}_{2} \vec{\tilde{S}}_{1}\cdot\vec{\tilde{S}}_{2}
	 - 3 \tilde{S}_{1}^2 \tilde{S}_{2}^2 \nl
	 + 30 \tilde{S}_{1}^2 ( \vec{n}\cdot\vec{\tilde{S}}_{2} )^{2}
	 - 6 ( \vec{\tilde{S}}_{1}\cdot\vec{\tilde{S}}_{2} )^{2}
	 - 105 ( \vec{n}\cdot\vec{\tilde{S}}_{1} )^{2} ( \vec{n}\cdot\vec{\tilde{S}}_{2} )^{2} \big]
\bigg\} + [ 1 \leftrightarrow 2 ] .
\end{align}
The spin-dependent center-of-mass Hamiltonians up to 3.5PN order, 
consistent with our conventions and gauge choices, can be found in 
\cite{Levi:2015uxa}. The above expressions complete this knowledge to 4PN order.

\section{Conserved integrals of motion: The Poincar\'e algebra}
\label{poincare}

The conservative action is invariant under global Poincar\'e 
transformations. This gives
rise to conserved integrals of motion, which were constructed for a 
non-spinning PN Lagrangian containing higher-order time derivatives in 
\cite{deAndrade:2000gf}.
On phase space the integrals of motion form a representation of the 
Poincar\'e algebra,
which was studied in the PN approximation for the non-spinning 
\cite{Damour:2000kk} and spinning cases
\cite{Damour:2007nc, Steinhoff:2008zr, Hergt:2008jn, Hartung:2013dza}.

Here we construct the conserved quantities on phase space to 4PN order 
from an ansatz. These conserved quantities generate the
Poincar\'e transformations on phase space and hence must fulfill the 
Poincar{\'e} algebra. It turns out that the Poincar{\'e} algebra 
uniquely fixes the ansatz for the conserved quantities.

The Poincar{\'e} algebra reads
\begin{align} \label{poincare1}
 \{P^i, H\} &= 0 , &
 \{J^i, H\} &= 0 , &
 \{J^i, P^j\} &= \epsilon_{ijk} P^k , &
 \{J^i, J^j\} &= \epsilon_{ijk} J^k , \\ \label{poincare2}
 \{J^i, G^j\} &= \epsilon_{ijk} G^k , &
 \{G^i, H\} &= P^i , &
 \{G^i, P^j\} &= \delta_{ij} H , &
 \{G^i, G^j\} &= - \epsilon_{ijk} J^k ,
\end{align}
where $\vec{P}$ is the total linear momentum, $\vec{J}$ is the total 
angular momentum, $\vec{G}$ is the center of mass, and $H$ is the full 
Hamiltonian (including rest-mass terms).
In order to arrive at this form the explicit time dependence of the 
boost generator was split as $K^i = G^i - t P^i$.
Since $\vec{P}$ and $\vec{J}$ are the generators of translations and 
rotations, respectively,
and our gauge choices do not affect these symmetries, one obtains the 
simple representations
\begin{align}
 \vec{P} = \vec{p}_1 + \vec{p}_2 , \qquad
 \vec{J} = \vec{y}_1 \times \vec{p}_1 + \vec{y}_2 \times \vec{p}_2 + \vec{S}_1 + \vec{S}_2 ,
\end{align}
see \cite{Damour:2000kk, Damour:2007nc}. A useful ansatz for the
center of mass is obtained by considering
\begin{equation}
 \{G^i, P^j\} = \delta_{ij} H ,
\end{equation}
which is solved for a $\vec{G}$ of the form \cite{Damour:2000kk}
\begin{equation}
\vec{G} = h_1 \vec{y}_1 + h_2 \vec{y}_2 + \vec{Y} , \qquad
h_1 + h_2 = H ,
\end{equation}
where $h_A$ and $\vec{Y}$ are translation invariant, 
i.e., $\{ h_A, P^j\} = 0$ and $\{ Y^i, P^j\} = 0$.
Due to the condition $h_1 + h_2 = H$ it is useful to write
\begin{equation}\label{hansatz}
h_A = \frac{H}{2} + h_A^\text{PM} + h_A^\text{SO} + h_A^\text{S$_1$S$_2$} + h_A^\text{SS} ,
\end{equation}
where no cubic- and quartic-in-spin contributions are needed. 
The condition $h_1 + h_2 = H$ is now equivalent to the antisymmetry of the point-mass part,
$h_1^\text{PM}$, and the subsequent parts, under exchange of the objects. 
In contrast, the ansatz for $\vec{Y}$ must be symmetric under exchange 
of the objects, and we write it in the form
\begin{equation}\label{Yansatz}
\vec{Y} = \vec{Y}^\text{PM} + \vec{Y}^\text{SO} + \vec{Y}^\text{S$_1$S$_2$} + 
\vec{Y}^\text{SS} ,
\end{equation}
where again no cubic- and quartic-in-spin contributions are needed.
Finally, the ansatz for $\vec{G}$ is uniquely fixed only by
\begin{equation}
 \{G^i, H\} = P^i ,
\end{equation}
where all the other relations in eq.~\eqref{poincare2} are verified to automatically be
fulfilled.

The unique solution for $h_A$ in terms of eq.~(\ref{hansatz}) is given 
by
\begin{align}
h_1^{\text{PM}} =& \frac{m_1}{2} + \frac{1}{4} m_{1} \hat{p}_{1}^2
- \frac{1}{16} m_{1} \hat{p}_{1}^{4}
+ \frac{G^2 m_{1}^2 m_{2}}{4 r^2}
	 - (1 \leftrightarrow 2) ,
\end{align}
\begin{align}
h_1^{\text{SO}} =&
- \frac{G m_{2}}{4 r^2} \vec{S}_{1}\times\vec{n}\cdot\vec{\hat{p}}_{1}
- \frac{G m_{2}}{16 r^2} \Big[ \vec{S}_{1}\times\vec{n}\cdot\vec{\hat{p}}_{1} \big( 5 \hat{p}_{1}^2
	 - 16 \vec{\hat{p}}_{1}\cdot\vec{\hat{p}}_{2}
	 + 8 \hat{p}_{2}^2 -6 ( \vec{\hat{p}}_{2}\cdot\vec{n} )^{2} \big) \nl
	 - 4 \vec{S}_{1}\times\vec{\hat{p}}_{1}\cdot\vec{\hat{p}}_{2} \big( 4 \vec{\hat{p}}_{1}\cdot\vec{n}
	 - 3 \vec{\hat{p}}_{2}\cdot\vec{n} \big) \Big]
+ \frac{G^2 m_{1} m_{2}}{4 r^3} \Big[ 31 \vec{S}_{1}\times\vec{n}\cdot\vec{\hat{p}}_{1}
	 - 4 \vec{S}_{1}\times\vec{n}\cdot\vec{\hat{p}}_{2} \Big] \nl
+ \frac{G^2 m_{2}^2}{4 r^3} \Big[ 15 \vec{S}_{1}\times\vec{n}\cdot\vec{\hat{p}}_{1}
	 + 19 \vec{S}_{1}\times\vec{n}\cdot\vec{\hat{p}}_{2} \Big]
	 - (1 \leftrightarrow 2) ,
\end{align}
\begin{align}
h_1^{\text{S$_1$S$_2$}} =&
- \frac{G}{2 r^3} \Big[ \vec{\hat{p}}_{1}\cdot\vec{S}_{1} \vec{\hat{p}}_{1}\cdot\vec{S}_{2}
	 - \hat{p}_{1}^2 \vec{S}_{1}\cdot\vec{S}_{2}
	 + 3 \hat{p}_{1}^2 \vec{S}_{1}\cdot\vec{n} \vec{S}_{2}\cdot\vec{n}
	 - 3 \vec{\hat{p}}_{1}\cdot\vec{n} \vec{\hat{p}}_{1}\cdot\vec{S}_{1} \vec{S}_{2}\cdot\vec{n} \nl
	 - 3 \vec{\hat{p}}_{2}\cdot\vec{n} \vec{S}_{1}\cdot\vec{n} \vec{\hat{p}}_{1}\cdot\vec{S}_{2}
	 + 3 \vec{\hat{p}}_{1}\cdot\vec{n} \vec{\hat{p}}_{2}\cdot\vec{n} \vec{S}_{1}\cdot\vec{S}_{2} \Big]
+ \frac{G^2 m_{2}}{r^4} \Big[ \vec{S}_{1}\cdot\vec{S}_{2} -2 \vec{S}_{1}\cdot\vec{n} \vec{S}_{2}\cdot\vec{n} \Big] \nl
	 - (1 \leftrightarrow 2) ,
\end{align}
\begin{align}
h_1^{\text{SS}} =& \frac{3 G m_{2}}{4 m_{1} r^3} \Big[ \vec{\hat{p}}_{1}\cdot\vec{S}_{1} \vec{\hat{p}}_{2}\cdot\vec{S}_{1}
	 + \hat{p}_{1}^2 S_{1}^2
	 - \vec{\hat{p}}_{1}\cdot\vec{\hat{p}}_{2} S_{1}^2
	 - ( \vec{\hat{p}}_{1}\cdot\vec{S}_{1} )^{2}
	 + 3 \vec{\hat{p}}_{1}\cdot\vec{n} \vec{S}_{1}\cdot\vec{n} \vec{\hat{p}}_{1}\cdot\vec{S}_{1} \nl
	 - \vec{\hat{p}}_{2}\cdot\vec{n} \vec{S}_{1}\cdot\vec{n} \vec{\hat{p}}_{1}\cdot\vec{S}_{1}
	 - 2 \vec{\hat{p}}_{1}\cdot\vec{n} \vec{S}_{1}\cdot\vec{n} \vec{\hat{p}}_{2}\cdot\vec{S}_{1}
	 + \vec{\hat{p}}_{1}\cdot\vec{n} \vec{\hat{p}}_{2}\cdot\vec{n} S_{1}^2
	 - S_{1}^2 ( \vec{\hat{p}}_{1}\cdot\vec{n} )^{2}
	 - 2 \hat{p}_{1}^2 ( \vec{S}_{1}\cdot\vec{n} )^{2} \nl
	 + 2 \vec{\hat{p}}_{1}\cdot\vec{\hat{p}}_{2} ( \vec{S}_{1}\cdot\vec{n} )^{2} \Big]
+ \frac{G^2 C_{1(\text{ES}^2)} m_{2}}{4 r^4} \Big[ S_{1}^2 -3 ( \vec{S}_{1}\cdot\vec{n} )^{2} \Big]
	 - \frac{G^2 m_{2}^2}{m_{1} r^4} \Big[ S_{1}^2
	 - ( \vec{S}_{1}\cdot\vec{n} )^{2} \Big] \nl
	 - \frac{G^2 C_{1(\text{ES}^2)} m_{2}^2}{2 m_{1} r^4} \Big[ 5 S_{1}^2 -6 ( \vec{S}_{1}\cdot\vec{n} )^{2} \Big]
+ \frac{5 G^2 m_{2}}{2 r^4} ( \vec{S}_{1}\cdot\vec{n} )^{2}
	 - (1 \leftrightarrow 2) ,
\end{align}
and similarly for $h_2$. 
The solution for $\vec{Y}$ in terms of eq.~(\ref{Yansatz}) is given by
\begin{align}
\vec{Y}^{\text{PM}} =&
- \frac{1}{4} G m_{1} m_{2} \vec{\hat{p}}_{2}\cdot\vec{n} \vec{\hat{p}}_{1}
	 + (1 \leftrightarrow 2) ,
\end{align}
\begin{align}
\vec{Y}^{\text{SO}} =&
- \frac{1}{2} \vec{S}_{1}\times\vec{\hat{p}}_{1}
+ \frac{1}{8} \hat{p}_{1}^2 \vec{S}_{1}\times\vec{\hat{p}}_{1}
- \frac{G m_{2}}{4 r} \Big[ \vec{S}_{1}\times\vec{n} \vec{\hat{p}}_{2}\cdot\vec{n}
	 - 10 \vec{S}_{1}\times\vec{\hat{p}}_{1}
	 + 11 \vec{S}_{1}\times\vec{\hat{p}}_{2} \Big] \nl
- \frac{1}{16} \vec{S}_{1}\times\vec{\hat{p}}_{1} \hat{p}_{1}^{4}
+ \frac{G m_{2}}{16 r} \Big[ 8 \vec{S}_{1}\times\vec{n}\cdot\vec{\hat{p}}_{1} \big( \vec{\hat{p}}_{2}\cdot\vec{n} \vec{\hat{p}}_{1}
	 - \vec{\hat{p}}_{1}\cdot\vec{n} \vec{\hat{p}}_{2} \big)
	 - 8 \vec{S}_{1}\times\vec{n}\cdot\vec{\hat{p}}_{2} \big( \vec{\hat{p}}_{2}\cdot\vec{n} \vec{\hat{p}}_{1} \nl
	 - \vec{\hat{p}}_{1}\cdot\vec{n} \vec{\hat{p}}_{2} \big)
	 - 8 \vec{S}_{1}\times\vec{\hat{p}}_{1}\cdot\vec{\hat{p}}_{2} \vec{\hat{p}}_{2}
	 - \vec{S}_{1}\times\vec{n} \big( 5 \hat{p}_{1}^2 \vec{\hat{p}}_{2}\cdot\vec{n}
	 - 14 \vec{\hat{p}}_{2}\cdot\vec{n} \vec{\hat{p}}_{1}\cdot\vec{\hat{p}}_{2}
	 + \vec{\hat{p}}_{1}\cdot\vec{n} \hat{p}_{2}^2
	 + 7 \vec{\hat{p}}_{2}\cdot\vec{n} \hat{p}_{2}^2 \nl
         - 3 \vec{\hat{p}}_{1}\cdot\vec{n} ( \vec{\hat{p}}_{2}\cdot\vec{n} )^{2}
	 - 3 ( \vec{\hat{p}}_{2}\cdot\vec{n} )^{3} \big)
	 - \vec{S}_{1}\times\vec{\hat{p}}_{1} \big( 2 \hat{p}_{1}^2
	 + 34 \vec{\hat{p}}_{1}\cdot\vec{\hat{p}}_{2}
	 - 25 \hat{p}_{2}^2
	 + 6 \vec{\hat{p}}_{1}\cdot\vec{n} \vec{\hat{p}}_{2}\cdot\vec{n} \nl
	 + 13 ( \vec{\hat{p}}_{2}\cdot\vec{n} )^{2} \big)
	 - \vec{S}_{1}\times\vec{\hat{p}}_{2} \big( 3 \hat{p}_{1}^2
	 - 26 \vec{\hat{p}}_{1}\cdot\vec{\hat{p}}_{2}
	 + 9 \hat{p}_{2}^2 -6 \vec{\hat{p}}_{1}\cdot\vec{n} \vec{\hat{p}}_{2}\cdot\vec{n}
	 - 13 ( \vec{\hat{p}}_{2}\cdot\vec{n} )^{2} \big) \Big] \nl
+ \frac{G^2 m_{2}^2}{8 r^2} \Big[ \vec{S}_{1}\times\vec{n} \big( 12 \vec{\hat{p}}_{1}\cdot\vec{n}
	 + 59 \vec{\hat{p}}_{2}\cdot\vec{n} \big)
	 - 20 \vec{S}_{1}\times\vec{\hat{p}}_{1}
	 + 30 \vec{S}_{1}\times\vec{\hat{p}}_{2} \Big] \nl
+ \frac{G^2 m_{1} m_{2}}{8 r^2} \Big[ 3 \vec{S}_{1}\times\vec{n} \big( 14 \vec{\hat{p}}_{1}\cdot\vec{n}
	 + 3 \vec{\hat{p}}_{2}\cdot\vec{n} \big)
	 - 43 \vec{S}_{1}\times\vec{\hat{p}}_{1}
	 + 34 \vec{S}_{1}\times\vec{\hat{p}}_{2} \Big]
	 + (1 \leftrightarrow 2) ,
\end{align}
\begin{align}
\vec{Y}^{\text{S$_1$S$_2$}} =& \frac{3 G}{2 r^2} \Big[ \vec{S}_{1}\cdot\vec{S}_{2} \vec{n}
	 - \vec{S}_{1}\cdot\vec{n} \vec{S}_{2} \Big]
+ \frac{G}{8 r^2} \Big[ 4 \vec{\hat{p}}_{1}\cdot\vec{S}_{1} \vec{\hat{p}}_{1}\cdot\vec{S}_{2} \vec{n}
	 + 16 \vec{\hat{p}}_{2}\cdot\vec{S}_{1} \vec{\hat{p}}_{1}\cdot\vec{S}_{2} \vec{n}
	 + 8 \vec{\hat{p}}_{1}\cdot\vec{S}_{1} \vec{\hat{p}}_{2}\cdot\vec{S}_{2} \vec{n} \nl
	 - 6 \hat{p}_{1}^2 \vec{S}_{1}\cdot\vec{S}_{2} \vec{n}
	 - 34 \vec{\hat{p}}_{1}\cdot\vec{\hat{p}}_{2} \vec{S}_{1}\cdot\vec{S}_{2} \vec{n}
	 - 14 \vec{\hat{p}}_{1}\cdot\vec{S}_{1} \vec{S}_{2}\cdot\vec{n} \vec{\hat{p}}_{1}
	 + 11 \vec{\hat{p}}_{2}\cdot\vec{S}_{1} \vec{S}_{2}\cdot\vec{n} \vec{\hat{p}}_{1} \nl
	 + 12 \vec{S}_{1}\cdot\vec{n} \vec{\hat{p}}_{1}\cdot\vec{S}_{2} \vec{\hat{p}}_{1}
	 - 8 \vec{S}_{1}\cdot\vec{n} \vec{\hat{p}}_{2}\cdot\vec{S}_{2} \vec{\hat{p}}_{1}
	 - 4 \vec{\hat{p}}_{1}\cdot\vec{n} \vec{S}_{1}\cdot\vec{S}_{2} \vec{\hat{p}}_{1}
	 + 4 \vec{\hat{p}}_{2}\cdot\vec{n} \vec{S}_{1}\cdot\vec{S}_{2} \vec{\hat{p}}_{1} \nl
	 - 9 \vec{\hat{p}}_{1}\cdot\vec{S}_{1} \vec{S}_{2}\cdot\vec{n} \vec{\hat{p}}_{2}
	 - 14 \vec{S}_{1}\cdot\vec{n} \vec{\hat{p}}_{1}\cdot\vec{S}_{2} \vec{\hat{p}}_{2}
	 + 26 \vec{\hat{p}}_{1}\cdot\vec{n} \vec{S}_{1}\cdot\vec{S}_{2} \vec{\hat{p}}_{2}
	 - 12 \vec{\hat{p}}_{2}\cdot\vec{n} \vec{S}_{1}\cdot\vec{S}_{2} \vec{\hat{p}}_{2} \nl
	 + 18 \hat{p}_{1}^2 \vec{S}_{2}\cdot\vec{n} \vec{S}_{1}
	 - 8 \vec{\hat{p}}_{1}\cdot\vec{n} \vec{\hat{p}}_{1}\cdot\vec{S}_{2} \vec{S}_{1}
	 + \vec{\hat{p}}_{2}\cdot\vec{n} \vec{\hat{p}}_{1}\cdot\vec{S}_{2} \vec{S}_{1}
	 - 12 \hat{p}_{1}^2 \vec{S}_{1}\cdot\vec{n} \vec{S}_{2}
	 + 34 \vec{\hat{p}}_{1}\cdot\vec{\hat{p}}_{2} \vec{S}_{1}\cdot\vec{n} \vec{S}_{2} \nl
	 + 10 \vec{\hat{p}}_{1}\cdot\vec{n} \vec{\hat{p}}_{1}\cdot\vec{S}_{1} \vec{S}_{2}
	 - 7 \vec{\hat{p}}_{2}\cdot\vec{n} \vec{\hat{p}}_{1}\cdot\vec{S}_{1} \vec{S}_{2}
	 - 30 \vec{\hat{p}}_{1}\cdot\vec{n} \vec{\hat{p}}_{2}\cdot\vec{S}_{1} \vec{S}_{2}
	 + 12 \vec{\hat{p}}_{2}\cdot\vec{n} \vec{\hat{p}}_{2}\cdot\vec{S}_{1} \vec{S}_{2} \nl
	 + 12 \vec{\hat{p}}_{2}\cdot\vec{n} \vec{\hat{p}}_{1}\cdot\vec{S}_{1} \vec{S}_{2}\cdot\vec{n} \vec{n}
	 + 3 \vec{\hat{p}}_{1}\cdot\vec{n} \vec{\hat{p}}_{2}\cdot\vec{S}_{1} \vec{S}_{2}\cdot\vec{n} \vec{n}
	 - 12 \vec{\hat{p}}_{2}\cdot\vec{n} \vec{\hat{p}}_{2}\cdot\vec{S}_{1} \vec{S}_{2}\cdot\vec{n} \vec{n} \nl
	 - 21 \vec{\hat{p}}_{1}\cdot\vec{n} \vec{\hat{p}}_{2}\cdot\vec{n} \vec{S}_{1}\cdot\vec{S}_{2} \vec{n}
	 - 6 \vec{\hat{p}}_{2}\cdot\vec{n} \vec{S}_{1}\cdot\vec{n} \vec{S}_{2}\cdot\vec{n} \vec{\hat{p}}_{1}
	 - 3 \vec{\hat{p}}_{1}\cdot\vec{n} \vec{S}_{1}\cdot\vec{n} \vec{S}_{2}\cdot\vec{n} \vec{\hat{p}}_{2} \nl
	 + 12 \vec{\hat{p}}_{2}\cdot\vec{n} \vec{S}_{1}\cdot\vec{n} \vec{S}_{2}\cdot\vec{n} \vec{\hat{p}}_{2}
	 + 21 \vec{\hat{p}}_{1}\cdot\vec{n} \vec{\hat{p}}_{2}\cdot\vec{n} \vec{S}_{1}\cdot\vec{n} \vec{S}_{2}
	 + 12 \vec{S}_{1}\cdot\vec{S}_{2} \vec{n} ( \vec{\hat{p}}_{1}\cdot\vec{n} )^{2} \nl
	 - 12 \vec{S}_{2}\cdot\vec{n} \vec{S}_{1} ( \vec{\hat{p}}_{1}\cdot\vec{n} )^{2}
	 + 12 \vec{S}_{1}\cdot\vec{S}_{2} \vec{n} ( \vec{\hat{p}}_{2}\cdot\vec{n} )^{2}
	 - 12 \vec{S}_{1}\cdot\vec{n} \vec{S}_{2} ( \vec{\hat{p}}_{2}\cdot\vec{n} )^{2} \Big] \nl
- \frac{G^2 m_{2}}{8 r^3} \Big[ \vec{S}_{1}\cdot\vec{S}_{2} \vec{n}
	 + 39 \vec{S}_{2}\cdot\vec{n} \vec{S}_{1}
	 - 40 \vec{S}_{1}\cdot\vec{n} \vec{S}_{2} \Big]
	 + (1 \leftrightarrow 2) ,
\end{align}
\begin{align}
\vec{Y}^{\text{SS}} =& \frac{G m_{2}}{2 m_{1} r^2} \Big[ S_{1}^2 \vec{n}
	 - \vec{S}_{1}\cdot\vec{n} \vec{S}_{1} \Big]
+ \frac{G C_{1(\text{ES}^2)} m_{2}}{2 m_{1} r^2} \Big[ S_{1}^2 \vec{n}
	 - \vec{S}_{1}\cdot\vec{n} \vec{S}_{1} \Big]
+ \frac{G C_{1(\text{ES}^2)} m_{2}}{8 m_{1} r^2} \Big[ 6 \hat{p}_{1}^2 S_{1}^2 \vec{n} \nl
	 - 14 \vec{\hat{p}}_{1}\cdot\vec{\hat{p}}_{2} S_{1}^2 \vec{n}
	 + 6 \hat{p}_{2}^2 S_{1}^2 \vec{n}
	 + 2 \vec{S}_{1}\cdot\vec{n} \vec{\hat{p}}_{1}\cdot\vec{S}_{1} \vec{\hat{p}}_{1}
	 - 2 \vec{S}_{1}\cdot\vec{n} \vec{\hat{p}}_{2}\cdot\vec{S}_{1} \vec{\hat{p}}_{1}
	 - 4 \vec{\hat{p}}_{1}\cdot\vec{n} S_{1}^2 \vec{\hat{p}}_{1} \nl
	 + 3 \vec{\hat{p}}_{2}\cdot\vec{n} S_{1}^2 \vec{\hat{p}}_{1}
	 + 2 \vec{S}_{1}\cdot\vec{n} \vec{\hat{p}}_{1}\cdot\vec{S}_{1} \vec{\hat{p}}_{2}
	 + \vec{\hat{p}}_{1}\cdot\vec{n} S_{1}^2 \vec{\hat{p}}_{2}
	 - 6 \hat{p}_{1}^2 \vec{S}_{1}\cdot\vec{n} \vec{S}_{1}
	 + 14 \vec{\hat{p}}_{1}\cdot\vec{\hat{p}}_{2} \vec{S}_{1}\cdot\vec{n} \vec{S}_{1} \nl
	 - 6 \hat{p}_{2}^2 \vec{S}_{1}\cdot\vec{n} \vec{S}_{1}
	 + 2 \vec{\hat{p}}_{1}\cdot\vec{n} \vec{\hat{p}}_{1}\cdot\vec{S}_{1} \vec{S}_{1}
	 - 2 \vec{\hat{p}}_{2}\cdot\vec{n} \vec{\hat{p}}_{1}\cdot\vec{S}_{1} \vec{S}_{1}
	 - 2 \vec{\hat{p}}_{1}\cdot\vec{n} \vec{\hat{p}}_{2}\cdot\vec{S}_{1} \vec{S}_{1} -6 \vec{\hat{p}}_{1}\cdot\vec{n} \vec{\hat{p}}_{2}\cdot\vec{n} S_{1}^2 \vec{n} \nl
	 + 6 \vec{\hat{p}}_{1}\cdot\vec{n} \vec{\hat{p}}_{2}\cdot\vec{n} \vec{S}_{1}\cdot\vec{n} \vec{S}_{1}
	 + 3 \vec{\hat{p}}_{2}\cdot\vec{n} \vec{\hat{p}}_{1} ( \vec{S}_{1}\cdot\vec{n} )^{2}
	 - 3 \vec{\hat{p}}_{1}\cdot\vec{n} \vec{\hat{p}}_{2} ( \vec{S}_{1}\cdot\vec{n} )^{2} \Big] \nl
+ \frac{G m_{2}}{16 m_{1} r^2} \Big[ 6 \vec{\hat{p}}_{1}\cdot\vec{S}_{1} \vec{\hat{p}}_{2}\cdot\vec{S}_{1} \vec{n}
	 + \hat{p}_{1}^2 S_{1}^2 \vec{n}
	 - 6 \vec{\hat{p}}_{1}\cdot\vec{\hat{p}}_{2} S_{1}^2 \vec{n}
	 + 8 \hat{p}_{2}^2 S_{1}^2 \vec{n}
	 - 28 \vec{S}_{1}\cdot\vec{n} \vec{\hat{p}}_{2}\cdot\vec{S}_{1} \vec{\hat{p}}_{1} \nl
	 + 4 \vec{\hat{p}}_{1}\cdot\vec{n} S_{1}^2 \vec{\hat{p}}_{1}
	 + 40 \vec{\hat{p}}_{2}\cdot\vec{n} S_{1}^2 \vec{\hat{p}}_{1}
	 + 22 \vec{S}_{1}\cdot\vec{n} \vec{\hat{p}}_{1}\cdot\vec{S}_{1} \vec{\hat{p}}_{2}
	 + 8 \vec{S}_{1}\cdot\vec{n} \vec{\hat{p}}_{2}\cdot\vec{S}_{1} \vec{\hat{p}}_{2}
	 - 22 \vec{\hat{p}}_{1}\cdot\vec{n} S_{1}^2 \vec{\hat{p}}_{2} \nl
	 - 16 \vec{\hat{p}}_{2}\cdot\vec{n} S_{1}^2 \vec{\hat{p}}_{2}
	 - \hat{p}_{1}^2 \vec{S}_{1}\cdot\vec{n} \vec{S}_{1}
	 + 6 \vec{\hat{p}}_{1}\cdot\vec{\hat{p}}_{2} \vec{S}_{1}\cdot\vec{n} \vec{S}_{1}
	 - 8 \hat{p}_{2}^2 \vec{S}_{1}\cdot\vec{n} \vec{S}_{1}
	 - 2 \vec{\hat{p}}_{1}\cdot\vec{n} \vec{\hat{p}}_{1}\cdot\vec{S}_{1} \vec{S}_{1} \nl
	 - 40 \vec{\hat{p}}_{2}\cdot\vec{n} \vec{\hat{p}}_{1}\cdot\vec{S}_{1} \vec{S}_{1}
	 + 22 \vec{\hat{p}}_{1}\cdot\vec{n} \vec{\hat{p}}_{2}\cdot\vec{S}_{1} \vec{S}_{1}
	 + 16 \vec{\hat{p}}_{2}\cdot\vec{n} \vec{\hat{p}}_{2}\cdot\vec{S}_{1} \vec{S}_{1}
	 - 2 \vec{n} ( \vec{\hat{p}}_{1}\cdot\vec{S}_{1} )^{2}
	 - 8 \vec{n} ( \vec{\hat{p}}_{2}\cdot\vec{S}_{1} )^{2} \nl
	 + 30 \vec{\hat{p}}_{2}\cdot\vec{n} \vec{S}_{1}\cdot\vec{n} \vec{\hat{p}}_{1}\cdot\vec{S}_{1} \vec{n}
	 - 24 \vec{\hat{p}}_{2}\cdot\vec{n} \vec{S}_{1}\cdot\vec{n} \vec{\hat{p}}_{2}\cdot\vec{S}_{1} \vec{n}
	 - 30 \vec{\hat{p}}_{1}\cdot\vec{n} \vec{\hat{p}}_{2}\cdot\vec{n} S_{1}^2 \vec{n} \nl
	 + 30 \vec{\hat{p}}_{1}\cdot\vec{n} \vec{\hat{p}}_{2}\cdot\vec{n} \vec{S}_{1}\cdot\vec{n} \vec{S}_{1}
	 + 12 S_{1}^2 \vec{n} ( \vec{\hat{p}}_{2}\cdot\vec{n} )^{2}
	 - 12 \vec{S}_{1}\cdot\vec{n} \vec{S}_{1} ( \vec{\hat{p}}_{2}\cdot\vec{n} )^{2}
	 - 30 \vec{\hat{p}}_{2}\cdot\vec{n} \vec{\hat{p}}_{1} ( \vec{S}_{1}\cdot\vec{n} )^{2} \nl
	 + 24 \vec{\hat{p}}_{2}\cdot\vec{n} \vec{\hat{p}}_{2} ( \vec{S}_{1}\cdot\vec{n} )^{2} \Big]
- \frac{3 G^2 m_{2}}{2 r^3} \Big[ S_{1}^2 \vec{n}
	 - \vec{S}_{1}\cdot\vec{n} \vec{S}_{1} \Big]
	 - \frac{G^2 C_{1(\text{ES}^2)} m_{2}}{2 r^3} \Big[ S_{1}^2 \vec{n}
	 - \vec{S}_{1}\cdot\vec{n} \vec{S}_{1} \Big] \nl
	 - \frac{4 G^2 m_{2}^2}{m_{1} r^3} \Big[ S_{1}^2 \vec{n}
	 - \vec{S}_{1}\cdot\vec{n} \vec{S}_{1} \Big]
+ \frac{23 G^2 C_{1(\text{ES}^2)} m_{2}^2}{2 m_{1} r^3} \Big[ S_{1}^2 \vec{n}
	 - \vec{S}_{1}\cdot\vec{n} \vec{S}_{1} \Big]
	 + (1 \leftrightarrow 2) .
\end{align}
This completes the construction of the integrals of motion and the Poincar{\'e} algebra, 
which provides a strong check for our new NNLO spin-squared Hamiltonian.



\section{Complete gauge-invariant relations to 4PN order with spins}
\label{GI}

We proceed with a completion of the gauge-invariant relations to 4PN 
order with spins following section 8 in \cite{Levi:2014sba}, 
and section 5 in \cite{Levi:2015uxa}.
These relations express the binding energy in terms of the orbital angular momentum or the
orbital frequency. The relations hold for circular orbits and spins aligned with the orbital
angular momentum. The spin-dependent relations to 3.5PN order can be found 
in \cite{Levi:2014sba, Levi:2015uxa}.

Let us briefly recall the necessary derivations from section 8 in \cite{Levi:2014sba}. For 
circular orbits we have that
\begin{equation}\label{circularcondition}
\dot{\tilde{p}}_r = - \frac{\partial \tilde{H}(\tilde{r}, \tilde{L})}{\partial \tilde{r}}=0, 
\end{equation}
where $\tilde{H}\equiv H/\mu$, and this relation can be solved to yield the constant radius 
$\tilde{r}$ in terms of $\tilde{L}$, $\tilde{r}(\tilde{L})$.
For the orbital angular frequency, $\tilde{\omega}$, we have
\begin{equation}\label{orbitalfrequency}
\frac{d\phi}{d\tilde{t}}\equiv\tilde{\omega} = \frac{\partial \tilde{H}(\tilde{r}, \tilde{L})}{\partial \tilde{L}} ,
\end{equation}
where $\phi$ is the orbital phase, and $\tilde{t}\equiv t/Gm$.

In order to obtain the gauge-invariant binding energy, the gauge-dependent radial coordinate 
must be substituted, where its expansion for a circular orbit from 
eq.~\eqref{circularcondition} reads
\begin{align} \label{rfuncofL}
\frac{1}{\tilde{r}} =& \dots
+ \frac{1}{\tilde{L}^{10}} \bigg\{
         \bigg[ (100403 - 6755 \nu) \frac{\nu^2}{112}
             + \left(\frac{538}{7} \nu^2 - 3 \nu^3\right) C_{1(\text{ES}^2)}
         \bigg] \tilde{S}_1^2 \nl
         + \bigg[ 10747 + 9751 \nu - 909 \nu^2
	     + (5064 + 7 \nu - 48 \nu^2) C_{1(\text{ES}^2)}
         \bigg] \frac{\nu \tilde{S}_1^2}{16 q} \nl
         + \bigg[ \frac{11349}{4} - \frac{665}{8} \nu - 7 \nu^2
         \bigg] \frac{\nu}{2} \tilde{S}_1 \tilde{S}_2
	 + \frac{9 \nu^2}{8} (8 + 9 C_{1(\text{ES}^2)} C_{2(\text{ES}^2)}) \tilde{S}_1^2 \tilde{S}_2^2 \nl
         + \frac{3 \nu^2}{8 q^2} (5 C_{1(\text{ES}^4)} + 12 C_{1(\text{ES}^2)}^2) \tilde{S}_1^4
	 + \frac{3 \nu^2}{2 q} (5 C_{1(\text{BS}^3)} + 12 C_{1(\text{ES}^2)}) \tilde{S}_1^3 \tilde{S}_2
         + [1 \leftrightarrow 2]
\bigg\} ,
\end{align}
and the dots denote the terms up to 3.5PN order given in eq.~(5.1) of \cite{Levi:2015uxa}.
Now, using eq.~\eqref{orbitalfrequency}, switching to the parameter 
$x\equiv\tilde{\omega}^{2/3}$, and plugging in $\tilde{r}(\tilde{L})$ from 
eq.~\eqref{rfuncofL}, we get the relation $x(\tilde{L})$. We then invert it to get 
$\tilde{L}(x)$, and similarly, 
extend eq.~(5.2) of \cite{Levi:2015uxa} to 4PN order, which leads to
\begin{align} \label{xfuncofL}
\frac{1}{\tilde{L}^2} =& \dots
+ x^5 \bigg\{
\bigg[\frac{65}{24} (549 - 98 \nu)
	 + (117 + 518 \nu) C_{1(\text{ES}^2)} \bigg] \frac{\nu^2 \tilde{S}_1^2}{63} \nl
- \bigg[ \frac{1}{12} (6777 + 2964 \nu + 419 \nu^2)
	 + (29 - 44 \nu) \nu C_{1(\text{ES}^2)} \bigg] \frac{\nu \tilde{S}_1^2}{6 q} \nl
+ \bigg[ \frac{155 \nu^2}{108}-\frac{8177 \nu}{72}-\frac{313}{2} \bigg] \frac{\nu}{2} \tilde{S}_1 \tilde{S}_2
+ \frac{\nu^2}{2} (22 - 4 C_{1(\text{ES}^2)} C_{2(\text{ES}^2)}) \tilde{S}_1^2 \tilde{S}_2^2 \nl
+ (11 C_{1(\text{ES}^2)}^2 - 5 C_{1(\text{ES}^4)}) \frac{\nu^2 \tilde{S}_1^4}{2 q^2}
+ (22 C_{1(\text{ES}^2)} - 10 C_{1(\text{BS}^3)}) \frac{\nu^2}{q} \tilde{S}_1^3 \tilde{S}_2
+ [1 \leftrightarrow 2]
\bigg\} .
\end{align}
This relation between the orbital angular momentum and the orbital frequency is in fact also 
gauge invariant for the configuration considered.
Again, using $\tilde{r}(\tilde{L})$ from eq.~\eqref{rfuncofL}, the binding energy as a 
function of the orbital angular momentum is defined as 
$e(\tilde{L})=\tilde{H}(\tilde{r}(\tilde{L}),\tilde{L})$.
The gauge-invariant relations for the binding energy, using also $\tilde{L}(x)$ from 
eq.~\eqref{xfuncofL}, then read
\begin{align}
e_\text{spin}^\text{4PN}(\tilde{L}) =& \frac{1}{\tilde{L}^{10}} \bigg\{
- \frac{\nu^2 \tilde{S}_1^2}{28} \bigg[ \frac{13}{4} (1380 + 7 \nu)
	 + (429 + 14 \nu) C_{1(\text{ES}^2)} \bigg] \nl
+ \frac{\nu \tilde{S}_1^2}{16 q} \bigg[ 3 (-672 - 637 \nu + 3 \nu^2)
	 - (819 + 165 \nu + 11 \nu^2) C_{1(\text{ES}^2)} \bigg] \nl
- \nu  \tilde{S}_1 \tilde{S}_2
         \bigg[ \frac{25 \nu^2}{16}+\frac{507 \nu}{32}+\frac{4041}{16} \bigg]
- \frac{9 \nu^2}{4} \tilde{S}_1^2 \tilde{S}_2^2 (1 + C_{1(\text{ES}^2)} C_{2(\text{ES}^2)}) \nl
- \frac{3 \nu^2}{8 q^2} \tilde{S}_1^4 (C_{1(\text{ES}^4)} + 3 C_{1(\text{ES}^2)}^2)
- \frac{3 \nu^2}{2 q} \tilde{S}_1^3 \tilde{S}_2 (C_{1(\text{BS}^3)} + 3 C_{1(\text{ES}^2)})
+ [1 \leftrightarrow 2] \bigg\} ,
\end{align}
\begin{align}
e_\text{spin}^\text{4PN}(x) =& x^5 \bigg\{
\frac{\nu^2 \tilde{S}_1^2}{36} \bigg[ \frac{1}{3} (360 - 749 \nu)
	 + (279 - 70 \nu) C_{1(\text{ES}^2)}\bigg] \nl
- \frac{7 \nu \tilde{S}_1^2}{12 q} \bigg[ \frac{1}{3} (27 + 6 \nu + 31 \nu^2)
	 + \frac{1}{4} (-27 - 11 \nu + 13 \nu^2) C_{1(\text{ES}^2)} \bigg] \nl
+ \frac{7 \nu}{432} \tilde{S}_1 \tilde{S}_2 (135 - 429 \nu - 53 \nu^2)
+ \frac{7 \nu^2}{4} \tilde{S}_1^2 \tilde{S}_2^2 (C_{1(\text{ES}^2)} C_{2(\text{ES}^2)} - 1) \nl
+ \frac{7 \nu^2}{8 q^2} \tilde{S}_1^4 (C_{1(\text{ES}^4)} - C_{1(\text{ES}^2)}^2)
+ \frac{7 \nu^2}{2 q} \tilde{S}_1^3 \tilde{S}_2 (C_{1(\text{BS}^3)} - C_{1(\text{ES}^2)})
+ [1 \leftrightarrow 2] \bigg\} ,
\end{align}
with the spin-dependent lower orders given in eqs.~(5.3) and (5.4) of \cite{Levi:2015uxa}, 
and the spin-independent part given in \cite{Damour:2014jta}. The relation $e(x)$ can be 
directly applied to improve the phasing of gravitational waveforms for the GW detectors.

\section{Conclusions} 
\label{theendmyfriend}

In this paper we completed the spin-dependent conservative dynamics of 
inspiralling compact binaries at the 4PN order, and in 
particular the derivation of the NNLO spin-squared interaction 
potential \cite{Levi:2015ixa}. These high PN orders, in particular 
taking into account the spins of the binary 
constituents, will enable to gain more accurate information from 
even more sensitive GW detections to come. 

We have derived the physical EOMs of the position and the spin, and the quadratic-in-spin 
Hamiltonians, as well as their expressions in the center-of-mass frame. We have constructed 
the conserved integrals of motion, which form the Poincar\'e algebra. As there is 
currently no other independent derivation of the NNLO spin-squared 
interaction other than \cite{Levi:2015ixa}, this construction actually 
provided a crucial consistency check for the validity of our result. 
Finally, we provide complete gauge-invariant relations 
among the binding energy, angular momentum, and orbital frequency of an 
inspiralling binary with generic compact spinning components to the 4PN order. 

We note that this research article appeared first as preprint ArXiv ePrint: 1607.04252v1. 
The present version 
contains minor revisions to this preprint's version, made only to enhance readability, but no 
changes whatsoever have been introduced in the research content. Moreover, to date this article 
still constitutes the only work that obtained the complete conservative dynamics for generic 
inspiralling compact binaries with spins at the state of the art in PN theory, i.e.~at the 4PN 
order, including the complete NNLO quadratic-in-spin sector.

Let us thus note subsequent applications and further checks of our results, which were
based on our earlier public preprint of this paper. An effective-one-body waveform model
matched to our aligned-spin binding energy at the 4PN order was derived in 
\cite{Nagar:2018plt}, 
and the 4PN spin-squared corrections in eq.~(5.6) were found to have a non-negligible effect 
on the waveform of binary neutron star inspirals.
Initial studies to extend such waveform models to precessing spins, including all
conservative spin effects at the 4PN order, have been performed more recently 
in~\cite{Khalil:2020mmr}. 

The 4PN spin-squared interactions in the center-of-mass Hamiltonian presented here were
partially checked against the redshift derived from the gravitational self-force on a small
rotating black hole for circular orbits around a non-rotating black hole~\cite{Bini:2020zqy}. 
They were also partially checked against the gravitational scattering calculated to 
NLO in a weak-field (post-Minkowskian) approximation based on scattering 
amplitudes \cite{Kosmopoulos:2021zoq}. We highlight that these are recent independent, but so 
far partial, confirmations of our
results and applications thereof, within various frameworks, i.e.~effective-one-body,
self-force, scattering amplitudes.

\acknowledgments

ML is supported by the European Research Council under the European 
Community's Seventh Framework Programme (FP7/2007-2013 Grant Agreement 
No.~307934, NIRG project). 
This work has been done within the LABEX ILP (reference ANR-10-LABX-63) 
part of the Idex SUPER, and received  financial French state aid managed 
by the Agence Nationale de la Recherche, as part of the programme 
Investissements d'Avenir under the reference ANR-11-IDEX-0004-02.

\appendix

\section{Equivalent NNLO spin1-spin2 potential}
\label{nnlos1s2}

The NNLO spin1-spin2 sector was already completed in earlier work 
\cite{Levi:2011eq, Hartung:2011ea}.
Yet, we want to complete here the line of work from
\cite{Levi:2015msa, Levi:2015uxa, Levi:2015ixa}, 
which follows specific different conventions and gauge choices. 
For completeness we therefore present in this appendix the NNLO 
spin1-spin2 potential obtained within the formulation and gauge choices of 
\cite{Levi:2015msa}. This is needed to complete all spin potentials to 4PN order in a 
consistent manner.

Due to the length of the potential we split it according to the number of higher-order 
time derivatives as follows:
\begin{equation}
V_{\text{NNLO}}^{\text{S$_1$S$_2$}} = \stackrel{(0)}{V} + \stackrel{(1)}{V} + \stackrel{(2)}{V} + \stackrel{(3)}{V} + \stackrel{(4)}{V} ,
\end{equation}
and
\begin{equation}
\stackrel{(1)}{V} = V_a + V_{\dot{S}} .
\end{equation}
These specific parts are given then by
\begin{align}
&\stackrel{(0)}{V} = \frac{G}{8 r^3} \Big[ 15 \vec{\tilde{S}}_{1}\cdot\vec{v}_{1} \vec{\tilde{S}}_{2}\cdot\vec{v}_{1} v_{1}^2
	 - 14 \vec{\tilde{S}}_{2}\cdot\vec{v}_{1} v_{1}^2 \vec{\tilde{S}}_{1}\cdot\vec{v}_{2}
	 - 2 v_{1}^2 \vec{\tilde{S}}_{1}\cdot\vec{v}_{2} \vec{\tilde{S}}_{2}\cdot\vec{v}_{2}
	 - 4 \vec{\tilde{S}}_{1}\cdot\vec{v}_{1} \vec{\tilde{S}}_{2}\cdot\vec{v}_{1} \vec{v}_{1}\cdot\vec{v}_{2} \nl
	 + 18 \vec{\tilde{S}}_{1}\cdot\vec{\tilde{S}}_{2} v_{1}^2 \vec{v}_{1}\cdot\vec{v}_{2}
	 + 18 \vec{\tilde{S}}_{2}\cdot\vec{v}_{1} \vec{\tilde{S}}_{1}\cdot\vec{v}_{2} \vec{v}_{1}\cdot\vec{v}_{2}
	 - 8 \vec{\tilde{S}}_{1}\cdot\vec{v}_{1} \vec{\tilde{S}}_{2}\cdot\vec{v}_{2} \vec{v}_{1}\cdot\vec{v}_{2}
	 - 4 \vec{\tilde{S}}_{1}\cdot\vec{v}_{2} \vec{\tilde{S}}_{2}\cdot\vec{v}_{2} \vec{v}_{1}\cdot\vec{v}_{2} \nl
	 - 2 \vec{\tilde{S}}_{1}\cdot\vec{v}_{1} \vec{\tilde{S}}_{2}\cdot\vec{v}_{1} v_{2}^2
	 + 3 \vec{\tilde{S}}_{1}\cdot\vec{\tilde{S}}_{2} v_{1}^2 v_{2}^2
	 - 14 \vec{\tilde{S}}_{2}\cdot\vec{v}_{1} \vec{\tilde{S}}_{1}\cdot\vec{v}_{2} v_{2}^2
	 + 15 \vec{\tilde{S}}_{1}\cdot\vec{v}_{2} \vec{\tilde{S}}_{2}\cdot\vec{v}_{2} v_{2}^2
	 + 18 \vec{\tilde{S}}_{1}\cdot\vec{\tilde{S}}_{2} \vec{v}_{1}\cdot\vec{v}_{2} v_{2}^2 \nl
	 - 8 \vec{\tilde{S}}_{1}\cdot\vec{\tilde{S}}_{2} ( \vec{v}_{1}\cdot\vec{v}_{2} )^{2}
	 - 15 \vec{\tilde{S}}_{1}\cdot\vec{\tilde{S}}_{2} v_{1}^{4}
	 - 15 \vec{\tilde{S}}_{1}\cdot\vec{\tilde{S}}_{2} v_{2}^{4}
         - 21 \vec{\tilde{S}}_{2}\cdot\vec{n} \vec{v}_{1}\cdot\vec{n} \vec{\tilde{S}}_{1}\cdot\vec{v}_{1} v_{1}^2 \nl
	 - 24 \vec{\tilde{S}}_{1}\cdot\vec{n} \vec{v}_{1}\cdot\vec{n} \vec{\tilde{S}}_{2}\cdot\vec{v}_{1} v_{1}^2
	 - 60 \vec{v}_{1}\cdot\vec{n} \vec{\tilde{S}}_{1}\cdot\vec{v}_{1} \vec{\tilde{S}}_{2}\cdot\vec{v}_{1} \vec{v}_{2}\cdot\vec{n}
	 + 42 \vec{\tilde{S}}_{1}\cdot\vec{\tilde{S}}_{2} \vec{v}_{1}\cdot\vec{n} v_{1}^2 \vec{v}_{2}\cdot\vec{n} \nl
	 + 18 \vec{\tilde{S}}_{1}\cdot\vec{n} \vec{\tilde{S}}_{2}\cdot\vec{v}_{1} v_{1}^2 \vec{v}_{2}\cdot\vec{n}
	 + 42 \vec{\tilde{S}}_{2}\cdot\vec{n} \vec{v}_{1}\cdot\vec{n} v_{1}^2 \vec{\tilde{S}}_{1}\cdot\vec{v}_{2}
	 + 102 \vec{v}_{1}\cdot\vec{n} \vec{\tilde{S}}_{2}\cdot\vec{v}_{1} \vec{v}_{2}\cdot\vec{n} \vec{\tilde{S}}_{1}\cdot\vec{v}_{2} \nl
	 + 12 \vec{\tilde{S}}_{2}\cdot\vec{n} v_{1}^2 \vec{v}_{2}\cdot\vec{n} \vec{\tilde{S}}_{1}\cdot\vec{v}_{2}
	 + 6 \vec{\tilde{S}}_{1}\cdot\vec{n} \vec{v}_{1}\cdot\vec{n} v_{1}^2 \vec{\tilde{S}}_{2}\cdot\vec{v}_{2}
	 + 36 \vec{v}_{1}\cdot\vec{n} \vec{\tilde{S}}_{1}\cdot\vec{v}_{1} \vec{v}_{2}\cdot\vec{n} \vec{\tilde{S}}_{2}\cdot\vec{v}_{2} \nl
	 + 18 \vec{\tilde{S}}_{1}\cdot\vec{n} v_{1}^2 \vec{v}_{2}\cdot\vec{n} \vec{\tilde{S}}_{2}\cdot\vec{v}_{2}
	 - 60 \vec{v}_{1}\cdot\vec{n} \vec{v}_{2}\cdot\vec{n} \vec{\tilde{S}}_{1}\cdot\vec{v}_{2} \vec{\tilde{S}}_{2}\cdot\vec{v}_{2}
	 - 12 \vec{\tilde{S}}_{2}\cdot\vec{n} \vec{v}_{1}\cdot\vec{n} \vec{\tilde{S}}_{1}\cdot\vec{v}_{1} \vec{v}_{1}\cdot\vec{v}_{2} \nl
	 - 30 \vec{\tilde{S}}_{1}\cdot\vec{n} \vec{\tilde{S}}_{2}\cdot\vec{n} v_{1}^2 \vec{v}_{1}\cdot\vec{v}_{2}
	 - 120 \vec{\tilde{S}}_{1}\cdot\vec{\tilde{S}}_{2} \vec{v}_{1}\cdot\vec{n} \vec{v}_{2}\cdot\vec{n} \vec{v}_{1}\cdot\vec{v}_{2}
	 + 6 \vec{\tilde{S}}_{2}\cdot\vec{n} \vec{\tilde{S}}_{1}\cdot\vec{v}_{1} \vec{v}_{2}\cdot\vec{n} \vec{v}_{1}\cdot\vec{v}_{2} \nl
	 - 42 \vec{\tilde{S}}_{1}\cdot\vec{n} \vec{\tilde{S}}_{2}\cdot\vec{v}_{1} \vec{v}_{2}\cdot\vec{n} \vec{v}_{1}\cdot\vec{v}_{2}
	 - 42 \vec{\tilde{S}}_{2}\cdot\vec{n} \vec{v}_{1}\cdot\vec{n} \vec{\tilde{S}}_{1}\cdot\vec{v}_{2} \vec{v}_{1}\cdot\vec{v}_{2}
	 + 6 \vec{\tilde{S}}_{1}\cdot\vec{n} \vec{v}_{1}\cdot\vec{n} \vec{\tilde{S}}_{2}\cdot\vec{v}_{2} \vec{v}_{1}\cdot\vec{v}_{2} \nl
	 - 12 \vec{\tilde{S}}_{1}\cdot\vec{n} \vec{v}_{2}\cdot\vec{n} \vec{\tilde{S}}_{2}\cdot\vec{v}_{2} \vec{v}_{1}\cdot\vec{v}_{2}
	 + 18 \vec{\tilde{S}}_{2}\cdot\vec{n} \vec{v}_{1}\cdot\vec{n} \vec{\tilde{S}}_{1}\cdot\vec{v}_{1} v_{2}^2
	 + 12 \vec{\tilde{S}}_{1}\cdot\vec{n} \vec{v}_{1}\cdot\vec{n} \vec{\tilde{S}}_{2}\cdot\vec{v}_{1} v_{2}^2 \nl
	 + 3 \vec{\tilde{S}}_{1}\cdot\vec{n} \vec{\tilde{S}}_{2}\cdot\vec{n} v_{1}^2 v_{2}^2
	 + 42 \vec{\tilde{S}}_{1}\cdot\vec{\tilde{S}}_{2} \vec{v}_{1}\cdot\vec{n} \vec{v}_{2}\cdot\vec{n} v_{2}^2
	 + 6 \vec{\tilde{S}}_{2}\cdot\vec{n} \vec{\tilde{S}}_{1}\cdot\vec{v}_{1} \vec{v}_{2}\cdot\vec{n} v_{2}^2
	 + 42 \vec{\tilde{S}}_{1}\cdot\vec{n} \vec{\tilde{S}}_{2}\cdot\vec{v}_{1} \vec{v}_{2}\cdot\vec{n} v_{2}^2 \nl
	 + 18 \vec{\tilde{S}}_{2}\cdot\vec{n} \vec{v}_{1}\cdot\vec{n} \vec{\tilde{S}}_{1}\cdot\vec{v}_{2} v_{2}^2
	 - 24 \vec{\tilde{S}}_{2}\cdot\vec{n} \vec{v}_{2}\cdot\vec{n} \vec{\tilde{S}}_{1}\cdot\vec{v}_{2} v_{2}^2
	 - 21 \vec{\tilde{S}}_{1}\cdot\vec{n} \vec{v}_{2}\cdot\vec{n} \vec{\tilde{S}}_{2}\cdot\vec{v}_{2} v_{2}^2 \nl
	 - 30 \vec{\tilde{S}}_{1}\cdot\vec{n} \vec{\tilde{S}}_{2}\cdot\vec{n} \vec{v}_{1}\cdot\vec{v}_{2} v_{2}^2
	 + 24 \vec{\tilde{S}}_{1}\cdot\vec{\tilde{S}}_{2} v_{1}^2 ( \vec{v}_{1}\cdot\vec{n} )^{2}
	 - 24 \vec{\tilde{S}}_{2}\cdot\vec{v}_{1} \vec{\tilde{S}}_{1}\cdot\vec{v}_{2} ( \vec{v}_{1}\cdot\vec{n} )^{2} \nl 
	 - 6 \vec{\tilde{S}}_{1}\cdot\vec{v}_{1} \vec{\tilde{S}}_{2}\cdot\vec{v}_{2} ( \vec{v}_{1}\cdot\vec{n} )^{2}
	 + 24 \vec{\tilde{S}}_{1}\cdot\vec{v}_{2} \vec{\tilde{S}}_{2}\cdot\vec{v}_{2} ( \vec{v}_{1}\cdot\vec{n} )^{2}
	 + 24 \vec{\tilde{S}}_{1}\cdot\vec{\tilde{S}}_{2} \vec{v}_{1}\cdot\vec{v}_{2} ( \vec{v}_{1}\cdot\vec{n} )^{2} \nl
	 - 27 \vec{\tilde{S}}_{1}\cdot\vec{\tilde{S}}_{2} v_{2}^2 ( \vec{v}_{1}\cdot\vec{n} )^{2}
	 + 24 \vec{\tilde{S}}_{1}\cdot\vec{v}_{1} \vec{\tilde{S}}_{2}\cdot\vec{v}_{1} ( \vec{v}_{2}\cdot\vec{n} )^{2}
	 - 27 \vec{\tilde{S}}_{1}\cdot\vec{\tilde{S}}_{2} v_{1}^2 ( \vec{v}_{2}\cdot\vec{n} )^{2} \nl
	 - 24 \vec{\tilde{S}}_{2}\cdot\vec{v}_{1} \vec{\tilde{S}}_{1}\cdot\vec{v}_{2} ( \vec{v}_{2}\cdot\vec{n} )^{2}
	 - 6 \vec{\tilde{S}}_{1}\cdot\vec{v}_{1} \vec{\tilde{S}}_{2}\cdot\vec{v}_{2} ( \vec{v}_{2}\cdot\vec{n} )^{2}
	 + 24 \vec{\tilde{S}}_{1}\cdot\vec{\tilde{S}}_{2} \vec{v}_{1}\cdot\vec{v}_{2} ( \vec{v}_{2}\cdot\vec{n} )^{2} \nl
	 + 12 \vec{\tilde{S}}_{1}\cdot\vec{n} \vec{\tilde{S}}_{2}\cdot\vec{n} ( \vec{v}_{1}\cdot\vec{v}_{2} )^{2}
	 + 24 \vec{\tilde{S}}_{1}\cdot\vec{\tilde{S}}_{2} v_{2}^2 ( \vec{v}_{2}\cdot\vec{n} )^{2}
	 + 21 \vec{\tilde{S}}_{1}\cdot\vec{n} \vec{\tilde{S}}_{2}\cdot\vec{n} v_{1}^{4}
	 + 21 \vec{\tilde{S}}_{1}\cdot\vec{n} \vec{\tilde{S}}_{2}\cdot\vec{n} v_{2}^{4} \nl
         - 90 \vec{\tilde{S}}_{1}\cdot\vec{n} \vec{\tilde{S}}_{2}\cdot\vec{n} \vec{v}_{1}\cdot\vec{n} v_{1}^2 \vec{v}_{2}\cdot\vec{n}
	 + 180 \vec{\tilde{S}}_{1}\cdot\vec{n} \vec{\tilde{S}}_{2}\cdot\vec{n} \vec{v}_{1}\cdot\vec{n} \vec{v}_{2}\cdot\vec{n} \vec{v}_{1}\cdot\vec{v}_{2}
	 - 90 \vec{\tilde{S}}_{1}\cdot\vec{n} \vec{\tilde{S}}_{2}\cdot\vec{n} \vec{v}_{1}\cdot\vec{n} \vec{v}_{2}\cdot\vec{n} v_{2}^2 \nl
	 - 120 \vec{\tilde{S}}_{1}\cdot\vec{\tilde{S}}_{2} \vec{v}_{2}\cdot\vec{n} ( \vec{v}_{1}\cdot\vec{n} )^{3}
	 + 90 \vec{\tilde{S}}_{2}\cdot\vec{n} \vec{\tilde{S}}_{1}\cdot\vec{v}_{1} \vec{v}_{2}\cdot\vec{n} ( \vec{v}_{1}\cdot\vec{n} )^{2}
	 + 120 \vec{\tilde{S}}_{1}\cdot\vec{n} \vec{\tilde{S}}_{2}\cdot\vec{v}_{1} \vec{v}_{2}\cdot\vec{n} ( \vec{v}_{1}\cdot\vec{n} )^{2} \nl
	 - 150 \vec{\tilde{S}}_{2}\cdot\vec{n} \vec{v}_{2}\cdot\vec{n} \vec{\tilde{S}}_{1}\cdot\vec{v}_{2} ( \vec{v}_{1}\cdot\vec{n} )^{2}
	 - 90 \vec{\tilde{S}}_{1}\cdot\vec{n} \vec{v}_{2}\cdot\vec{n} \vec{\tilde{S}}_{2}\cdot\vec{v}_{2} ( \vec{v}_{1}\cdot\vec{n} )^{2}
	 - 15 \vec{\tilde{S}}_{1}\cdot\vec{n} \vec{\tilde{S}}_{2}\cdot\vec{n} v_{2}^2 ( \vec{v}_{1}\cdot\vec{n} )^{2} \nl
	 + 225 \vec{\tilde{S}}_{1}\cdot\vec{\tilde{S}}_{2} ( \vec{v}_{1}\cdot\vec{n} )^{2} ( \vec{v}_{2}\cdot\vec{n} )^{2}
	 - 90 \vec{\tilde{S}}_{2}\cdot\vec{n} \vec{v}_{1}\cdot\vec{n} \vec{\tilde{S}}_{1}\cdot\vec{v}_{1} ( \vec{v}_{2}\cdot\vec{n} )^{2}
	 - 150 \vec{\tilde{S}}_{1}\cdot\vec{n} \vec{v}_{1}\cdot\vec{n} \vec{\tilde{S}}_{2}\cdot\vec{v}_{1} ( \vec{v}_{2}\cdot\vec{n} )^{2} \nl
	 - 15 \vec{\tilde{S}}_{1}\cdot\vec{n} \vec{\tilde{S}}_{2}\cdot\vec{n} v_{1}^2 ( \vec{v}_{2}\cdot\vec{n} )^{2}
	 - 120 \vec{\tilde{S}}_{1}\cdot\vec{\tilde{S}}_{2} \vec{v}_{1}\cdot\vec{n} ( \vec{v}_{2}\cdot\vec{n} )^{3}
	 + 120 \vec{\tilde{S}}_{2}\cdot\vec{n} \vec{v}_{1}\cdot\vec{n} \vec{\tilde{S}}_{1}\cdot\vec{v}_{2} ( \vec{v}_{2}\cdot\vec{n} )^{2} \nl
	 + 90 \vec{\tilde{S}}_{1}\cdot\vec{n} \vec{v}_{1}\cdot\vec{n} \vec{\tilde{S}}_{2}\cdot\vec{v}_{2} ( \vec{v}_{2}\cdot\vec{n} )^{2}
	 + 105 \vec{\tilde{S}}_{1}\cdot\vec{n} \vec{\tilde{S}}_{2}\cdot\vec{n} ( \vec{v}_{1}\cdot\vec{n} )^{2} ( \vec{v}_{2}\cdot\vec{n} )^{2} \Big] \nl
+ \frac{G^2 m_{1}}{r^4} \Big[ 5 \vec{\tilde{S}}_{1}\cdot\vec{v}_{1} \vec{\tilde{S}}_{2}\cdot\vec{v}_{1}
	 - 4 \vec{\tilde{S}}_{1}\cdot\vec{\tilde{S}}_{2} v_{1}^2
	 - 8 \vec{\tilde{S}}_{2}\cdot\vec{v}_{1} \vec{\tilde{S}}_{1}\cdot\vec{v}_{2}
	 - 2 \vec{\tilde{S}}_{1}\cdot\vec{v}_{1} \vec{\tilde{S}}_{2}\cdot\vec{v}_{2}
	 + 5 \vec{\tilde{S}}_{1}\cdot\vec{v}_{2} \vec{\tilde{S}}_{2}\cdot\vec{v}_{2} \nl
	 + 8 \vec{\tilde{S}}_{1}\cdot\vec{\tilde{S}}_{2} \vec{v}_{1}\cdot\vec{v}_{2}
	 - 5 \vec{\tilde{S}}_{1}\cdot\vec{\tilde{S}}_{2} v_{2}^2 -12 \vec{\tilde{S}}_{2}\cdot\vec{n} \vec{v}_{1}\cdot\vec{n} \vec{\tilde{S}}_{1}\cdot\vec{v}_{1}
	 - 16 \vec{\tilde{S}}_{1}\cdot\vec{n} \vec{v}_{1}\cdot\vec{n} \vec{\tilde{S}}_{2}\cdot\vec{v}_{1}
	 + 2 \vec{\tilde{S}}_{1}\cdot\vec{n} \vec{\tilde{S}}_{2}\cdot\vec{n} v_{1}^2 \nl
	 - 16 \vec{\tilde{S}}_{1}\cdot\vec{\tilde{S}}_{2} \vec{v}_{1}\cdot\vec{n} \vec{v}_{2}\cdot\vec{n}
	 + 6 \vec{\tilde{S}}_{2}\cdot\vec{n} \vec{\tilde{S}}_{1}\cdot\vec{v}_{1} \vec{v}_{2}\cdot\vec{n}
	 + 16 \vec{\tilde{S}}_{1}\cdot\vec{n} \vec{\tilde{S}}_{2}\cdot\vec{v}_{1} \vec{v}_{2}\cdot\vec{n}
	 + 10 \vec{\tilde{S}}_{2}\cdot\vec{n} \vec{v}_{1}\cdot\vec{n} \vec{\tilde{S}}_{1}\cdot\vec{v}_{2} \nl
	 - 8 \vec{\tilde{S}}_{2}\cdot\vec{n} \vec{v}_{2}\cdot\vec{n} \vec{\tilde{S}}_{1}\cdot\vec{v}_{2}
	 + 2 \vec{\tilde{S}}_{1}\cdot\vec{n} \vec{v}_{1}\cdot\vec{n} \vec{\tilde{S}}_{2}\cdot\vec{v}_{2}
	 - 6 \vec{\tilde{S}}_{1}\cdot\vec{n} \vec{v}_{2}\cdot\vec{n} \vec{\tilde{S}}_{2}\cdot\vec{v}_{2}
	 - 4 \vec{\tilde{S}}_{1}\cdot\vec{n} \vec{\tilde{S}}_{2}\cdot\vec{n} \vec{v}_{1}\cdot\vec{v}_{2} \nl
	 + 6 \vec{\tilde{S}}_{1}\cdot\vec{n} \vec{\tilde{S}}_{2}\cdot\vec{n} v_{2}^2
	 + 4 \vec{\tilde{S}}_{1}\cdot\vec{\tilde{S}}_{2} ( \vec{v}_{1}\cdot\vec{n} )^{2}
	 + 8 \vec{\tilde{S}}_{1}\cdot\vec{\tilde{S}}_{2} ( \vec{v}_{2}\cdot\vec{n} )^{2}
	 + 24 \vec{\tilde{S}}_{1}\cdot\vec{n} \vec{\tilde{S}}_{2}\cdot\vec{n} ( \vec{v}_{1}\cdot\vec{n} )^{2} \Big] \nl
+ \frac{G^2 m_{2}}{r^4} \Big[ 5 \vec{\tilde{S}}_{1}\cdot\vec{v}_{1} \vec{\tilde{S}}_{2}\cdot\vec{v}_{1}
	 - 5 \vec{\tilde{S}}_{1}\cdot\vec{\tilde{S}}_{2} v_{1}^2
	 - 8 \vec{\tilde{S}}_{2}\cdot\vec{v}_{1} \vec{\tilde{S}}_{1}\cdot\vec{v}_{2}
	 - 2 \vec{\tilde{S}}_{1}\cdot\vec{v}_{1} \vec{\tilde{S}}_{2}\cdot\vec{v}_{2}
	 + 5 \vec{\tilde{S}}_{1}\cdot\vec{v}_{2} \vec{\tilde{S}}_{2}\cdot\vec{v}_{2} \nl
	 + 8 \vec{\tilde{S}}_{1}\cdot\vec{\tilde{S}}_{2} \vec{v}_{1}\cdot\vec{v}_{2}
	 - 4 \vec{\tilde{S}}_{1}\cdot\vec{\tilde{S}}_{2} v_{2}^2 -6 \vec{\tilde{S}}_{2}\cdot\vec{n} \vec{v}_{1}\cdot\vec{n} \vec{\tilde{S}}_{1}\cdot\vec{v}_{1}
	 - 8 \vec{\tilde{S}}_{1}\cdot\vec{n} \vec{v}_{1}\cdot\vec{n} \vec{\tilde{S}}_{2}\cdot\vec{v}_{1}
	 + 6 \vec{\tilde{S}}_{1}\cdot\vec{n} \vec{\tilde{S}}_{2}\cdot\vec{n} v_{1}^2 \nl
	 - 16 \vec{\tilde{S}}_{1}\cdot\vec{\tilde{S}}_{2} \vec{v}_{1}\cdot\vec{n} \vec{v}_{2}\cdot\vec{n}
	 + 2 \vec{\tilde{S}}_{2}\cdot\vec{n} \vec{\tilde{S}}_{1}\cdot\vec{v}_{1} \vec{v}_{2}\cdot\vec{n}
	 + 10 \vec{\tilde{S}}_{1}\cdot\vec{n} \vec{\tilde{S}}_{2}\cdot\vec{v}_{1} \vec{v}_{2}\cdot\vec{n}
	 + 16 \vec{\tilde{S}}_{2}\cdot\vec{n} \vec{v}_{1}\cdot\vec{n} \vec{\tilde{S}}_{1}\cdot\vec{v}_{2} \nl
	 - 16 \vec{\tilde{S}}_{2}\cdot\vec{n} \vec{v}_{2}\cdot\vec{n} \vec{\tilde{S}}_{1}\cdot\vec{v}_{2}
	 + 6 \vec{\tilde{S}}_{1}\cdot\vec{n} \vec{v}_{1}\cdot\vec{n} \vec{\tilde{S}}_{2}\cdot\vec{v}_{2}
	 - 12 \vec{\tilde{S}}_{1}\cdot\vec{n} \vec{v}_{2}\cdot\vec{n} \vec{\tilde{S}}_{2}\cdot\vec{v}_{2}
	 - 4 \vec{\tilde{S}}_{1}\cdot\vec{n} \vec{\tilde{S}}_{2}\cdot\vec{n} \vec{v}_{1}\cdot\vec{v}_{2} \nl
	 + 2 \vec{\tilde{S}}_{1}\cdot\vec{n} \vec{\tilde{S}}_{2}\cdot\vec{n} v_{2}^2
	 + 8 \vec{\tilde{S}}_{1}\cdot\vec{\tilde{S}}_{2} ( \vec{v}_{1}\cdot\vec{n} )^{2}
	 + 4 \vec{\tilde{S}}_{1}\cdot\vec{\tilde{S}}_{2} ( \vec{v}_{2}\cdot\vec{n} )^{2}
	 + 24 \vec{\tilde{S}}_{1}\cdot\vec{n} \vec{\tilde{S}}_{2}\cdot\vec{n} ( \vec{v}_{2}\cdot\vec{n} )^{2} \Big] \nl
- \frac{G^3 m_{1}^2}{2 r^5} \Big[ 7 \vec{\tilde{S}}_{1}\cdot\vec{\tilde{S}}_{2} -31 \vec{\tilde{S}}_{1}\cdot\vec{n} \vec{\tilde{S}}_{2}\cdot\vec{n} \Big]
	 - \frac{G^3 m_{2}^2}{2 r^5} \Big[ 7 \vec{\tilde{S}}_{1}\cdot\vec{\tilde{S}}_{2} -31 \vec{\tilde{S}}_{1}\cdot\vec{n} \vec{\tilde{S}}_{2}\cdot\vec{n} \Big] \nl
	 - \frac{6 G^3 m_{1} m_{2}}{r^5} \Big[ 3 \vec{\tilde{S}}_{1}\cdot\vec{\tilde{S}}_{2} -13 \vec{\tilde{S}}_{1}\cdot\vec{n} \vec{\tilde{S}}_{2}\cdot\vec{n} \Big] ,
\end{align}
\begin{align}
& V_a = \frac{G}{8 r^2} \Big[ 16 \vec{\tilde{S}}_{1}\cdot\vec{v}_{1} \vec{\tilde{S}}_{2}\cdot\vec{v}_{1} \vec{a}_{1}\cdot\vec{n}
	 - 24 \vec{\tilde{S}}_{1}\cdot\vec{\tilde{S}}_{2} v_{1}^2 \vec{a}_{1}\cdot\vec{n}
	 - 16 \vec{v}_{1}\cdot\vec{n} \vec{\tilde{S}}_{1}\cdot\vec{v}_{1} \vec{\tilde{S}}_{2}\cdot\vec{a}_{1}
	 + 24 \vec{\tilde{S}}_{1}\cdot\vec{n} v_{1}^2 \vec{\tilde{S}}_{2}\cdot\vec{a}_{1} \nl
	 + 10 \vec{\tilde{S}}_{2}\cdot\vec{v}_{1} \vec{\tilde{S}}_{1}\cdot\vec{a}_{1} \vec{v}_{2}\cdot\vec{n}
	 + 26 \vec{\tilde{S}}_{1}\cdot\vec{v}_{1} \vec{\tilde{S}}_{2}\cdot\vec{a}_{1} \vec{v}_{2}\cdot\vec{n}
	 - 36 \vec{\tilde{S}}_{1}\cdot\vec{\tilde{S}}_{2} \vec{v}_{1}\cdot\vec{a}_{1} \vec{v}_{2}\cdot\vec{n}
	 - 8 \vec{\tilde{S}}_{2}\cdot\vec{v}_{1} \vec{a}_{1}\cdot\vec{n} \vec{\tilde{S}}_{1}\cdot\vec{v}_{2} \nl
	 + 32 \vec{v}_{1}\cdot\vec{n} \vec{\tilde{S}}_{2}\cdot\vec{a}_{1} \vec{\tilde{S}}_{1}\cdot\vec{v}_{2}
	 - 12 \vec{\tilde{S}}_{2}\cdot\vec{n} \vec{v}_{1}\cdot\vec{a}_{1} \vec{\tilde{S}}_{1}\cdot\vec{v}_{2}
	 - 42 \vec{\tilde{S}}_{2}\cdot\vec{a}_{1} \vec{v}_{2}\cdot\vec{n} \vec{\tilde{S}}_{1}\cdot\vec{v}_{2}
	 - 10 \vec{\tilde{S}}_{1}\cdot\vec{v}_{1} \vec{a}_{1}\cdot\vec{n} \vec{\tilde{S}}_{2}\cdot\vec{v}_{2} \nl
	 - 2 \vec{v}_{1}\cdot\vec{n} \vec{\tilde{S}}_{1}\cdot\vec{a}_{1} \vec{\tilde{S}}_{2}\cdot\vec{v}_{2}
	 + 12 \vec{\tilde{S}}_{1}\cdot\vec{n} \vec{v}_{1}\cdot\vec{a}_{1} \vec{\tilde{S}}_{2}\cdot\vec{v}_{2}
	 - 6 \vec{\tilde{S}}_{1}\cdot\vec{a}_{1} \vec{v}_{2}\cdot\vec{n} \vec{\tilde{S}}_{2}\cdot\vec{v}_{2}
	 + 10 \vec{a}_{1}\cdot\vec{n} \vec{\tilde{S}}_{1}\cdot\vec{v}_{2} \vec{\tilde{S}}_{2}\cdot\vec{v}_{2} \nl
	 + 24 \vec{\tilde{S}}_{1}\cdot\vec{\tilde{S}}_{2} \vec{a}_{1}\cdot\vec{n} \vec{v}_{1}\cdot\vec{v}_{2}
	 + 6 \vec{\tilde{S}}_{2}\cdot\vec{n} \vec{\tilde{S}}_{1}\cdot\vec{a}_{1} \vec{v}_{1}\cdot\vec{v}_{2}
	 - 32 \vec{\tilde{S}}_{1}\cdot\vec{n} \vec{\tilde{S}}_{2}\cdot\vec{a}_{1} \vec{v}_{1}\cdot\vec{v}_{2}
	 - 24 \vec{\tilde{S}}_{1}\cdot\vec{\tilde{S}}_{2} \vec{v}_{1}\cdot\vec{n} \vec{a}_{1}\cdot\vec{v}_{2} \nl
	 + 6 \vec{\tilde{S}}_{2}\cdot\vec{n} \vec{\tilde{S}}_{1}\cdot\vec{v}_{1} \vec{a}_{1}\cdot\vec{v}_{2}
	 + 8 \vec{\tilde{S}}_{1}\cdot\vec{n} \vec{\tilde{S}}_{2}\cdot\vec{v}_{1} \vec{a}_{1}\cdot\vec{v}_{2}
	 + 46 \vec{\tilde{S}}_{1}\cdot\vec{\tilde{S}}_{2} \vec{v}_{2}\cdot\vec{n} \vec{a}_{1}\cdot\vec{v}_{2}
	 + 6 \vec{\tilde{S}}_{2}\cdot\vec{n} \vec{\tilde{S}}_{1}\cdot\vec{v}_{2} \vec{a}_{1}\cdot\vec{v}_{2} \nl
	 - 14 \vec{\tilde{S}}_{1}\cdot\vec{n} \vec{\tilde{S}}_{2}\cdot\vec{v}_{2} \vec{a}_{1}\cdot\vec{v}_{2}
	 - 9 \vec{\tilde{S}}_{1}\cdot\vec{\tilde{S}}_{2} \vec{a}_{1}\cdot\vec{n} v_{2}^2
	 - 5 \vec{\tilde{S}}_{2}\cdot\vec{n} \vec{\tilde{S}}_{1}\cdot\vec{a}_{1} v_{2}^2
	 + 15 \vec{\tilde{S}}_{1}\cdot\vec{n} \vec{\tilde{S}}_{2}\cdot\vec{a}_{1} v_{2}^2 \nl
	 - 10 \vec{\tilde{S}}_{1}\cdot\vec{v}_{1} \vec{\tilde{S}}_{2}\cdot\vec{v}_{1} \vec{a}_{2}\cdot\vec{n}
	 + 9 \vec{\tilde{S}}_{1}\cdot\vec{\tilde{S}}_{2} v_{1}^2 \vec{a}_{2}\cdot\vec{n}
	 + 8 \vec{\tilde{S}}_{2}\cdot\vec{v}_{1} \vec{\tilde{S}}_{1}\cdot\vec{v}_{2} \vec{a}_{2}\cdot\vec{n}
	 + 10 \vec{\tilde{S}}_{1}\cdot\vec{v}_{1} \vec{\tilde{S}}_{2}\cdot\vec{v}_{2} \vec{a}_{2}\cdot\vec{n} \nl
	 - 16 \vec{\tilde{S}}_{1}\cdot\vec{v}_{2} \vec{\tilde{S}}_{2}\cdot\vec{v}_{2} \vec{a}_{2}\cdot\vec{n}
	 - 24 \vec{\tilde{S}}_{1}\cdot\vec{\tilde{S}}_{2} \vec{v}_{1}\cdot\vec{v}_{2} \vec{a}_{2}\cdot\vec{n}
	 + 24 \vec{\tilde{S}}_{1}\cdot\vec{\tilde{S}}_{2} v_{2}^2 \vec{a}_{2}\cdot\vec{n}
	 + 42 \vec{v}_{1}\cdot\vec{n} \vec{\tilde{S}}_{2}\cdot\vec{v}_{1} \vec{\tilde{S}}_{1}\cdot\vec{a}_{2} \nl
	 - 15 \vec{\tilde{S}}_{2}\cdot\vec{n} v_{1}^2 \vec{\tilde{S}}_{1}\cdot\vec{a}_{2}
	 - 32 \vec{\tilde{S}}_{2}\cdot\vec{v}_{1} \vec{v}_{2}\cdot\vec{n} \vec{\tilde{S}}_{1}\cdot\vec{a}_{2}
	 - 26 \vec{v}_{1}\cdot\vec{n} \vec{\tilde{S}}_{2}\cdot\vec{v}_{2} \vec{\tilde{S}}_{1}\cdot\vec{a}_{2}
	 + 16 \vec{v}_{2}\cdot\vec{n} \vec{\tilde{S}}_{2}\cdot\vec{v}_{2} \vec{\tilde{S}}_{1}\cdot\vec{a}_{2} \nl
	 + 32 \vec{\tilde{S}}_{2}\cdot\vec{n} \vec{v}_{1}\cdot\vec{v}_{2} \vec{\tilde{S}}_{1}\cdot\vec{a}_{2}
	 - 24 \vec{\tilde{S}}_{2}\cdot\vec{n} v_{2}^2 \vec{\tilde{S}}_{1}\cdot\vec{a}_{2}
	 + 6 \vec{v}_{1}\cdot\vec{n} \vec{\tilde{S}}_{1}\cdot\vec{v}_{1} \vec{\tilde{S}}_{2}\cdot\vec{a}_{2}
	 + 5 \vec{\tilde{S}}_{1}\cdot\vec{n} v_{1}^2 \vec{\tilde{S}}_{2}\cdot\vec{a}_{2} \nl
	 + 2 \vec{\tilde{S}}_{1}\cdot\vec{v}_{1} \vec{v}_{2}\cdot\vec{n} \vec{\tilde{S}}_{2}\cdot\vec{a}_{2}
	 - 10 \vec{v}_{1}\cdot\vec{n} \vec{\tilde{S}}_{1}\cdot\vec{v}_{2} \vec{\tilde{S}}_{2}\cdot\vec{a}_{2}
	 - 6 \vec{\tilde{S}}_{1}\cdot\vec{n} \vec{v}_{1}\cdot\vec{v}_{2} \vec{\tilde{S}}_{2}\cdot\vec{a}_{2}
	 - 46 \vec{\tilde{S}}_{1}\cdot\vec{\tilde{S}}_{2} \vec{v}_{1}\cdot\vec{n} \vec{v}_{1}\cdot\vec{a}_{2} \nl
	 + 14 \vec{\tilde{S}}_{2}\cdot\vec{n} \vec{\tilde{S}}_{1}\cdot\vec{v}_{1} \vec{v}_{1}\cdot\vec{a}_{2}
	 - 6 \vec{\tilde{S}}_{1}\cdot\vec{n} \vec{\tilde{S}}_{2}\cdot\vec{v}_{1} \vec{v}_{1}\cdot\vec{a}_{2}
	 + 24 \vec{\tilde{S}}_{1}\cdot\vec{\tilde{S}}_{2} \vec{v}_{2}\cdot\vec{n} \vec{v}_{1}\cdot\vec{a}_{2}
	 - 8 \vec{\tilde{S}}_{2}\cdot\vec{n} \vec{\tilde{S}}_{1}\cdot\vec{v}_{2} \vec{v}_{1}\cdot\vec{a}_{2} \nl
	 - 6 \vec{\tilde{S}}_{1}\cdot\vec{n} \vec{\tilde{S}}_{2}\cdot\vec{v}_{2} \vec{v}_{1}\cdot\vec{a}_{2}
	 + 36 \vec{\tilde{S}}_{1}\cdot\vec{\tilde{S}}_{2} \vec{v}_{1}\cdot\vec{n} \vec{v}_{2}\cdot\vec{a}_{2}
	 - 12 \vec{\tilde{S}}_{2}\cdot\vec{n} \vec{\tilde{S}}_{1}\cdot\vec{v}_{1} \vec{v}_{2}\cdot\vec{a}_{2}
	 + 12 \vec{\tilde{S}}_{1}\cdot\vec{n} \vec{\tilde{S}}_{2}\cdot\vec{v}_{1} \vec{v}_{2}\cdot\vec{a}_{2} \nl
	 + 72 \vec{\tilde{S}}_{1}\cdot\vec{\tilde{S}}_{2} \vec{v}_{1}\cdot\vec{n} \vec{a}_{1}\cdot\vec{n} \vec{v}_{2}\cdot\vec{n}
	 - 18 \vec{\tilde{S}}_{2}\cdot\vec{n} \vec{\tilde{S}}_{1}\cdot\vec{v}_{1} \vec{a}_{1}\cdot\vec{n} \vec{v}_{2}\cdot\vec{n}
	 - 24 \vec{\tilde{S}}_{1}\cdot\vec{n} \vec{\tilde{S}}_{2}\cdot\vec{v}_{1} \vec{a}_{1}\cdot\vec{n} \vec{v}_{2}\cdot\vec{n} \nl
	 - 18 \vec{\tilde{S}}_{2}\cdot\vec{n} \vec{v}_{1}\cdot\vec{n} \vec{\tilde{S}}_{1}\cdot\vec{a}_{1} \vec{v}_{2}\cdot\vec{n}
	 - 48 \vec{\tilde{S}}_{1}\cdot\vec{n} \vec{v}_{1}\cdot\vec{n} \vec{\tilde{S}}_{2}\cdot\vec{a}_{1} \vec{v}_{2}\cdot\vec{n}
	 + 36 \vec{\tilde{S}}_{1}\cdot\vec{n} \vec{\tilde{S}}_{2}\cdot\vec{n} \vec{v}_{1}\cdot\vec{a}_{1} \vec{v}_{2}\cdot\vec{n} \nl
	 + 30 \vec{\tilde{S}}_{2}\cdot\vec{n} \vec{a}_{1}\cdot\vec{n} \vec{v}_{2}\cdot\vec{n} \vec{\tilde{S}}_{1}\cdot\vec{v}_{2}
	 + 18 \vec{\tilde{S}}_{1}\cdot\vec{n} \vec{a}_{1}\cdot\vec{n} \vec{v}_{2}\cdot\vec{n} \vec{\tilde{S}}_{2}\cdot\vec{v}_{2}
	 - 42 \vec{\tilde{S}}_{1}\cdot\vec{n} \vec{\tilde{S}}_{2}\cdot\vec{n} \vec{v}_{2}\cdot\vec{n} \vec{a}_{1}\cdot\vec{v}_{2} \nl
	 + 3 \vec{\tilde{S}}_{1}\cdot\vec{n} \vec{\tilde{S}}_{2}\cdot\vec{n} \vec{a}_{1}\cdot\vec{n} v_{2}^2
	 - 18 \vec{\tilde{S}}_{2}\cdot\vec{n} \vec{v}_{1}\cdot\vec{n} \vec{\tilde{S}}_{1}\cdot\vec{v}_{1} \vec{a}_{2}\cdot\vec{n}
	 - 30 \vec{\tilde{S}}_{1}\cdot\vec{n} \vec{v}_{1}\cdot\vec{n} \vec{\tilde{S}}_{2}\cdot\vec{v}_{1} \vec{a}_{2}\cdot\vec{n} \nl
	 - 3 \vec{\tilde{S}}_{1}\cdot\vec{n} \vec{\tilde{S}}_{2}\cdot\vec{n} v_{1}^2 \vec{a}_{2}\cdot\vec{n}
	 - 72 \vec{\tilde{S}}_{1}\cdot\vec{\tilde{S}}_{2} \vec{v}_{1}\cdot\vec{n} \vec{v}_{2}\cdot\vec{n} \vec{a}_{2}\cdot\vec{n}
	 + 24 \vec{\tilde{S}}_{2}\cdot\vec{n} \vec{v}_{1}\cdot\vec{n} \vec{\tilde{S}}_{1}\cdot\vec{v}_{2} \vec{a}_{2}\cdot\vec{n} \nl
	 + 18 \vec{\tilde{S}}_{1}\cdot\vec{n} \vec{v}_{1}\cdot\vec{n} \vec{\tilde{S}}_{2}\cdot\vec{v}_{2} \vec{a}_{2}\cdot\vec{n}
	 + 48 \vec{\tilde{S}}_{2}\cdot\vec{n} \vec{v}_{1}\cdot\vec{n} \vec{v}_{2}\cdot\vec{n} \vec{\tilde{S}}_{1}\cdot\vec{a}_{2}
	 + 18 \vec{\tilde{S}}_{1}\cdot\vec{n} \vec{v}_{1}\cdot\vec{n} \vec{v}_{2}\cdot\vec{n} \vec{\tilde{S}}_{2}\cdot\vec{a}_{2} \nl
	 + 42 \vec{\tilde{S}}_{1}\cdot\vec{n} \vec{\tilde{S}}_{2}\cdot\vec{n} \vec{v}_{1}\cdot\vec{n} \vec{v}_{1}\cdot\vec{a}_{2}
	 - 36 \vec{\tilde{S}}_{1}\cdot\vec{n} \vec{\tilde{S}}_{2}\cdot\vec{n} \vec{v}_{1}\cdot\vec{n} \vec{v}_{2}\cdot\vec{a}_{2}
	 + 45 \vec{\tilde{S}}_{1}\cdot\vec{\tilde{S}}_{2} \vec{a}_{2}\cdot\vec{n} ( \vec{v}_{1}\cdot\vec{n} )^{2} \nl
	 - 27 \vec{\tilde{S}}_{2}\cdot\vec{n} \vec{\tilde{S}}_{1}\cdot\vec{a}_{2} ( \vec{v}_{1}\cdot\vec{n} )^{2}
	 - 15 \vec{\tilde{S}}_{1}\cdot\vec{n} \vec{\tilde{S}}_{2}\cdot\vec{a}_{2} ( \vec{v}_{1}\cdot\vec{n} )^{2}
	 - 45 \vec{\tilde{S}}_{1}\cdot\vec{\tilde{S}}_{2} \vec{a}_{1}\cdot\vec{n} ( \vec{v}_{2}\cdot\vec{n} )^{2} \nl
	 + 15 \vec{\tilde{S}}_{2}\cdot\vec{n} \vec{\tilde{S}}_{1}\cdot\vec{a}_{1} ( \vec{v}_{2}\cdot\vec{n} )^{2}
	 + 27 \vec{\tilde{S}}_{1}\cdot\vec{n} \vec{\tilde{S}}_{2}\cdot\vec{a}_{1} ( \vec{v}_{2}\cdot\vec{n} )^{2}
	 + 15 \vec{\tilde{S}}_{1}\cdot\vec{n} \vec{\tilde{S}}_{2}\cdot\vec{n} \vec{a}_{2}\cdot\vec{n} ( \vec{v}_{1}\cdot\vec{n} )^{2} \nl
	 - 15 \vec{\tilde{S}}_{1}\cdot\vec{n} \vec{\tilde{S}}_{2}\cdot\vec{n} \vec{a}_{1}\cdot\vec{n} ( \vec{v}_{2}\cdot\vec{n} )^{2} \Big] \nl
- \frac{2 G^2 m_{1}}{r^3} \Big[ \vec{\tilde{S}}_{1}\cdot\vec{\tilde{S}}_{2} \vec{a}_{1}\cdot\vec{n}
	 - 2 \vec{\tilde{S}}_{2}\cdot\vec{n} \vec{\tilde{S}}_{1}\cdot\vec{a}_{1}
	 + \vec{\tilde{S}}_{1}\cdot\vec{n} \vec{\tilde{S}}_{2}\cdot\vec{a}_{1}
	 - \vec{\tilde{S}}_{1}\cdot\vec{\tilde{S}}_{2} \vec{a}_{2}\cdot\vec{n}
	 + \vec{\tilde{S}}_{2}\cdot\vec{n} \vec{\tilde{S}}_{1}\cdot\vec{a}_{2} \Big] \nl
- \frac{2 G^2 m_{2}}{r^3} \Big[ \vec{\tilde{S}}_{1}\cdot\vec{\tilde{S}}_{2} \vec{a}_{1}\cdot\vec{n}
	 - \vec{\tilde{S}}_{1}\cdot\vec{n} \vec{\tilde{S}}_{2}\cdot\vec{a}_{1}
	 - \vec{\tilde{S}}_{1}\cdot\vec{\tilde{S}}_{2} \vec{a}_{2}\cdot\vec{n}
	 - \vec{\tilde{S}}_{2}\cdot\vec{n} \vec{\tilde{S}}_{1}\cdot\vec{a}_{2}
	 + 2 \vec{\tilde{S}}_{1}\cdot\vec{n} \vec{\tilde{S}}_{2}\cdot\vec{a}_{2} \Big] ,
\end{align}
\begin{align}
& V_{\dot{S}} = - \frac{G}{4 r^2} \Big[ 8 \vec{v}_{1}\cdot\vec{n} \vec{\tilde{S}}_{1}\cdot\vec{v}_{1} \dot{\vec{S}}_{2}\cdot\vec{v}_{1}
	 + 4 \dot{\vec{S}}_{1}\cdot\vec{\tilde{S}}_{2} \vec{v}_{1}\cdot\vec{n} v_{1}^2
	 - 9 \vec{\tilde{S}}_{1}\cdot\dot{\vec{S}}_{2} \vec{v}_{1}\cdot\vec{n} v_{1}^2
	 - 4 \dot{\vec{S}}_{1}\cdot\vec{n} \vec{\tilde{S}}_{2}\cdot\vec{v}_{1} v_{1}^2 \nl
	 + \vec{\tilde{S}}_{1}\cdot\vec{n} \dot{\vec{S}}_{2}\cdot\vec{v}_{1} v_{1}^2
	 - 11 \dot{\vec{S}}_{1}\cdot\vec{v}_{1} \vec{\tilde{S}}_{2}\cdot\vec{v}_{1} \vec{v}_{2}\cdot\vec{n}
	 + 4 \vec{\tilde{S}}_{1}\cdot\vec{v}_{1} \dot{\vec{S}}_{2}\cdot\vec{v}_{1} \vec{v}_{2}\cdot\vec{n}
	 + 11 \dot{\vec{S}}_{1}\cdot\vec{\tilde{S}}_{2} v_{1}^2 \vec{v}_{2}\cdot\vec{n} \nl
	 - 3 \vec{\tilde{S}}_{1}\cdot\dot{\vec{S}}_{2} v_{1}^2 \vec{v}_{2}\cdot\vec{n}
	 - 22 \vec{v}_{1}\cdot\vec{n} \dot{\vec{S}}_{2}\cdot\vec{v}_{1} \vec{\tilde{S}}_{1}\cdot\vec{v}_{2}
	 + 7 \dot{\vec{S}}_{2}\cdot\vec{n} v_{1}^2 \vec{\tilde{S}}_{1}\cdot\vec{v}_{2}
	 + 8 \dot{\vec{S}}_{2}\cdot\vec{v}_{1} \vec{v}_{2}\cdot\vec{n} \vec{\tilde{S}}_{1}\cdot\vec{v}_{2} \nl
	 - 8 \vec{v}_{1}\cdot\vec{n} \vec{\tilde{S}}_{2}\cdot\vec{v}_{1} \dot{\vec{S}}_{1}\cdot\vec{v}_{2}
	 + 3 \vec{\tilde{S}}_{2}\cdot\vec{n} v_{1}^2 \dot{\vec{S}}_{1}\cdot\vec{v}_{2}
	 + 22 \vec{\tilde{S}}_{2}\cdot\vec{v}_{1} \vec{v}_{2}\cdot\vec{n} \dot{\vec{S}}_{1}\cdot\vec{v}_{2}
	 + 3 \vec{v}_{1}\cdot\vec{n} \dot{\vec{S}}_{1}\cdot\vec{v}_{1} \vec{\tilde{S}}_{2}\cdot\vec{v}_{2} \nl
	 - 3 \dot{\vec{S}}_{1}\cdot\vec{n} v_{1}^2 \vec{\tilde{S}}_{2}\cdot\vec{v}_{2}
	 + 4 \dot{\vec{S}}_{1}\cdot\vec{v}_{1} \vec{v}_{2}\cdot\vec{n} \vec{\tilde{S}}_{2}\cdot\vec{v}_{2}
	 - 4 \vec{v}_{1}\cdot\vec{n} \dot{\vec{S}}_{1}\cdot\vec{v}_{2} \vec{\tilde{S}}_{2}\cdot\vec{v}_{2}
	 - 8 \vec{v}_{2}\cdot\vec{n} \dot{\vec{S}}_{1}\cdot\vec{v}_{2} \vec{\tilde{S}}_{2}\cdot\vec{v}_{2} \nl
	 - 4 \vec{v}_{1}\cdot\vec{n} \vec{\tilde{S}}_{1}\cdot\vec{v}_{1} \dot{\vec{S}}_{2}\cdot\vec{v}_{2}
	 - 3 \vec{\tilde{S}}_{1}\cdot\vec{n} v_{1}^2 \dot{\vec{S}}_{2}\cdot\vec{v}_{2}
	 - 3 \vec{\tilde{S}}_{1}\cdot\vec{v}_{1} \vec{v}_{2}\cdot\vec{n} \dot{\vec{S}}_{2}\cdot\vec{v}_{2}
	 + 11 \vec{v}_{1}\cdot\vec{n} \vec{\tilde{S}}_{1}\cdot\vec{v}_{2} \dot{\vec{S}}_{2}\cdot\vec{v}_{2} \nl
	 + 4 \dot{\vec{S}}_{1}\cdot\vec{\tilde{S}}_{2} \vec{v}_{1}\cdot\vec{n} \vec{v}_{1}\cdot\vec{v}_{2}
	 + 26 \vec{\tilde{S}}_{1}\cdot\dot{\vec{S}}_{2} \vec{v}_{1}\cdot\vec{n} \vec{v}_{1}\cdot\vec{v}_{2}
	 - 8 \dot{\vec{S}}_{2}\cdot\vec{n} \vec{\tilde{S}}_{1}\cdot\vec{v}_{1} \vec{v}_{1}\cdot\vec{v}_{2}
	 - 3 \vec{\tilde{S}}_{2}\cdot\vec{n} \dot{\vec{S}}_{1}\cdot\vec{v}_{1} \vec{v}_{1}\cdot\vec{v}_{2} \nl
	 + 4 \dot{\vec{S}}_{1}\cdot\vec{n} \vec{\tilde{S}}_{2}\cdot\vec{v}_{1} \vec{v}_{1}\cdot\vec{v}_{2}
	 + 2 \vec{\tilde{S}}_{1}\cdot\vec{n} \dot{\vec{S}}_{2}\cdot\vec{v}_{1} \vec{v}_{1}\cdot\vec{v}_{2}
	 - 26 \dot{\vec{S}}_{1}\cdot\vec{\tilde{S}}_{2} \vec{v}_{2}\cdot\vec{n} \vec{v}_{1}\cdot\vec{v}_{2}
	 - 4 \vec{\tilde{S}}_{1}\cdot\dot{\vec{S}}_{2} \vec{v}_{2}\cdot\vec{n} \vec{v}_{1}\cdot\vec{v}_{2} \nl
	 - 4 \dot{\vec{S}}_{2}\cdot\vec{n} \vec{\tilde{S}}_{1}\cdot\vec{v}_{2} \vec{v}_{1}\cdot\vec{v}_{2}
	 - 2 \vec{\tilde{S}}_{2}\cdot\vec{n} \dot{\vec{S}}_{1}\cdot\vec{v}_{2} \vec{v}_{1}\cdot\vec{v}_{2}
	 + 8 \dot{\vec{S}}_{1}\cdot\vec{n} \vec{\tilde{S}}_{2}\cdot\vec{v}_{2} \vec{v}_{1}\cdot\vec{v}_{2}
	 + 3 \vec{\tilde{S}}_{1}\cdot\vec{n} \dot{\vec{S}}_{2}\cdot\vec{v}_{2} \vec{v}_{1}\cdot\vec{v}_{2} \nl
	 + 3 \dot{\vec{S}}_{1}\cdot\vec{\tilde{S}}_{2} \vec{v}_{1}\cdot\vec{n} v_{2}^2
	 - 11 \vec{\tilde{S}}_{1}\cdot\dot{\vec{S}}_{2} \vec{v}_{1}\cdot\vec{n} v_{2}^2
	 + 3 \dot{\vec{S}}_{2}\cdot\vec{n} \vec{\tilde{S}}_{1}\cdot\vec{v}_{1} v_{2}^2
	 + 3 \vec{\tilde{S}}_{2}\cdot\vec{n} \dot{\vec{S}}_{1}\cdot\vec{v}_{1} v_{2}^2
	 - 7 \dot{\vec{S}}_{1}\cdot\vec{n} \vec{\tilde{S}}_{2}\cdot\vec{v}_{1} v_{2}^2 \nl
	 - 3 \vec{\tilde{S}}_{1}\cdot\vec{n} \dot{\vec{S}}_{2}\cdot\vec{v}_{1} v_{2}^2
	 + 9 \dot{\vec{S}}_{1}\cdot\vec{\tilde{S}}_{2} \vec{v}_{2}\cdot\vec{n} v_{2}^2
	 - 4 \vec{\tilde{S}}_{1}\cdot\dot{\vec{S}}_{2} \vec{v}_{2}\cdot\vec{n} v_{2}^2
	 + 4 \dot{\vec{S}}_{2}\cdot\vec{n} \vec{\tilde{S}}_{1}\cdot\vec{v}_{2} v_{2}^2
	 - \vec{\tilde{S}}_{2}\cdot\vec{n} \dot{\vec{S}}_{1}\cdot\vec{v}_{2} v_{2}^2 \nl
	 + 9 \vec{\tilde{S}}_{1}\cdot\vec{n} \dot{\vec{S}}_{2}\cdot\vec{n} \vec{v}_{1}\cdot\vec{n} v_{1}^2
	 + 12 \dot{\vec{S}}_{2}\cdot\vec{n} \vec{v}_{1}\cdot\vec{n} \vec{\tilde{S}}_{1}\cdot\vec{v}_{1} \vec{v}_{2}\cdot\vec{n}
	 + 9 \vec{\tilde{S}}_{2}\cdot\vec{n} \vec{v}_{1}\cdot\vec{n} \dot{\vec{S}}_{1}\cdot\vec{v}_{1} \vec{v}_{2}\cdot\vec{n} \nl
	 + 24 \dot{\vec{S}}_{1}\cdot\vec{n} \vec{v}_{1}\cdot\vec{n} \vec{\tilde{S}}_{2}\cdot\vec{v}_{1} \vec{v}_{2}\cdot\vec{n}
	 + 18 \vec{\tilde{S}}_{1}\cdot\vec{n} \vec{v}_{1}\cdot\vec{n} \dot{\vec{S}}_{2}\cdot\vec{v}_{1} \vec{v}_{2}\cdot\vec{n}
	 - 9 \dot{\vec{S}}_{1}\cdot\vec{n} \vec{\tilde{S}}_{2}\cdot\vec{n} v_{1}^2 \vec{v}_{2}\cdot\vec{n} \nl
	 + 3 \vec{\tilde{S}}_{1}\cdot\vec{n} \dot{\vec{S}}_{2}\cdot\vec{n} v_{1}^2 \vec{v}_{2}\cdot\vec{n}
	 - 24 \dot{\vec{S}}_{2}\cdot\vec{n} \vec{v}_{1}\cdot\vec{n} \vec{v}_{2}\cdot\vec{n} \vec{\tilde{S}}_{1}\cdot\vec{v}_{2}
	 - 18 \vec{\tilde{S}}_{2}\cdot\vec{n} \vec{v}_{1}\cdot\vec{n} \vec{v}_{2}\cdot\vec{n} \dot{\vec{S}}_{1}\cdot\vec{v}_{2} \nl
	 - 12 \dot{\vec{S}}_{1}\cdot\vec{n} \vec{v}_{1}\cdot\vec{n} \vec{v}_{2}\cdot\vec{n} \vec{\tilde{S}}_{2}\cdot\vec{v}_{2}
	 - 9 \vec{\tilde{S}}_{1}\cdot\vec{n} \vec{v}_{1}\cdot\vec{n} \vec{v}_{2}\cdot\vec{n} \dot{\vec{S}}_{2}\cdot\vec{v}_{2}
	 - 18 \vec{\tilde{S}}_{1}\cdot\vec{n} \dot{\vec{S}}_{2}\cdot\vec{n} \vec{v}_{1}\cdot\vec{n} \vec{v}_{1}\cdot\vec{v}_{2} \nl
	 + 18 \dot{\vec{S}}_{1}\cdot\vec{n} \vec{\tilde{S}}_{2}\cdot\vec{n} \vec{v}_{2}\cdot\vec{n} \vec{v}_{1}\cdot\vec{v}_{2}
	 - 3 \dot{\vec{S}}_{1}\cdot\vec{n} \vec{\tilde{S}}_{2}\cdot\vec{n} \vec{v}_{1}\cdot\vec{n} v_{2}^2
	 + 9 \vec{\tilde{S}}_{1}\cdot\vec{n} \dot{\vec{S}}_{2}\cdot\vec{n} \vec{v}_{1}\cdot\vec{n} v_{2}^2 \nl
	 - 9 \dot{\vec{S}}_{1}\cdot\vec{n} \vec{\tilde{S}}_{2}\cdot\vec{n} \vec{v}_{2}\cdot\vec{n} v_{2}^2
	 + 12 \vec{\tilde{S}}_{1}\cdot\dot{\vec{S}}_{2} ( \vec{v}_{1}\cdot\vec{n} )^{3}
	 - 9 \dot{\vec{S}}_{2}\cdot\vec{n} \vec{\tilde{S}}_{1}\cdot\vec{v}_{1} ( \vec{v}_{1}\cdot\vec{n} )^{2}
	 - 12 \vec{\tilde{S}}_{1}\cdot\vec{n} \dot{\vec{S}}_{2}\cdot\vec{v}_{1} ( \vec{v}_{1}\cdot\vec{n} )^{2} \nl
	 - 24 \dot{\vec{S}}_{1}\cdot\vec{\tilde{S}}_{2} \vec{v}_{2}\cdot\vec{n} ( \vec{v}_{1}\cdot\vec{n} )^{2}
	 - 27 \vec{\tilde{S}}_{1}\cdot\dot{\vec{S}}_{2} \vec{v}_{2}\cdot\vec{n} ( \vec{v}_{1}\cdot\vec{n} )^{2}
	 + 15 \dot{\vec{S}}_{2}\cdot\vec{n} \vec{\tilde{S}}_{1}\cdot\vec{v}_{2} ( \vec{v}_{1}\cdot\vec{n} )^{2} \nl
	 + 9 \vec{\tilde{S}}_{1}\cdot\vec{n} \dot{\vec{S}}_{2}\cdot\vec{v}_{2} ( \vec{v}_{1}\cdot\vec{n} )^{2}
	 + 27 \dot{\vec{S}}_{1}\cdot\vec{\tilde{S}}_{2} \vec{v}_{1}\cdot\vec{n} ( \vec{v}_{2}\cdot\vec{n} )^{2}
	 + 24 \vec{\tilde{S}}_{1}\cdot\dot{\vec{S}}_{2} \vec{v}_{1}\cdot\vec{n} ( \vec{v}_{2}\cdot\vec{n} )^{2} \nl
	 - 9 \vec{\tilde{S}}_{2}\cdot\vec{n} \dot{\vec{S}}_{1}\cdot\vec{v}_{1} ( \vec{v}_{2}\cdot\vec{n} )^{2}
	 - 15 \dot{\vec{S}}_{1}\cdot\vec{n} \vec{\tilde{S}}_{2}\cdot\vec{v}_{1} ( \vec{v}_{2}\cdot\vec{n} )^{2}
	 - 12 \dot{\vec{S}}_{1}\cdot\vec{\tilde{S}}_{2} ( \vec{v}_{2}\cdot\vec{n} )^{3}
	 + 12 \vec{\tilde{S}}_{2}\cdot\vec{n} \dot{\vec{S}}_{1}\cdot\vec{v}_{2} ( \vec{v}_{2}\cdot\vec{n} )^{2} \nl
	 + 9 \dot{\vec{S}}_{1}\cdot\vec{n} \vec{\tilde{S}}_{2}\cdot\vec{v}_{2} ( \vec{v}_{2}\cdot\vec{n} )^{2} -15 \vec{\tilde{S}}_{1}\cdot\vec{n} \dot{\vec{S}}_{2}\cdot\vec{n} \vec{v}_{2}\cdot\vec{n} ( \vec{v}_{1}\cdot\vec{n} )^{2}
	 + 15 \dot{\vec{S}}_{1}\cdot\vec{n} \vec{\tilde{S}}_{2}\cdot\vec{n} \vec{v}_{1}\cdot\vec{n} ( \vec{v}_{2}\cdot\vec{n} )^{2} \Big] \nl
+ \frac{G^2 m_{1}}{r^3} \Big[ 6 \dot{\vec{S}}_{1}\cdot\vec{\tilde{S}}_{2} \vec{v}_{1}\cdot\vec{n}
	 - \vec{\tilde{S}}_{1}\cdot\dot{\vec{S}}_{2} \vec{v}_{1}\cdot\vec{n}
	 + \dot{\vec{S}}_{2}\cdot\vec{n} \vec{\tilde{S}}_{1}\cdot\vec{v}_{1}
	 + 4 \vec{\tilde{S}}_{2}\cdot\vec{n} \dot{\vec{S}}_{1}\cdot\vec{v}_{1}
	 + 6 \dot{\vec{S}}_{1}\cdot\vec{n} \vec{\tilde{S}}_{2}\cdot\vec{v}_{1} \nl
	 + 2 \vec{\tilde{S}}_{1}\cdot\vec{n} \dot{\vec{S}}_{2}\cdot\vec{v}_{1}
	 - 4 \dot{\vec{S}}_{1}\cdot\vec{\tilde{S}}_{2} \vec{v}_{2}\cdot\vec{n}
	 + 2 \vec{\tilde{S}}_{1}\cdot\dot{\vec{S}}_{2} \vec{v}_{2}\cdot\vec{n}
	 - 2 \dot{\vec{S}}_{2}\cdot\vec{n} \vec{\tilde{S}}_{1}\cdot\vec{v}_{2}
	 - 4 \vec{\tilde{S}}_{2}\cdot\vec{n} \dot{\vec{S}}_{1}\cdot\vec{v}_{2} \nl
	 - 4 \dot{\vec{S}}_{1}\cdot\vec{n} \vec{\tilde{S}}_{2}\cdot\vec{v}_{2} -24 \dot{\vec{S}}_{1}\cdot\vec{n} \vec{\tilde{S}}_{2}\cdot\vec{n} \vec{v}_{1}\cdot\vec{n}
	 + 16 \dot{\vec{S}}_{1}\cdot\vec{n} \vec{\tilde{S}}_{2}\cdot\vec{n} \vec{v}_{2}\cdot\vec{n} \Big] \nl
- \frac{G^2 m_{2}}{r^3} \Big[ 2 \dot{\vec{S}}_{1}\cdot\vec{\tilde{S}}_{2} \vec{v}_{1}\cdot\vec{n}
	 - 4 \vec{\tilde{S}}_{1}\cdot\dot{\vec{S}}_{2} \vec{v}_{1}\cdot\vec{n}
	 - 4 \dot{\vec{S}}_{2}\cdot\vec{n} \vec{\tilde{S}}_{1}\cdot\vec{v}_{1}
	 - 2 \dot{\vec{S}}_{1}\cdot\vec{n} \vec{\tilde{S}}_{2}\cdot\vec{v}_{1}
	 - 4 \vec{\tilde{S}}_{1}\cdot\vec{n} \dot{\vec{S}}_{2}\cdot\vec{v}_{1} \nl
	 - \dot{\vec{S}}_{1}\cdot\vec{\tilde{S}}_{2} \vec{v}_{2}\cdot\vec{n}
	 + 6 \vec{\tilde{S}}_{1}\cdot\dot{\vec{S}}_{2} \vec{v}_{2}\cdot\vec{n}
	 + 6 \dot{\vec{S}}_{2}\cdot\vec{n} \vec{\tilde{S}}_{1}\cdot\vec{v}_{2}
	 + 2 \vec{\tilde{S}}_{2}\cdot\vec{n} \dot{\vec{S}}_{1}\cdot\vec{v}_{2}
	 + \dot{\vec{S}}_{1}\cdot\vec{n} \vec{\tilde{S}}_{2}\cdot\vec{v}_{2}
	 + 4 \vec{\tilde{S}}_{1}\cdot\vec{n} \dot{\vec{S}}_{2}\cdot\vec{v}_{2} \nl
	 + 16 \vec{\tilde{S}}_{1}\cdot\vec{n} \dot{\vec{S}}_{2}\cdot\vec{n} \vec{v}_{1}\cdot\vec{n}
	 - 24 \vec{\tilde{S}}_{1}\cdot\vec{n} \dot{\vec{S}}_{2}\cdot\vec{n} \vec{v}_{2}\cdot\vec{n} \Big] ,
\end{align}
\begin{align}
& \stackrel{(2)}{V} = - \frac{G}{8 r} \Big[ 8 \big( \vec{\tilde{S}}_{2}\cdot\dot{\vec{a}}_{1} \vec{\tilde{S}}_{1}\cdot\vec{v}_{2}
	 - \vec{\tilde{S}}_{1}\cdot\vec{\tilde{S}}_{2} \dot{\vec{a}}_{1}\cdot\vec{v}_{2}
	 + \vec{\tilde{S}}_{2}\cdot\vec{v}_{1} \vec{\tilde{S}}_{1}\cdot\dot{\vec{a}}_{2}
	 - \vec{\tilde{S}}_{1}\cdot\vec{\tilde{S}}_{2} \vec{v}_{1}\cdot\dot{\vec{a}}_{2}
	 + \vec{\tilde{S}}_{1}\cdot\vec{\tilde{S}}_{2} \dot{\vec{a}}_{1}\cdot\vec{n} \vec{v}_{2}\cdot\vec{n} \nl
	 - \vec{\tilde{S}}_{1}\cdot\vec{n} \vec{\tilde{S}}_{2}\cdot\dot{\vec{a}}_{1} \vec{v}_{2}\cdot\vec{n}
	 + \vec{\tilde{S}}_{1}\cdot\vec{\tilde{S}}_{2} \vec{v}_{1}\cdot\vec{n} \dot{\vec{a}}_{2}\cdot\vec{n}
	 - \vec{\tilde{S}}_{2}\cdot\vec{n} \vec{v}_{1}\cdot\vec{n} \vec{\tilde{S}}_{1}\cdot\dot{\vec{a}}_{2} \big)
	 - \big( 2 \vec{\tilde{S}}_{1}\cdot\vec{v}_{1} \ddot{\vec{S}}_{2}\cdot\vec{v}_{1}
	 - 3 \vec{\tilde{S}}_{1}\cdot\ddot{\vec{S}}_{2} v_{1}^2 \nl
	 - 8 \ddot{\vec{S}}_{2}\cdot\vec{v}_{1} \vec{\tilde{S}}_{1}\cdot\vec{v}_{2}
	 - 8 \vec{\tilde{S}}_{2}\cdot\vec{v}_{1} \ddot{\vec{S}}_{1}\cdot\vec{v}_{2}
	 + 2 \ddot{\vec{S}}_{1}\cdot\vec{v}_{2} \vec{\tilde{S}}_{2}\cdot\vec{v}_{2}
	 + 8 \ddot{\vec{S}}_{1}\cdot\vec{\tilde{S}}_{2} \vec{v}_{1}\cdot\vec{v}_{2}
	 + 8 \vec{\tilde{S}}_{1}\cdot\ddot{\vec{S}}_{2} \vec{v}_{1}\cdot\vec{v}_{2} \nl
	 - 3 \ddot{\vec{S}}_{1}\cdot\vec{\tilde{S}}_{2} v_{2}^2 -2 \ddot{\vec{S}}_{2}\cdot\vec{n} \vec{v}_{1}\cdot\vec{n} \vec{\tilde{S}}_{1}\cdot\vec{v}_{1}
	 - 2 \vec{\tilde{S}}_{1}\cdot\vec{n} \vec{v}_{1}\cdot\vec{n} \ddot{\vec{S}}_{2}\cdot\vec{v}_{1}
	 - \vec{\tilde{S}}_{1}\cdot\vec{n} \ddot{\vec{S}}_{2}\cdot\vec{n} v_{1}^2
	 - 8 \ddot{\vec{S}}_{1}\cdot\vec{\tilde{S}}_{2} \vec{v}_{1}\cdot\vec{n} \vec{v}_{2}\cdot\vec{n} \nl
	 - 8 \vec{\tilde{S}}_{1}\cdot\ddot{\vec{S}}_{2} \vec{v}_{1}\cdot\vec{n} \vec{v}_{2}\cdot\vec{n}
	 + 8 \ddot{\vec{S}}_{1}\cdot\vec{n} \vec{\tilde{S}}_{2}\cdot\vec{v}_{1} \vec{v}_{2}\cdot\vec{n}
	 + 8 \ddot{\vec{S}}_{2}\cdot\vec{n} \vec{v}_{1}\cdot\vec{n} \vec{\tilde{S}}_{1}\cdot\vec{v}_{2}
	 - 2 \vec{\tilde{S}}_{2}\cdot\vec{n} \vec{v}_{2}\cdot\vec{n} \ddot{\vec{S}}_{1}\cdot\vec{v}_{2} \nl
	 - 2 \ddot{\vec{S}}_{1}\cdot\vec{n} \vec{v}_{2}\cdot\vec{n} \vec{\tilde{S}}_{2}\cdot\vec{v}_{2}
	 - \ddot{\vec{S}}_{1}\cdot\vec{n} \vec{\tilde{S}}_{2}\cdot\vec{n} v_{2}^2
	 + 3 \vec{\tilde{S}}_{1}\cdot\ddot{\vec{S}}_{2} ( \vec{v}_{1}\cdot\vec{n} )^{2}
	 + 3 \ddot{\vec{S}}_{1}\cdot\vec{\tilde{S}}_{2} ( \vec{v}_{2}\cdot\vec{n} )^{2} \nl
	 + 3 \vec{\tilde{S}}_{1}\cdot\vec{n} \ddot{\vec{S}}_{2}\cdot\vec{n} ( \vec{v}_{1}\cdot\vec{n} )^{2}
	 + 3 \ddot{\vec{S}}_{1}\cdot\vec{n} \vec{\tilde{S}}_{2}\cdot\vec{n} ( \vec{v}_{2}\cdot\vec{n} )^{2} \big)
	 + \big( 33 \vec{\tilde{S}}_{2}\cdot\vec{a}_{1} \vec{\tilde{S}}_{1}\cdot\vec{a}_{2}
	 + \vec{\tilde{S}}_{1}\cdot\vec{a}_{1} \vec{\tilde{S}}_{2}\cdot\vec{a}_{2} \nl
	 - 31 \vec{\tilde{S}}_{1}\cdot\vec{\tilde{S}}_{2} \vec{a}_{1}\cdot\vec{a}_{2}
	 + 15 \vec{\tilde{S}}_{1}\cdot\vec{\tilde{S}}_{2} \vec{a}_{1}\cdot\vec{n} \vec{a}_{2}\cdot\vec{n}
	 - 5 \vec{\tilde{S}}_{2}\cdot\vec{n} \vec{\tilde{S}}_{1}\cdot\vec{a}_{1} \vec{a}_{2}\cdot\vec{n}
	 - 9 \vec{\tilde{S}}_{1}\cdot\vec{n} \vec{\tilde{S}}_{2}\cdot\vec{a}_{1} \vec{a}_{2}\cdot\vec{n} \nl
	 - 9 \vec{\tilde{S}}_{2}\cdot\vec{n} \vec{a}_{1}\cdot\vec{n} \vec{\tilde{S}}_{1}\cdot\vec{a}_{2}
	 - 5 \vec{\tilde{S}}_{1}\cdot\vec{n} \vec{a}_{1}\cdot\vec{n} \vec{\tilde{S}}_{2}\cdot\vec{a}_{2}
	 + 15 \vec{\tilde{S}}_{1}\cdot\vec{n} \vec{\tilde{S}}_{2}\cdot\vec{n} \vec{a}_{1}\cdot\vec{a}_{2}
	 + 3 \vec{\tilde{S}}_{1}\cdot\vec{n} \vec{\tilde{S}}_{2}\cdot\vec{n} \vec{a}_{1}\cdot\vec{n} \vec{a}_{2}\cdot\vec{n} \big) \nl
	 - 2 \big( \dot{\vec{S}}_{2}\cdot\vec{v}_{1} \vec{\tilde{S}}_{1}\cdot\vec{a}_{1}
	 + 5 \vec{\tilde{S}}_{1}\cdot\vec{v}_{1} \dot{\vec{S}}_{2}\cdot\vec{a}_{1}
	 - 6 \vec{\tilde{S}}_{1}\cdot\dot{\vec{S}}_{2} \vec{v}_{1}\cdot\vec{a}_{1}
	 - 17 \dot{\vec{S}}_{2}\cdot\vec{a}_{1} \vec{\tilde{S}}_{1}\cdot\vec{v}_{2}
	 - 8 \vec{\tilde{S}}_{2}\cdot\vec{a}_{1} \dot{\vec{S}}_{1}\cdot\vec{v}_{2} \nl
	 - \vec{\tilde{S}}_{1}\cdot\vec{a}_{1} \dot{\vec{S}}_{2}\cdot\vec{v}_{2}
	 + 8 \dot{\vec{S}}_{1}\cdot\vec{\tilde{S}}_{2} \vec{a}_{1}\cdot\vec{v}_{2}
	 + 17 \vec{\tilde{S}}_{1}\cdot\dot{\vec{S}}_{2} \vec{a}_{1}\cdot\vec{v}_{2}
	 - 8 \dot{\vec{S}}_{2}\cdot\vec{v}_{1} \vec{\tilde{S}}_{1}\cdot\vec{a}_{2}
	 - 17 \vec{\tilde{S}}_{2}\cdot\vec{v}_{1} \dot{\vec{S}}_{1}\cdot\vec{a}_{2} \nl
	 + 5 \vec{\tilde{S}}_{2}\cdot\vec{v}_{2} \dot{\vec{S}}_{1}\cdot\vec{a}_{2}
	 - \dot{\vec{S}}_{1}\cdot\vec{v}_{1} \vec{\tilde{S}}_{2}\cdot\vec{a}_{2}
	 + \dot{\vec{S}}_{1}\cdot\vec{v}_{2} \vec{\tilde{S}}_{2}\cdot\vec{a}_{2}
	 + 17 \dot{\vec{S}}_{1}\cdot\vec{\tilde{S}}_{2} \vec{v}_{1}\cdot\vec{a}_{2}
	 + 8 \vec{\tilde{S}}_{1}\cdot\dot{\vec{S}}_{2} \vec{v}_{1}\cdot\vec{a}_{2} \nl
	 - 6 \dot{\vec{S}}_{1}\cdot\vec{\tilde{S}}_{2} \vec{v}_{2}\cdot\vec{a}_{2}
	 + 12 \vec{\tilde{S}}_{1}\cdot\dot{\vec{S}}_{2} \vec{v}_{1}\cdot\vec{n} \vec{a}_{1}\cdot\vec{n}
	 - 3 \dot{\vec{S}}_{2}\cdot\vec{n} \vec{\tilde{S}}_{1}\cdot\vec{v}_{1} \vec{a}_{1}\cdot\vec{n}
	 - 4 \vec{\tilde{S}}_{1}\cdot\vec{n} \dot{\vec{S}}_{2}\cdot\vec{v}_{1} \vec{a}_{1}\cdot\vec{n} \nl
	 - 3 \dot{\vec{S}}_{2}\cdot\vec{n} \vec{v}_{1}\cdot\vec{n} \vec{\tilde{S}}_{1}\cdot\vec{a}_{1}
	 - 8 \vec{\tilde{S}}_{1}\cdot\vec{n} \vec{v}_{1}\cdot\vec{n} \dot{\vec{S}}_{2}\cdot\vec{a}_{1}
	 + 6 \vec{\tilde{S}}_{1}\cdot\vec{n} \dot{\vec{S}}_{2}\cdot\vec{n} \vec{v}_{1}\cdot\vec{a}_{1}
	 - 8 \dot{\vec{S}}_{1}\cdot\vec{\tilde{S}}_{2} \vec{a}_{1}\cdot\vec{n} \vec{v}_{2}\cdot\vec{n} \nl
	 - 9 \vec{\tilde{S}}_{1}\cdot\dot{\vec{S}}_{2} \vec{a}_{1}\cdot\vec{n} \vec{v}_{2}\cdot\vec{n}
	 + 3 \dot{\vec{S}}_{2}\cdot\vec{n} \vec{\tilde{S}}_{1}\cdot\vec{a}_{1} \vec{v}_{2}\cdot\vec{n}
	 + 8 \dot{\vec{S}}_{1}\cdot\vec{n} \vec{\tilde{S}}_{2}\cdot\vec{a}_{1} \vec{v}_{2}\cdot\vec{n}
	 + 5 \vec{\tilde{S}}_{1}\cdot\vec{n} \dot{\vec{S}}_{2}\cdot\vec{a}_{1} \vec{v}_{2}\cdot\vec{n} \nl
	 + 5 \dot{\vec{S}}_{2}\cdot\vec{n} \vec{a}_{1}\cdot\vec{n} \vec{\tilde{S}}_{1}\cdot\vec{v}_{2}
	 + 3 \vec{\tilde{S}}_{1}\cdot\vec{n} \vec{a}_{1}\cdot\vec{n} \dot{\vec{S}}_{2}\cdot\vec{v}_{2}
	 - 7 \vec{\tilde{S}}_{1}\cdot\vec{n} \dot{\vec{S}}_{2}\cdot\vec{n} \vec{a}_{1}\cdot\vec{v}_{2}
	 - 9 \dot{\vec{S}}_{1}\cdot\vec{\tilde{S}}_{2} \vec{v}_{1}\cdot\vec{n} \vec{a}_{2}\cdot\vec{n} \nl
	 - 8 \vec{\tilde{S}}_{1}\cdot\dot{\vec{S}}_{2} \vec{v}_{1}\cdot\vec{n} \vec{a}_{2}\cdot\vec{n}
	 + 3 \vec{\tilde{S}}_{2}\cdot\vec{n} \dot{\vec{S}}_{1}\cdot\vec{v}_{1} \vec{a}_{2}\cdot\vec{n}
	 + 5 \dot{\vec{S}}_{1}\cdot\vec{n} \vec{\tilde{S}}_{2}\cdot\vec{v}_{1} \vec{a}_{2}\cdot\vec{n}
	 + 12 \dot{\vec{S}}_{1}\cdot\vec{\tilde{S}}_{2} \vec{v}_{2}\cdot\vec{n} \vec{a}_{2}\cdot\vec{n} \nl
	 - 4 \vec{\tilde{S}}_{2}\cdot\vec{n} \dot{\vec{S}}_{1}\cdot\vec{v}_{2} \vec{a}_{2}\cdot\vec{n}
	 - 3 \dot{\vec{S}}_{1}\cdot\vec{n} \vec{\tilde{S}}_{2}\cdot\vec{v}_{2} \vec{a}_{2}\cdot\vec{n}
	 + 8 \dot{\vec{S}}_{2}\cdot\vec{n} \vec{v}_{1}\cdot\vec{n} \vec{\tilde{S}}_{1}\cdot\vec{a}_{2}
	 + 5 \vec{\tilde{S}}_{2}\cdot\vec{n} \vec{v}_{1}\cdot\vec{n} \dot{\vec{S}}_{1}\cdot\vec{a}_{2} \nl
	 - 8 \vec{\tilde{S}}_{2}\cdot\vec{n} \vec{v}_{2}\cdot\vec{n} \dot{\vec{S}}_{1}\cdot\vec{a}_{2}
	 + 3 \dot{\vec{S}}_{1}\cdot\vec{n} \vec{v}_{1}\cdot\vec{n} \vec{\tilde{S}}_{2}\cdot\vec{a}_{2}
	 - 3 \dot{\vec{S}}_{1}\cdot\vec{n} \vec{v}_{2}\cdot\vec{n} \vec{\tilde{S}}_{2}\cdot\vec{a}_{2}
	 - 7 \dot{\vec{S}}_{1}\cdot\vec{n} \vec{\tilde{S}}_{2}\cdot\vec{n} \vec{v}_{1}\cdot\vec{a}_{2} \nl
	 + 6 \dot{\vec{S}}_{1}\cdot\vec{n} \vec{\tilde{S}}_{2}\cdot\vec{n} \vec{v}_{2}\cdot\vec{a}_{2} -3 \vec{\tilde{S}}_{1}\cdot\vec{n} \dot{\vec{S}}_{2}\cdot\vec{n} \vec{a}_{1}\cdot\vec{n} \vec{v}_{2}\cdot\vec{n}
	 - 3 \dot{\vec{S}}_{1}\cdot\vec{n} \vec{\tilde{S}}_{2}\cdot\vec{n} \vec{v}_{1}\cdot\vec{n} \vec{a}_{2}\cdot\vec{n} \big)
	 - 2 \big( 5 \dot{\vec{S}}_{1}\cdot\vec{v}_{1} \dot{\vec{S}}_{2}\cdot\vec{v}_{1} \nl
	 - 5 \dot{\vec{S}}_{1}\cdot\dot{\vec{S}}_{2} v_{1}^2
	 - 18 \dot{\vec{S}}_{2}\cdot\vec{v}_{1} \dot{\vec{S}}_{1}\cdot\vec{v}_{2}
	 - 2 \dot{\vec{S}}_{1}\cdot\vec{v}_{1} \dot{\vec{S}}_{2}\cdot\vec{v}_{2}
	 + 5 \dot{\vec{S}}_{1}\cdot\vec{v}_{2} \dot{\vec{S}}_{2}\cdot\vec{v}_{2}
	 + 20 \dot{\vec{S}}_{1}\cdot\dot{\vec{S}}_{2} \vec{v}_{1}\cdot\vec{v}_{2} \nl
	 - 5 \dot{\vec{S}}_{1}\cdot\dot{\vec{S}}_{2} v_{2}^2 -3 \dot{\vec{S}}_{2}\cdot\vec{n} \vec{v}_{1}\cdot\vec{n} \dot{\vec{S}}_{1}\cdot\vec{v}_{1}
	 - 8 \dot{\vec{S}}_{1}\cdot\vec{n} \vec{v}_{1}\cdot\vec{n} \dot{\vec{S}}_{2}\cdot\vec{v}_{1}
	 + 3 \dot{\vec{S}}_{1}\cdot\vec{n} \dot{\vec{S}}_{2}\cdot\vec{n} v_{1}^2 \nl
	 - 12 \dot{\vec{S}}_{1}\cdot\dot{\vec{S}}_{2} \vec{v}_{1}\cdot\vec{n} \vec{v}_{2}\cdot\vec{n}
	 + 4 \dot{\vec{S}}_{2}\cdot\vec{n} \dot{\vec{S}}_{1}\cdot\vec{v}_{1} \vec{v}_{2}\cdot\vec{n}
	 + 6 \dot{\vec{S}}_{1}\cdot\vec{n} \dot{\vec{S}}_{2}\cdot\vec{v}_{1} \vec{v}_{2}\cdot\vec{n}
	 + 6 \dot{\vec{S}}_{2}\cdot\vec{n} \vec{v}_{1}\cdot\vec{n} \dot{\vec{S}}_{1}\cdot\vec{v}_{2} \nl
	 - 8 \dot{\vec{S}}_{2}\cdot\vec{n} \vec{v}_{2}\cdot\vec{n} \dot{\vec{S}}_{1}\cdot\vec{v}_{2}
	 + 4 \dot{\vec{S}}_{1}\cdot\vec{n} \vec{v}_{1}\cdot\vec{n} \dot{\vec{S}}_{2}\cdot\vec{v}_{2}
	 - 3 \dot{\vec{S}}_{1}\cdot\vec{n} \vec{v}_{2}\cdot\vec{n} \dot{\vec{S}}_{2}\cdot\vec{v}_{2}
	 - 6 \dot{\vec{S}}_{1}\cdot\vec{n} \dot{\vec{S}}_{2}\cdot\vec{n} \vec{v}_{1}\cdot\vec{v}_{2} \nl
	 + 3 \dot{\vec{S}}_{1}\cdot\vec{n} \dot{\vec{S}}_{2}\cdot\vec{n} v_{2}^2
	 + 8 \dot{\vec{S}}_{1}\cdot\dot{\vec{S}}_{2} ( \vec{v}_{1}\cdot\vec{n} )^{2}
	 + 8 \dot{\vec{S}}_{1}\cdot\dot{\vec{S}}_{2} ( \vec{v}_{2}\cdot\vec{n} )^{2} -6 \dot{\vec{S}}_{1}\cdot\vec{n} \dot{\vec{S}}_{2}\cdot\vec{n} \vec{v}_{1}\cdot\vec{n} \vec{v}_{2}\cdot\vec{n} \big) \Big] \nl
- \frac{G^2 m_{1}}{r^2} \Big[ 2 \dot{\vec{S}}_{1}\cdot\dot{\vec{S}}_{2} -5 \dot{\vec{S}}_{1}\cdot\vec{n} \dot{\vec{S}}_{2}\cdot\vec{n} \Big]
- \frac{G^2 m_{2}}{r^2} \Big[ 2 \dot{\vec{S}}_{1}\cdot\dot{\vec{S}}_{2} -5 \dot{\vec{S}}_{1}\cdot\vec{n} \dot{\vec{S}}_{2}\cdot\vec{n} \Big] ,
\end{align}
\begin{align}
& \stackrel{(3)}{V} = - \frac{1}{8} G \Big[ 8 \big( \vec{\tilde{S}}_{1}\cdot\dot{\vec{S}}_{2} \dot{\vec{a}}_{1}\cdot\vec{n}
	 - \vec{\tilde{S}}_{1}\cdot\vec{n} \dot{\vec{S}}_{2}\cdot\dot{\vec{a}}_{1}
	 - \dot{\vec{S}}_{1}\cdot\vec{\tilde{S}}_{2} \dot{\vec{a}}_{2}\cdot\vec{n}
	 + \vec{\tilde{S}}_{2}\cdot\vec{n} \dot{\vec{S}}_{1}\cdot\dot{\vec{a}}_{2} \big)
	 + \big( 3 \vec{\tilde{S}}_{1}\cdot\ddot{\vec{S}}_{2} \vec{a}_{1}\cdot\vec{n} \nl
	 - \ddot{\vec{S}}_{2}\cdot\vec{n} \vec{\tilde{S}}_{1}\cdot\vec{a}_{1}
	 - \vec{\tilde{S}}_{1}\cdot\vec{n} \ddot{\vec{S}}_{2}\cdot\vec{a}_{1}
	 - 3 \ddot{\vec{S}}_{1}\cdot\vec{\tilde{S}}_{2} \vec{a}_{2}\cdot\vec{n}
	 + \vec{\tilde{S}}_{2}\cdot\vec{n} \ddot{\vec{S}}_{1}\cdot\vec{a}_{2}
	 + \ddot{\vec{S}}_{1}\cdot\vec{n} \vec{\tilde{S}}_{2}\cdot\vec{a}_{2} \nl
	 + \vec{\tilde{S}}_{1}\cdot\vec{n} \ddot{\vec{S}}_{2}\cdot\vec{n} \vec{a}_{1}\cdot\vec{n}
	 - \ddot{\vec{S}}_{1}\cdot\vec{n} \vec{\tilde{S}}_{2}\cdot\vec{n} \vec{a}_{2}\cdot\vec{n} \big)
	 + 2 \big( 4 \ddot{\vec{S}}_{1}\cdot\dot{\vec{S}}_{2} \vec{v}_{1}\cdot\vec{n}
	 + 3 \dot{\vec{S}}_{1}\cdot\ddot{\vec{S}}_{2} \vec{v}_{1}\cdot\vec{n}
	 - \ddot{\vec{S}}_{2}\cdot\vec{n} \dot{\vec{S}}_{1}\cdot\vec{v}_{1} \nl
	 - 4 \ddot{\vec{S}}_{1}\cdot\vec{n} \dot{\vec{S}}_{2}\cdot\vec{v}_{1}
	 - \dot{\vec{S}}_{1}\cdot\vec{n} \ddot{\vec{S}}_{2}\cdot\vec{v}_{1}
	 - 3 \ddot{\vec{S}}_{1}\cdot\dot{\vec{S}}_{2} \vec{v}_{2}\cdot\vec{n}
	 - 4 \dot{\vec{S}}_{1}\cdot\ddot{\vec{S}}_{2} \vec{v}_{2}\cdot\vec{n}
	 + 4 \ddot{\vec{S}}_{2}\cdot\vec{n} \dot{\vec{S}}_{1}\cdot\vec{v}_{2}
	 + \dot{\vec{S}}_{2}\cdot\vec{n} \ddot{\vec{S}}_{1}\cdot\vec{v}_{2} \nl
	 + \ddot{\vec{S}}_{1}\cdot\vec{n} \dot{\vec{S}}_{2}\cdot\vec{v}_{2}
	 + \dot{\vec{S}}_{1}\cdot\vec{n} \ddot{\vec{S}}_{2}\cdot\vec{n} \vec{v}_{1}\cdot\vec{n}
	 - \ddot{\vec{S}}_{1}\cdot\vec{n} \dot{\vec{S}}_{2}\cdot\vec{n} \vec{v}_{2}\cdot\vec{n} \big)
	 + 16 \big( \dot{\vec{S}}_{1}\cdot\dot{\vec{S}}_{2} \vec{a}_{1}\cdot\vec{n}
	 - \dot{\vec{S}}_{1}\cdot\vec{n} \dot{\vec{S}}_{2}\cdot\vec{a}_{1} \nl
	 - \dot{\vec{S}}_{1}\cdot\dot{\vec{S}}_{2} \vec{a}_{2}\cdot\vec{n}
	 + \dot{\vec{S}}_{2}\cdot\vec{n} \dot{\vec{S}}_{1}\cdot\vec{a}_{2} \big) \Big] ,
\end{align}
\begin{align}
& \stackrel{(4)}{V} = - \frac{1}{8} G r \Big[ 3 \ddot{\vec{S}}_{1}\cdot\ddot{\vec{S}}_{2}
	 - \ddot{\vec{S}}_{1}\cdot\vec{n} \ddot{\vec{S}}_{2}\cdot\vec{n} \Big] .
\end{align}
The Hamiltonian resulting from this potential is given by eq.~(\ref{HNNLOS1S2}).
It is canonically equivalent to the NNLO spin1-spin2 Hamiltonians in 
\cite{Hartung:2011ea, Levi:2014sba}.
The generator of the canonical transformation, which connects it to the ADM-gauge Hamiltonian
\cite{Hartung:2011ea} is given by eq.~(7.5) of \cite{Levi:2014sba}, 
with the coefficients being equal to
\begin{align}
g_1 &= 0, &
g_2 &= - \frac{7}{4}, &
g_3 &= \frac{9}{12}, &
g_4 &= 0, &
g_5 &= \frac{9}{4},&
g_6 &= 0, &
g_7 &= \frac{9}{4}, \nonumber \\
g_8 &= -2, &
g_9 &= \frac{1}{4}, &
g_{10} &= 0, &
g_{11} &= - \frac{9}{2}, &
g_{12} &= 0, & 
g_{13} &= \frac{1}{4}, &
g_{14} &= \frac{1}{2}, \nonumber \\
g_{15} &= - \frac{1}{4}, &
g_{16} &= 0, &
g_{17} &= -3, &
g_{18} &= 0, & 
g_{19} &= - \frac{3}{2}, &
g_{20} &= \frac{9}{4}, &
g_{21} &= 0, \nonumber \\
g_{22} &= 0, &
g_{23} &= 0, &
g_{24} &= - \frac{3}{4}, & 
g_{25} &= \frac{3}{4}, &
g_{26} &= \frac{5}{2}, &
g_{27} &= 0, &
g_{28} &= - \frac{3}{2}, \nonumber \\
g_{29} &= -2, & 
g_{30} &= - \frac{11}{4}, &
g_{31} &= -2, &
g_{32} &= - \frac{13}{4}, &
g_{33} &= 0, &
g_{34} &= 4 .
\end{align}

\bibliographystyle{JHEP}
\bibliography{gwbibtex}

\end{document}